\documentclass[12pt]{article}

\usepackage{color}

\usepackage{amsfonts}
\usepackage{amsmath}
\usepackage{amssymb}
\usepackage{amsthm}

\usepackage{amsmath}
\usepackage{amsfonts}

\def\vs{\vspace{5mm}}

\def\x{x}
\newtheorem{theorem}{Theorem}
\newtheorem{definition}{Definition}
\newtheorem{corollary}{Corollary}
\newtheorem{Proposition}{Proposition}
\newtheorem{lemma}{Lemma}

\newcommand\kk{\Delta_{\theta}}
\newcommand{\bm}[1]{\mbox{\boldmath $#1$}}
\newcommand\ov[1]{\overline{#1}}

\def\y{x}
\def\v{\beta}

\def\defi{\stackrel{\mbox{\tiny \bf def}}{=}}

\def\D{{\cal D}}
\def\L{{\cal L}}
\def\M{{\cal M}}
\def\C{{\cal C}}
\def\I{{\cal I}}
\def\F{{\cal F}}

\def\W{{\cal W}}

\def\VV{{\cal V}}
\def\HH{{\cal H}}

\def\nablahat{\widehat{\nabla}}
\def\J{J}
\def\j{j}

\def\Jbar{\overline{\J}}
\def\Pbar{\overline{P}}
\def\Rbar{\overline{R}}

\def\chibar{\overline{\chi}}

\def\gradZ{\mbox{grad } Z}
\def\hminus{H}
\def\hminusdown{\hat{H}}
\def\hepsilon{{\mathfrak{h}}}
\def\hepsilondown{\hat{\mathfrak{h}}}
\def\be{\begin{equation}}
\def\ee{\end{equation}}
\def\bea{\begin{eqnarray}}
\def\eea{\end{eqnarray}}
\def\bean{\begin{eqnarray*}}
\def\eean{\end{eqnarray*}}

\newcounter{mnotecount}[section]

\renewcommand{\themnotecount}{\thesection.\arabic{mnotecount}}
\newcommand{\mnote}[1]%{}
{\protect{\stepcounter{mnotecount}}$^{\mbox{\footnotesize
$%\!\!\!\!\!\!\,
\bullet$\themnotecount}}$ \marginpar{%\color{red}%
\raggedright\tiny\em
$\!\!\!\!\!\!\,\bullet$\themnotecount: #1} }

\begin{document}

\title{A spacetime characterization of the Kerr-NUT-(A)de Sitter and related metrics}
\author{Marc Mars$^{1}$ and Jos\'e M. M. Senovilla$^{2}$ \\
\\
$^{1}$ Instituto de F\'{\i}sica Fundamental y Matem\'aticas, Universidad de Salamanca\\
Plaza de la Merced s/n, 37008 Salamanca, Spain\\
marc@usal.es\\
\\
$^{2}$ F\'{\i}sica Te\'orica, Universidad del Pa\'{\i}s Vasco, \\
Apartado 644, 48080 Bilbao, Spain \\ 
josemm.senovilla@ehu.es}
\date{}
\maketitle

\begin{abstract} 
A characterization of the Kerr-NUT-(A)de Sitter metric among
four dimensional $\Lambda$-vacuum spacetimes admitting a Killing vector $\xi$
is obtained in terms of the proportionality of the self-dual Weyl tensor
and a natural self-dual double two-form constructed from the Killing vector.
This result recovers and extends a previous characterization of the
Kerr and Kerr-NUT metrics \cite{Mars1}.  The method of proof
is based on (i) the presence
of a second Killing vector field which is built in terms of geometric information 
arising from the Killing vector $\xi$ exclusively, and (ii) the existence of 
an interesting underlying geometric structure involving
a  Riemannian submersion of a conformally related metric, both of which may be of
independent interest. Other related metrics can also be similarly characterized, in particular the $\Lambda<0$ ``black branes'' recently used in AdS/CFT correspondence to describe via holography the physics of Quark-Gluon plasma.
\end{abstract} 

%PACS:

\section{Introduction}

The Kerr-NUT-(A)de Sitter spacetimes is a class of solutions of
the Einstein field 
equations with cosmological constant $\Lambda$ (of any value) 
which includes the class of Kerr and Kerr-NUT spacetimes. 
These spacetimes are widely believed to play an important role as stationary
endpoints of self-gravitating collapsing systems. Although we are very far from
any rigorous result in this direction, substantial work has been made on this 
problem at the linear
level, specially in the case of vanishing cosmological constant where
there is  a large
body of literature concerning boundedness and decay properties
of linear waves propagating in a Kerr
background (see \cite{LecturesDafermos} and references therein). 
For the case of positive cosmological constant, it has been shown that
linear  waves in the Kerr-de Sitter background decay 
exponentially in time outside and including the event horizon (see
\cite{Dyatlov} and references therein).
In the case of negative cosmological constant, 
Holzegel
has proven boundedness  \cite{Holzegel2010}
of the Klein-Gordon equation in 
Kerr-Anti de Sitter backgrounds 
and Holzegel and Smulevici \cite{HolzegelSmulevici2011} 
have shown  logarithmic decay of the solutions.
This very slow 
decay suggests non-linear instability of these spacetimes,
and this behaviour has been
conjectured for all asymptotically Anti-de Sitter spacetimes in
\cite{HolzegelSmulevici2011} (see, however, \cite{Dias} for heuristic
arguments in favor of non-linear stability). All these considerations show the
physical relevance of the class of 
Kerr-NUT-(A)de Sitter spacetimes,  which makes it of interest 
to try and find a geometric characterization for them.

In the case of the Kerr metric (and also Kerr-NUT metric),
a spacetime characterization 
within the class of vacuum spacetimes admitting a Killing vector $\xi$ was found in
\cite{Mars1} motivated by previous results \cite{Simon}
in the quotient manifold
of the Killing field in the region where the Killing is timelike.
This characterization was local in nature (see \cite{Mars3} for a discussion of
this point) and involved the proportionality
of the self-dual Weyl tensor and a tensor with the same algebraic properties
constructed out of the Killing vector $\xi$. \textcolor{blue}{The characterization assumed implicitly a property never spelled out, namely, that the Killing vector was nowhere orthogonal to the 2-plane spanned by the two principal directions of the Weyl tensor, see \cite{MSnull} where this omission was corrected and the accurate result explained in detail. Of course, the omitted assumption is automatically satisfied if the Killing vector is timelike somewhere}. Nevertheless, the characterization of Kerr in \cite{Mars1,Mars2} has played an interesting role in trying to extend the uniqueness of stationary black holes from the analytic setting to the general case \cite{Ionescu1},
\cite{Ionescu2}. 
It has also been used to characterize initial data for the Kerr metric in
\cite{ParradoJuan}  and, more recently, to define a quality factor
measuring the deviation of a given metric admitting a Killing vector
with respect to the Kerr metric \cite{ParradoJose}. A characterization of
the Kerr-Newman spacetime in the same spirit has been obtained
by Wong \cite{Wong}. 

Given the natural relation between the Kerr (and Kerr-NUT) metrics
and the  Kerr-NUT-(A)de Sitter metrics, it is most natural to ask whether
the characterization above can be extended to the larger 
class of spacetimes. This is
the problem we address in this paper. The derivations
in \cite{Mars1} and \cite{Wong} used extensively the Newman-Penrose formalism
to find local properties of the spacetimes under consideration. Here we 
follow a completely different approach which allows us to dispense
of the use of the Newman-Penrose formalism altogether. This has
the advantage that the underlying geometric structures become much
more clearly
exposed. As we will see, two such structures will play a fundamental role. 
The first one involves a Riemannian submersion (of codimension two) involving
a conformally related metric, where the conformal factor is determined
in terms of objects constructed from the Killing vector $\xi$. 
The second one is the presence
of a second Killing vector field in the spacetime. This Killing field
is again constructed solely in terms of geometric information 
arising from the Killing vector $\xi$. The explicit form
of the second Killing field may have potential 
interest in other areas of research, such as, for instance, in extending
the Hawking rigidity theorem for stationary black holes from the analytic to
the smooth setting (see \cite{Ionescu3}, \cite{Ionescu5},\cite{Ionescu4}
for interesting developments in 
this direction). We emphasize that the cosmological constant  $\Lambda$
is not assumed to be non-zero anywhere in the paper, so all our results
apply to the vacuum case as well.
As already mentioned,
the proof presented here is very 
different to the proofs in \cite{Mars1} and \cite{Mars2} which were based in a
tetrad approach. The method in \cite{Mars1} was based on constructing a
tetrad of commuting vector fields, two of which were Killing vector
fields of the spacetime. There is no simple parallelism between
this proof and the one presented here. The method in \cite{Mars2},
although still based on tetrads, already has some of the ingredients
we exploit here. More specifically a two-dimensional
distribution and its corresponding quotient space also plays
a fundamental role in integrating the field equations in \cite{Mars2} (cf.
Proposition 2 in this reference).
However, no isometric submersion is identified
in that paper. The geometrically much more transparent proof presented here
is not only able to deal with Einstein spaces
but also reveals interesting geometric
structure that helps understanding the vacuum characterization as well.

Given that the characterization condition we impose
is fully local in nature, it is clear that 
no global restrictions on the spacetime can be deduced from it.
Our aim is therefore to prove a local isometry to the 
Kerr-NUT-(A)de Sitter spacetime in the following precise sense.

\begin{definition}
\label{defiKNUTaDS}
Given real constants 
$\{\Lambda,m,a,l\}$ 
for which the function
$\kk := 1+\frac{\Lambda}{3} a\cos \theta (4l+a \cos \theta )$ is 
everywhere positive, a spacetime $(\M,g)$ is locally isometric 
to the Kerr-NUT-(A)de Sitter spacetime with parameters
$\{\Lambda,m,a,l\}$ 
if for any point $p\in \M$ where
no Killing vector of $(\M,g)$ vanishes
there is a neighbourhood $U_p$ of $p$ and
coordinates $\{u,r,\theta,\phi\}$ on $U_p$ such that $g$ takes the
local form
\begin{align}
g= & -\frac{\Delta-a^2 \sin^2 \theta  \kk}{\rho^2}
\left ( \frac{}{} du - (a \sin^2 \theta +4 l \sin^2 (\theta/2 )) d\phi 
\right )^2 +2 \left (dr -  a \sin^2 \theta \kk d\phi \right ) \times \nonumber \\ 
& \left ( \frac{}{} du -  (a \sin^2 \theta
+4 l \sin^2 (\theta/2)  ) d\phi \right )
+\rho^2 \left ( \frac{d\theta^2}{\kk}+ \kk \sin^2 \theta  d\phi^2 \right )
\label{KNUTaDS}
\end{align}
where
\begin{eqnarray*}
\rho^2:=r^2+(l+a \cos \theta )^2, \quad
\Delta:=a^2-l^2-2 m r+r^2- \frac{\Lambda}{3}
(3 l^2 (a^2-l^2)+(a^2+6 l^2) r^2+r^4).
\end{eqnarray*}
\end{definition}
Note that, at points where $dr$ is non-null, the metric (\ref{KNUTaDS}) can be transformed
into the more usual form (see e.g. \cite{GP3}):
\begin{align*}
g = &  -\frac{\Delta}{\rho^2} \left ( \frac{}{}
dt-(a \sin^2 \theta + 4 l \sin^2(\theta/2)) d\varphi \right )^2
+ \frac{\kk \sin^2 \theta}{\rho^2} 
\left ( \frac{}{} a dt-(r^2+(a+l)^2) d\varphi
\right )^2 + \\
& +\frac{\rho^2}{\Delta} dr^2 
+\frac{\rho^2}{\kk} d\theta^2
\end{align*}
by the coordinate transformation
\begin{eqnarray*}
dt=du- \frac{r^2+(a+l)^2}{\Delta} dr, \quad \quad \quad
d \varphi = d \phi - \frac{a}{\Delta} dr.
\end{eqnarray*}

A suitable combination of the main results in this paper can be stated as the following theorem (see below for definitions)

\begin{theorem}
\label{intro}
Let $(\M,g)$ be a $\Lambda$-vacuum spacetime admitting a Killing vector
$\xi$ with self-dual two-form  $\F_{\alpha\beta}$.
Assume there exists $Q \in C^{\infty}(\M,\mathbb{C})$
such that 
\begin{eqnarray}
\C_{\alpha\beta\mu\nu} = Q \left (\F_{\alpha\beta}
\F_{\mu\nu} - \frac{1}{3} \F^2 \I_{\alpha\beta\mu\nu} \right ),
\label{main1}
\end{eqnarray}
where $\C_{\alpha\beta\mu\nu}$ is the self-dual Weyl tensor of $g$ and
$\F^2 \defi \F_{\alpha\beta} \F^{\alpha\beta}$ and assume 
that $\exists \, p,p',\textcolor{blue}{p''} \in \M$ such that
$(Q \F^2) |_p \neq 0$, $(Q \F^2 - 4 \Lambda) |_{p'} \neq 0$, \textcolor{blue}{and $\xi|_{p''}$ is not orthogonal to the 2-plane generated by the two real null eigenvectors of $\F_{\alpha\beta}|_{p''}$}. Then 
$\F^2 \neq 0$  and $Q \F^2 - 4 \Lambda \neq 0$ everywhere and
there exist constants $b_1,b_2,c,k \in \mathbb{R}$ such that
\begin{align*}
36 Q ( \F^2)^{\frac{5}{2}} + \left ( b_2 - i b_1  \right )
( Q \F^2 - 4 \Lambda )^3 & = 0 \\
g (\xi,\xi) + \mbox{Re} \left ( \frac{6 \F^2 \left ( Q \F^2 + 2 \Lambda
\right )}{(Q \F^2 - 4 \Lambda )^2} \right ) + c & = 0 \\
- k + \textcolor{blue}{\left |
\frac{36 \F^2}{(Q \F^2 - 4 \Lambda)^2} \right |}
\nabla_{\alpha} Z \nabla^{\alpha} Z
- b_2 {\textcolor{blue}{Z}}+ c Z^2 + \frac{\Lambda}{3} Z^{\textcolor{blue}{4}} & =0
\end{align*}
where $Z = \mbox{Im} \left ( \frac{6 i \sqrt{\F^2}}{Q \F^2 - 4 \Lambda}
\right)$.
If these constants are such that the polynomial $V(\zeta) :=
k + b_2 \zeta - c \zeta^2 - \frac{\Lambda}{3} \zeta^4$ can be factored
as
\begin{equation}
V(\zeta) = \hat{V}(\zeta) (\zeta - \zeta_0 ) ( \zeta_1 - \zeta)
\end{equation}
with $\zeta_0 \leq \zeta_1$ and $\hat{V}(\zeta) > 0$ on $[\zeta_0,\zeta_1]$
and $Z : \M \rightarrow  [\zeta_0, \zeta_1]$ then
$(\M,g)$ is locally isometric 
to the
Kerr-NUT-(A)dS with parameters $\{\Lambda, m, a, l \}$ where
\begin{eqnarray*}
m = \frac{b_1}{2 v_0 \sqrt{v_0}}, \quad
\quad a = \frac{\zeta_1 - \zeta_0}{2 \sqrt{v_0}},
\quad  \quad l = 
\frac{\zeta_1 + \zeta_0}{2 \sqrt{v_0}}
\end{eqnarray*}
and $v_0 := \hat{V}(\frac{\zeta_0+ \zeta_1}{2})$.
\end{theorem}

In fact we obtain in Theorem \ref{full} 
a full characterization of $\Lambda$-vacuum 
spacetimes for which (\ref{main1}) holds and for which $\F^2$
is not identically zero.

Other local characterization results for Kerr-NUT-(A)de Sitter spacetimes
have been obtained based on the separability of the Hamilton-Jacobi equation. 
As it is well-known
from the Kerr case, the underlying reason for this separability is the presence
of hidden symmetries in the form of Killing tensors or, more generally,
in the form of conformal Killing-Yano tensors. The presence of such symmetries
of the Kerr-NUT-(A)de Sitter spacetime
(in fact, on the corresponding
class of metrics in arbitrary dimension, as given in \cite{Chen}) was found in \cite{Frolov}
(see also \cite{Chong} for the particular case when all rotation parameters
except one vanish). Results concerning the relationship between
the Kerr-NUT-(A)de Sitter spacetimes and $\Lambda$-vacuum spacetimes
admitting a closed conformal
Killing-Yano tensor have been obtained
in \cite{Houri_b} (in arbitrary dimension) 
based on previous work in \cite{Houri_a} and \cite{Krtous}. The
results in these papers and the characterization 
of Kerr-NUT-(A)de Sitter  presented here are a priori 
completely different. It is an interesting open problem
whether there is any way of
connecting both types of results. Besides its intrinsic interest, this
would open up the possibility of extending the characterization results
in this paper (which are restricted to four spacetime
dimensions) to the higher dimensional case. In this respect, it is worth
mentioning that a generalization \cite{Mars2}
of the spacetime characterization of
the Kerr spacetime in \cite{Mars1} has been used 
\cite{JuanThomas_a}  
(see also \cite{FerrandoSaez} for related statements) 
to obtain an  alternative characterization of this metric involving 
Killing spinors (and hence hidden symmetries). 

A final remark is in order. The hypotheses in our characterization 
theorems restrict the spacetime to be of Petrov type D
on a non-empty open subset.
However, our hypotheses
allow a priori that the Weyl tensor degenerates to type N or type O
elsewhere in the spacetime.
In the field of exact solutions,
$\Lambda$-vacuum spacetimes  of Petrov type D
have been thoroughly studied, and the most
general class of such metrics has been
found to be the so-called Pleba\'nski-Demia\'nski
metric and its limiting cases (this was analyzed by Deveber {\it at al}
\cite{Debever} and Garc\'{\i}a \cite{Garcia}, see also \cite{GP}). So, one
might think that a convenient strategy to prove the characterization results
in this paper 
would be to identify (within this general class) those metrics satisfying
the proportionality condition between the self-dual Weyl tensor and
the self-dual double two-form of the Killing vector. However, there are two
main reasons why this is not so.
First and foremost, in the field
of exact solutions the aim is to find explicit Lorentzian metrics
solving the Einstein field equations. Often
the equations split into several cases depending on whether certain
quantities
vanish or not. The method typically proceeds by assuming that either the
quantity is not zero or it vanishes {\it on an open set}. This is fine
in this field of research,
since the regions left out by the analysis are of empty interior.
However, it is not quite sufficient for our purposes, where we want
to identify the metric locally around {\it any} point in the given spacetime
(or at least keeping full control of the points that are left out). 
Studying the transition between a priori disjoint regions, which may or may
not belong to the same spacetime, is a difficult task in general. This makes
the use of the exact solution results of little use in the present context.
The second reason is related to the previous one. As already
mentioned, main hypotheses do not fix
a priori the Weyl tensor 
to  be of Petrov type D everywhere in the spacetime
so it becomes necessary to study what happens
at the boundary (if any) of  open set where the Petrov type is $D$.
For this, a semiglobal
analysis is required, which again precludes (or makes it difficult)
the use of fully local results, as those in the field of exact solutions.
All these considerations were also present in the case of vanishing
$\Lambda$ studied  in \cite{Mars1}.

The plan of the paper is as follows. In Section 
\ref{sec:preliminaries} we review a number of known results
for spacetimes admitting a Killing vector, especially those
concerning the so-called self-dual Killing form and
the Ernst one-form constructed from it. We first present 
results valid for general
spacetimes and then we concentrate, in Subsection \ref{sec:Lambda-vacuum},
on $\Lambda$-vacuum spacetimes. 
Algebraic and differential  consequences of the proportionality between
the self-dual Weyl tensor and the self-dual double Killing two-form
are obtained in Subsection \ref{sec:alignment} (further results on the
self-dual Killing form necessary for the paper are included in 
an Appendix).
In Section 
\ref{sec:non-null} we show that the Ernst one-form is globally exact,
and construct explicitly a second Killing vector in terms of the geometry
of $\xi$. The two Killing vectors span a two-dimensional Killing
algebra except in special cases, which are 
analyzed in detail in 
Subsection \ref{sec:special}, as they
require a separate treatment throughout the paper. In Section \ref{sec:solving}
the field equations are solved by exploiting 
the local
Riemannian submersion in a conformally related spacetime mentioned above.
Our main results here are Theorems \ref{non-orthogonal} and \ref{th:orthogonal}
where local forms
of the metric are obtained around any point in the spacetime, except
fixed points of the Killing vector $\xi$. Section \ref{sec:semiglobal}
is devoted to showing first that the Petrov type must remain 
constant throughout the spacetime and understanding the relationship between
the special cases and the generic cases. For this, the Killing algebra
derived in Sect. \ref{sec:non-null} plays a fundamental role.
Our last Section \ref{sec:characterization} is devoted to identifying the
metrics obtained in the previous sections, and hence writing down our
characterization results. First of all we summarize in Theorem \ref{full}
the main results
of the previous sections in a self-contained form, so as to avoid 
the need to refer to previous definitions in the reminder of the paper.
Based on this result we obtain characterizations of the Pleba\'nski, 
uncharged Bertotti-Robinson and Nariai metrics in Theorem \ref{Plebanski}.
 Finally, the Kerr-NUT-(A)de Sitter metric is fully characterized 
in Theorem \ref{Kerr-NUT-AdS}, and we also discuss similar characterizations of other related interesting metrics, such as the ``black branes'' (also called ``planar'' black holes) found in \cite{KMV} and recently discussed in connection with the AdS/CFT holographic description of the Quark-Gluon plasma \cite{Mc1,Mc2,McT}.

\section{Preliminaries}
\label{sec:preliminaries}

Throughout this paper, a spacetime $(\M,g)$ is 
a smooth, orientable, four-dimensional, connected
manifold endowed with a smooth\footnote{Smoothness 
is assumed only for notational simplicity. All the results
hold if the metric is merely $C^3$.}
 metric $g$ of
Lorentzian signature.
We assume further that the spacetime
is time orientable and time oriented. The Levi-Civita covariant
derivative of $g$ is denoted by $\nabla$ and the volume form
by $\eta_{\alpha\beta\gamma\delta}$.

Our basic assumption is that the spacetime possesses a Killing vector field
$\xi$,
whose norm is denoted by $N\defi -\xi^\mu\xi_\mu$. The 2-form $F_{\mu\nu}\defi \nabla_\mu\xi_\nu$ defines the complex self-dual {\em Killing 2-form}:
$$
\F_{\alpha \beta} \defi F_{\alpha\beta} + i F^{\star}_{\alpha\beta} , \hspace{1cm} \F_{\alpha \beta}^\star =-i \F_{\alpha \beta},
$$
where $\star$ is the Hodge dual operator. Well known identities for $\F$ (see
e.g. \cite{Israel1970}) are
\be
\F_{\mu\rho}\F_{\nu}{}^\rho = \frac{1}{4} \F^2 g_{\mu\nu}, \hspace{1cm} \F^2 \defi \F_{\alpha\beta} \F^{\alpha\beta} , \hspace{1cm}  \F_{\mu\rho}\overline\F^{\mu\rho} =0 \label{fsquare}
\ee
where overbars denote complex conjugation. The Ernst one-form is defined by
\be 
\chi_{\beta} \defi 2 \xi^{\alpha} \F_{\alpha\beta}, \label{defchi} \hspace{1cm} \xi^\beta\chi_\beta =0 \hspace{1cm}
\chi_{\alpha} \chi^{\alpha} = - N \F^2 
\ee
the last relation being a consequence of (\ref{fsquare})  by contraction
with $\xi$. Actually, by the same route one gets from (\ref{fsquare})
\be
\chi^\rho \F_{\rho\mu} =-\frac{1}{2} \F^2 \xi_\mu, 
%\hspace{1cm} \F_{\mu\nu}\chi^\mu\chi^\nu =0, 
\hspace{1cm}  \F_{\mu\nu}\chi^\mu\overline\chi^\nu =0.
%\hspace{1cm}  \F_{\mu\nu}\overline\chi^\mu\overline\chi^\nu =0. 
\label{propchi}
\ee
The real part of the Ernst one-form satisfies
$$
\chi_\mu +\overline\chi_\mu = 4\xi^\rho F_{\rho\mu}=-4\xi^\rho \nabla_\mu \xi_\rho =2\nabla_\mu N
$$
so that we can always write
\be
\chi_\mu =\nabla_\mu N +i \omega_\mu \label{DN+twist}
\ee
for some real one-form $\omega$ called the twist of $\xi$.
We also introduce the following real one-form
$$
\eta_{\mu}\defi \overline\chi^{\rho}\F_{\rho\mu}=\chi^{\rho}\overline\F_{\rho\mu} , \hspace{1cm} \overline\eta_{\mu}=\eta_{\mu}
$$
which satisfies
\be
\eta^{\rho}\F_{\rho\mu}=-\frac{1}{4}\F^{2}\overline\chi_{\mu}, \hspace{1cm}
\eta_{\mu}\chi^{\mu}=0, \hspace{1cm}
\eta_{\mu}\xi^{\mu}=-\frac{1}{2}\chi^{\rho}\overline\chi_{\rho}, \hspace{1cm}
\eta_{\mu}\eta^{\mu}=-\frac{N}{4}\F^{2}\overline\F^{2} . \label{etaprop}
\ee
This seems to be a better choice than the more ``natural'' real one-form $i\eta_{\alpha\beta\mu\nu}\overline\chi^{\beta}\xi^{\mu}\chi^{\nu}$, and in any case one has the following relations
$$
\bm\xi \wedge \bm\eta =\frac{i}{2}\star (\bm\chi \wedge \overline{\bm\chi}) \hspace{3mm} \Longrightarrow \hspace{3mm} i\eta_{\alpha\beta\mu\nu}\overline\chi^{\beta}\xi^{\mu}\chi^{\nu} = 2N \eta_{\alpha}-(\chi^{\rho}\overline\chi_{\rho})\xi_{\alpha} \, ,
$$
where we use bold symbols to refer to $p$-forms (in particular covectors)
when no abstract indices are used.
We will also use the following symmetric tensor
\be
t_{\mu\nu}\defi \frac{1}{2} \F_{\mu\rho}\overline\F_{\nu}{}^\rho \label{emt}
\ee
which is nothing but  the ``energy-momentum tensor'' of the 2-form $F_{\mu\nu}$. Well known properties of this  tensor are (cf. (\ref{fsquare}))
$$
t_{\mu\nu}=t_{\nu\mu}, \hspace{1cm} t^\rho{}_\rho =0, \hspace{1cm} t_{\mu\rho}t_{\nu}{}^{\rho} =\frac{1}{4} t_{\rho\sigma}t^{\rho\sigma} g_{\mu\nu}, \hspace{1cm} t_{\rho\sigma}t^{\rho\sigma}=\frac{1}{16}\F^2\overline\F^2.
$$
Note however that it is not divergence-free in general, as $\F$ does not necessarily satisfy the Maxwell equations in vacuum. 
% we will come back to this point presently. 

As $\xi$ is a Killing vector we obviously have
$$
\pounds_{\xi}\F_{\mu\nu}=0, \quad \pounds_{\xi}t_{\mu\nu}=0, \quad \xi(\F^{2})=0, \quad \xi(N)=0, \quad [\xi,\chi]=0,Ê\quad [\xi, \eta]=0 \, .
$$
Further properties of all these objects are collected in an Appendix.

A complex self-dual 2-form $\F_{\mu\nu}$ is said to be regular at $p\in \M$ if $\F^2|_p \neq 0$. If $\F^2|_p=0$ then $\F_{\mu\nu}$ is called singular.
In the regular case, there exist two different real (and necessarily null) eigenvectors $k_{\pm}$ of $\F_{\mu\nu}$ at $p$ with opposite eigenvalues
$\pm R$
%i.e. satisfying $\ell^\mu \F_{\mu[\nu}\ell_{\rho]}=0$ at $p$, 
$$
k_{\pm}^{\nu}\F_{\mu\nu}=\pm R k_{\pm\mu} \, , \quad \quad g(k_{\pm},k_{\pm})=0
$$
while there exists only one, $k_{+}=k_{-}$, with zero eigenvalue, in the singular case.
%These null eigenvectors are also eigenvectors with real eigenvalues of $t_{\mu\nu}$, the eigenvalue vanishing in the case of null $\F_{\mu\nu}$.
Then $k_{\pm}$ are also eigenvectors of $t_{\mu\nu}$ and we have
\be
k_{\pm}^{\mu}t_{\mu\nu}=-\frac{1}{2} R\overline R k^\pm_{\nu}, \hspace{1cm} \F^{2}=-4R^{2}
\label{eigen-t}
\ee
as well as
\be
g(k_{\pm},\chi) =\pm 2R\, g(\xi,k_{\pm}), \hspace{1cm} g(k_{\pm},\eta) =2R\overline R \, g(\xi,k_{\pm}). \label{contractions}
\ee

The relations (\ref{ks}) in the Appendix imply in general
\bean
\bm\eta +2R\overline R \bm{\xi} =-4R\overline R [g(k_-,\xi)\bm{k}_+ +g(k_+,\xi)\bm{k}_-]\\
R\overline{\bm{\chi}}+ \overline R \bm{\chi}=4R\overline R [g(k_-,\xi)\bm{k}_+ -g(k_+,\xi)\bm{k}_-]
\eean
from where some particular possibilities arise
\begin{enumerate}
\item Case with $g(k_{+},\xi)=0$. 
In this situation one easily obtains
$$
g(k_{+},\chi)=g(k_{+},\eta)=0, \hspace{1cm} g(\chi,\overline\chi)=-4NR\overline R
$$
and also
$$
-(\bm\eta +2R\overline R \bm{\xi} )=R\overline{\bm{\chi}}+ \overline R \bm{\chi} =4R\overline R \, g(k_{-},\xi) \bm k_+ \, .
$$
\item Case $g(k_{+},\xi)=g(k_{-},\xi)=0$. In this subcase one obviously has, in addition to the above, 
$$
g(k_{-},\chi)=g(k_{-},\eta)=0, \hspace{1cm} \bm\eta =-2R\overline R \bm{\xi} , \hspace{1cm} R\overline{\bm{\chi}}+ \overline R \bm{\chi} =0
$$
so that $\eta$, $\xi$ and $\chi$ are all eigenvectors of $t_{\mu\nu}$, the first two with eigenvalue $-R\overline R/2$, the last with the opposite eigenvalue.
\end{enumerate}

\subsection{$\Lambda$-vacuum}
\label{sec:Lambda-vacuum}

Throughout this paper we will assume that the spacetime satisfies 
the Einstein field equations for  vacuum with a cosmological constant $\Lambda$, that is to say, 
\be
R_{\alpha\beta} = \Lambda g_{\alpha\beta} \label{ric}
\ee
where $R_{\alpha\beta}$ is the Ricci tensor. In that case we will say that the metric $g$ is $\Lambda$-vacuum. 
%It is called vacuum ---also Ricci-flat--- if it is $\Lambda$-vacuum with $\Lambda=0$.
The Weyl curvature tensor of the spacetime is denoted by $C_{\alpha\beta\lambda\mu}$ and its (unique) Hodge dual by $C^\star_{\alpha\beta\lambda\mu}=\frac{1}{2}\eta_{\rho\sigma\lambda\mu}C_{\alpha\beta}{}^{\rho\sigma}$. Then, the complex self-dual Weyl tensor is
$$
\C_{\alpha\beta\lambda\mu}=C_{\alpha\beta\lambda\mu}+i C^\star_{\alpha\beta\lambda\mu}, \hspace{1cm}
\C^\star_{\alpha\beta\lambda\mu}=-i \C_{\alpha\beta\lambda\mu}.
$$

Under assumption (\ref{ric}) one has that the Riemann tensor becomes
\be
R_{\alpha\beta\lambda\mu}=C_{\alpha\beta\lambda\mu}+\frac{\Lambda}{3} \left(g_{\alpha\lambda}g_{\beta\mu}- g_{\alpha\mu}g_{\beta\lambda}\right) \label{RiemLam}
\ee
so that 
\be
\nabla^\alpha \C_{\alpha\beta\lambda\mu}=0, %\hspace{1cm} \nabla^\alpha T_{\alpha\beta\lambda\mu}=0
\label{divfree}
\ee
and the standard property of Killing vectors $\nabla_\beta\nabla_\lambda\xi_\mu =\xi^\alpha R_{\alpha\beta\lambda\mu}$ can be appropriately rewritten in the language of complex self-dual objects as
\be
\nabla_{\mu} \F_{\alpha\beta} = \xi^{\nu} \left ( \C_{\nu \mu \alpha\beta} +
\frac{4 \Lambda}{3}  \I_{\nu \mu\alpha\beta} \right ), \label{nablaF}
\ee
where 
$$
{\cal I}_{\alpha\beta\mu\nu} = \frac{1}{4} \left ( g_{\alpha\mu}
g_{\beta\nu} - g_{\alpha \nu} g_{\beta \mu} + i \eta_{\alpha\beta\mu\nu} \right ) 
$$
is the metric in the space of complex self-dual 2-forms, i.e. it
satisfies ${\cal I}_{\alpha\beta\mu\nu} {\cal U}^{\mu\nu} = {\cal U}_{\alpha\beta}$
for any self-dual two form.

Taking the covariant derivative of the first in (\ref{defchi}) one can obtain
\be
\nabla_\mu \chi_\nu =-\frac{\F^2}{4} g_{\mu\nu} - 2 t_{\mu\nu}+2\xi^\rho\xi^\sigma \C_{\rho\mu\sigma\nu}-\frac{2\Lambda}{3}\left(Ng_{\mu\nu} +\xi_\mu \xi_\nu \right) \label{derchi}
\ee
and therefore, in this situation the Ernst one-form is closed 
\cite{Papapetrou1966}
\be
\nabla_{[\mu}\chi_{\nu]} =0. \label{chiclosed}
\ee
For completeness, one can also derive an expression for the covariant derivative of $\eta$, which after a calculation becomes
\be
\nabla_\mu \eta_\nu = -\frac{\overline\F^2}{4} \F_{\mu\nu} -\frac{\Lambda}{3}\left(\xi_\mu \chi_\nu +2N\F_{\mu\nu} \right)+\xi^\sigma \overline\chi^\rho \C_{\sigma\mu\rho\nu} + \mbox{c.c.} \label{dereta}
\ee

Immediate consequences of (\ref{nablaF}) are
\begin{align}
\nabla_{\mu} \F^2 & = 2 \xi^{\nu} \F^{\alpha\beta} \C_{\nu\mu\alpha\beta}
+ \frac{4 \Lambda}{3} \chi_{\mu}\, , \label{nablaFsq}\\
\nabla_{\alpha} \chi^{\alpha} & = - \F^2 - 2 \Lambda N \, , \label{nablachi} \\
\nabla_{\mu} \nabla^{\mu} \F^2 & = - \C_{\alpha\beta\mu\nu} \F^{\alpha\beta} \F^{\mu\nu}
- \frac{4}{3} \Lambda \F^2 - \frac{N}{2} \left ( \C_{\alpha\beta\mu\nu}
\C^{\alpha\beta\mu\nu} + \frac{16 \Lambda^2}{3} \right ). \label{DeltaFsq}
\end{align}
%as well as
%\be
%\nabla_{[\mu} \F_{\alpha\beta]}=i  \frac{\Lambda}{3} \eta_{\rho\mu\alpha\beta}\xi^\rho \, , \hspace{1cm}
%\nabla^\rho\F_{\rho\mu} =-\Lambda \xi_\mu \, .\label{maxwell}
%\ee
%From these, and after a calculation, one can derive the expression for the divergence of the tensor (\ref{emt})
%\be
%\nabla^\mu t_{\mu\nu} =\frac{\Lambda}{4} (\chi_\nu +\overline\chi_\nu).\label{divt}
%\ee

\subsection{Alignment of $\C$ and $\F$}
\label{sec:alignment}

The self-dual $\C_{\alpha\beta\lambda\mu}$ defines an eigenvalue problem acting on self-dual 2-forms. The content of this problem leads to the important Petrov classification of the Weyl tensor \cite{Exact}. In the case under consideration, the spacetime has a distinguished self-dual 2-form, the Killing 2-form $\F_{\mu\nu}$. It seems natural to ask what are the implications of $\F_{\mu\nu}$ being an eigen-2-form of $\C_{\alpha\beta\lambda\mu}$
$$
\C_{\alpha\beta\mu\nu}\F^{\alpha\beta}\propto \F_{\mu\nu}
$$
and this was the essential assumption in the several unique characterizations of the Kerr and other relative spacetimes given in references \cite{Mars1,Mars2,Mars3,Mars4}.

The previous relation does not restrict the Petrov type of the spacetime in general. However, a particularly interesting case 
where that condition is achieved is by assuming
\be
\C_{\alpha\beta\mu\nu} = Q \left (\F_{\alpha\beta}
\F_{\mu\nu} - \frac{1}{3} \F^2 \I_{\alpha\beta\mu\nu} \right ) \label{C=FF}
\ee
which is actually the starting point in \cite{Mars1} ---also \cite{Mars4}.
We will assume this condition herein. If (\ref{C=FF}) holds, then at $p\in \M$ 
$$
k_\pm^\alpha k_\pm^\mu\C_{\alpha\beta\mu\nu} =\frac{2}{3} Q R^2 k^\pm_\beta k^\pm_\nu
$$
showing that the real null eigenvectors of $\F_{\mu\nu}$ are multiple principal null directions of the Weyl tensor, so that
the Petrov type is D if $Q\F^2|_p\neq 0$, N if $\F^2|_p=0$ and $Q|_p\neq 0$, or type 0 if $Q|_p=0$.

Under the assumption (\ref{C=FF})  equations 
(\ref{nablaF}), (\ref{nablaFsq}) and (\ref{DeltaFsq}) become, after using
(\ref{defchi}),
\begin{align}
& \nabla_{\mu} \F_{\alpha\beta} = \frac{1}{2} Q \chi_{\mu} \F_{\alpha\beta}
+ \frac{1}{3} 
\left ( 4 \Lambda - Q \F^2 \right ) \xi^{\nu} \I_{\nu\mu\alpha\beta},
\label{nablaF_2} \\
& \nabla_{\mu} \F^2 = \frac{2}{3}  \left(Q \F^2 +2\Lambda
\right ) \chi_{\mu},  \label{nablaFsq_2}  \\
& \nabla_{\mu} \nabla^{\mu} \F^2 =  -\frac{2}{3} \F^2 \left ( Q \F^2 + 2 \Lambda \right ) - \frac{N}{3} \left (Q^2 \left ( \F^2 \right )^2 + 8 \Lambda^2 \right ).
\label{DeltaFsq_2}
\end{align}
Similarly, equations (\ref{derchi}) and (\ref{dereta}) become, respectively
\begin{align}
&\nabla_\mu \chi_\nu =-\frac{\F^2}{4} g_{\mu\nu} - 2 t_{\mu\nu}+\frac{Q}{2}\chi_\mu \chi_\nu+\frac{1}{6}\left(Q\F^2-4\LambdaïŸ\right)\left(Ng_{\mu\nu} +\xi_\mu \xi_\nu \right) \label{derchi2}\\
& \nabla_\mu \eta_\nu = -\frac{\overline\F^2}{4} \F_{\mu\nu} +\frac{1}{12}\left(\overline Q\,  \overline\F^2 -4\Lambda \right)\left(\xi_\mu \chi_\nu +2N\F_{\mu\nu} \right)+\frac{1}{2}Q\chi_\mu \eta_\nu + \mbox{c.c.} \, .\label{dereta2}
\end{align}

%The explicit expression (\ref{nablaF_2}) allows us to compute a general expression for the covariant derivative of $t_{\mu\nu}$, which after a calculation can be written as
%\bea
%2\nabla_\lambda t_{\mu\nu}=(Q\chi_\lambda +\overline Q \overline\chi_\lambda)t_{\mu\nu}+\frac{1}{12}\left(4\Lambda -\overline Q \, \overline\F^2 \right)\left(2\xi_{(\mu} \F_{\nu)\lambda} +\chi_{(\mu}g_{\nu)\lambda}-\frac{1}{2} \chi_\lambda g_{\mu\nu}\right)\nonumber \\
%+\frac{1}{12}\left(4\Lambda - Q \F^2 \right)\left(2\xi_{(\mu} \overline\F_{\nu)\lambda} +\overline\chi_{(\mu}g_{\nu)\lambda}-\frac{1}{2} \overline\chi_\lambda g_{\mu\nu}\right). \label{nablat}
%\eea
%Formula (\ref{divt}) can be easily recovered from here.

\section{Existence of Ernst potential and a second Killing vector}
\label{sec:non-null}
In this section we will assume that the self-dual Killing form
$\F_{\alpha\beta}$ is regular 
everywhere, i.e. $\F^2 \neq 0$ everywhere on $\M$.
This implies the existence of 
a smooth
complex function $R : \M \longrightarrow
\mathbb{C}$ satisfying $\F^2 = - 4 R^2$.
Note that at each point $\pm R$
coincide with the eigenvalues
introduced in Section \ref{sec:preliminaries}, and that there are two
possible choices for the smooth function $R$, namely $R$ and $-R$. We assume from now on
that one such choice has been made.  
%Consider the eigenvalue problem
%\begin{eqnarray*}
%\F_{\alpha\beta} k^{\prime \, \, \beta} = q k^{\prime}_{\alpha}
%\end{eqnarray*}
%The eigenvalue $q$ necessarily satisfies $q^2 = R^2$ (simply contract
%this expression with $\F_{\mu}^{\,\,\,\alpha}$). 
Let $k=k_{+}$ be a (real) null eigenvector field
with eigenvalue $R$ and $\ell=k_{-}$ a (real) null eigenvector field with eigenvalue $-R$.
Being eigenvectors of
multiplicity one, they can be both chosen to be smooth on $\M$ and, without loss of generality, as future directed and satisfying $g(k,\ell)=-1$. The freedom left in the choice
is $k^{\prime} \longrightarrow A k$, $\ell^{\prime} \longrightarrow A^{-1} \ell$
with $A: \M \longrightarrow \mathbb{R}^{+}$ smooth. Note that,
had we chosen $-R$ instead of $R$, the null vector fields $\{k,\ell\}$
would have been interchanged.

Let us also introduce a complex one-form $P_{\alpha}$, a real one-form
$q_{\alpha}$ and a self-dual two-form $\W_{\alpha\beta}$ by means of
\be
\chi_{\alpha} \defi 2 R P_{\alpha}, \hspace{1cm} \eta_{\mu}\defi 2R\overline R\,  q_{\mu}, \hspace{1cm} \W_{\alpha\beta} \defi \frac{1}{R} \F_{\alpha\beta}\, .
\ee
These objects are all smooth and satisfy the following
properties as a consequence of the expressions in Section
\ref{sec:preliminaries}
\begin{align}
& \W_{\mu\rho}\W_{\nu}{}^\rho   = - g_{\mu\nu}, & &
 \W_{\alpha\beta} \W^{\alpha\beta}   =-4 , & &  
\W_{\mu\rho}\overline \W^{\mu\rho}  =0,  \label{Wsquare}\\
& P_{\alpha}  = \xi^{\beta} \W_{\beta\alpha}, & & 
 P^{\alpha} \W_{\alpha\beta}   = \xi_{\beta}, & &  
\overline P^{\alpha} \W_{\alpha\beta} = q_{\beta}, & & 
q^{\alpha} \W_{\alpha\beta} =  \overline P_{\beta}, \label{PxiW} \\
& \xi^{\alpha} P_{\alpha}  = 0, & &  
q^{\alpha} P_{\alpha}  = 0, & &  P^{\alpha} P_{\alpha}  = - q^{\alpha} q_{\alpha}  
= N, & & 
 q^{\mu}\xi_{\mu} =-P^{\mu}\overline P_{\mu}, \label{Psquare} \\
& t_{\mu\nu}   = \frac{R \ov{R}}{2} \W_{\mu \rho} \ov{\W}_{\nu}{}^{\rho}, & & 
P^{\alpha} t_{\alpha\beta}  =  -\frac{R \ov{R}}{2} \ov{P}_{\beta}, & & 
\xi^{\alpha} t_{\alpha\beta}  =  -\frac{R \ov{R}}{2} q_{\beta}, & & 
q^{\alpha} t_{\alpha\beta}  =  -\frac{R \ov{R}}{2} \xi_{\beta}, \label{eigen-t2}\\
& \xi(R)=0, & & \pounds_{\xi}\W_{\mu\nu}=0, & & [\xi,P]=0, & & [\xi,q]=0 , \label{Lie}
\end{align}
as well as
\be
\bm{\xi}\wedge \bm{q} =i \star \left(\bm{P} \wedge \overline{\bm{P}} \right) \quad \Longrightarrow \quad i\eta_{\alpha\beta\lambda\mu}\overline P^{\beta}\xi^{\lambda}P^{\mu}=N q_{\alpha}-(P_{\mu}\overline P^{\mu})\xi_{\alpha} . \label{xiwedgeq}
\ee
In combination with the null eigenvectors $k$ and $\ell$ we also have
\begin{align}
t_{\alpha\beta} & = \frac{R \ov{R}}{2} \left ( g_{\alpha\beta} + 2 k_{\alpha} \ell_{\beta}
+ 2 k_{\beta} \ell_{\alpha} \right ), \hspace{2cm} \label{EMTensor} \\
\W_{\alpha\beta} & = W^{-}_{\alpha\beta} - i W^{+}_{\alpha\beta}, 
\quad  \mbox{with} \quad
W^{-}_{\alpha\beta} = - k_{\alpha} \ell_{\beta} + k_{\beta} \ell_{\alpha}, \quad
W^{+}_{\alpha\beta}  = \eta_{\alpha\beta\mu\nu} k^{\mu} \ell^{\nu} \label{RealImW} \\
W^{-}_{\alpha\beta} W^{-}{}^{\alpha\beta} & =
- W^{+}_{\alpha\beta} W^{+}{}^{\alpha\beta} = -2, \quad \quad
W^{-}_{\alpha\beta} W^{+}{}^{\alpha\beta} =0, \label{squareW} \\
g(k,P) & =g(k,q)=g(k,\xi),\label{gk} \hspace{2cm}\\
-g(\ell,P)& =g(\ell,q)=g(\ell,\xi),\label{gell} \hspace{2cm}\\
\bm{\xi}+\bm{q} & =-2\left[g(\ell,\xi)\bm{k}+g(k,\xi)\bm{\ell} \right], 
\label{xi+q}\hspace{2cm}\\
\bm{P}+\ov{\bm{P}} & = 2\left[g(\ell,\xi)\bm{k}-g(k,\xi)\bm{\ell} \right] \, . \label{P+P}
\hspace{2cm}
\end{align}

In this section, it is useful to introduce a smooth function $\J$ by
\begin{eqnarray}
Q \defi \frac{3 J}{R} - \frac{\Lambda}{R^2}
\label{DefJ}
\end{eqnarray}
where recall that $Q$ is the proportionality factor in (\ref{C=FF}).
In terms of the new variables, equation 
(\ref{nablaFsq_2}) transforms into
\begin{eqnarray}
\nabla_{\mu} R = \left ( 2 \J R - \Lambda \right ) P_{\mu}
\label{nablaFsq_a}
\end{eqnarray}
which, inserted in (\ref{nablachi}), implies the following expression for the
divergence of $P_{\mu}$
\be
\nabla_{\mu} P^{\mu} = 2 \left (  R - \J N \right ). \label{divP}
\ee
More generally we have
\begin{lemma}
The tensors $\W_{\alpha\beta}$, $P_{\alpha}$, $q_\alpha$ and $\xi_{\alpha}$
satisfy the following equations
\begin{align}
& \nabla_{\mu} \W_{\alpha\beta}  = \J \left (  P_{\mu} \W_{\alpha\beta} 
+ 4  \xi^{\nu} \I_{\nu\mu\alpha\beta} \right ),  \label{nablaW_1} \\
& \nabla_{\mu} P_{\alpha}  = \frac{R}{2} g_{\mu\alpha} - \frac{1}{R} t_{\mu\alpha}
+ \J \left ( P_{\mu} P_{\alpha} - N g_{\mu\alpha} - \xi_{\mu}\xi_{\alpha} \right ), \label{nablaP}\\
&\nabla_\mu q_\alpha  = \left(\frac{\ov R}{2} -N \ov J \right)\W_{\mu\alpha} +\left(\frac{R}{2} - N J \right)\ov \W_{\mu\alpha}+q_\alpha (JP_\mu +\ov J \ov P_\mu)-\xi_\mu (\ov J P_\alpha +J \ov P_\alpha), \label{nablaq} \\
&\nabla_{\mu} \xi_{\alpha}  = \frac{1}{2} \left ( R \W_{\mu\alpha}
+ \ov{R}  \, \ov{\W}_{\mu\alpha} \right ). \label{nablaxi}
\end{align}
\end{lemma}

\noindent {\it Proof:} First note that (\ref{nablaF_2})  in terms
of the new variables becomes
\begin{eqnarray*}
\nabla_{\mu} \F_{\alpha\beta} = \left ( 3 \J R - \Lambda \right) P_{\mu} 
\W_{\alpha\beta} + 4 \J R \, \xi^{\nu} \I_{\nu\mu\alpha\beta}.
\end{eqnarray*}
which inserted into 
\begin{eqnarray}
\nabla_{\mu} \W_{\alpha\beta} = \frac{1}{R} \nabla_{\mu} \F_{\alpha\beta}
- \frac{1}{R^2} \F_{\alpha\beta} \nabla_{\mu} R 
\end{eqnarray}
gives (\ref{nablaW_1}). 
%For the second equation we compute
%\begin{align*}
%\nabla_{\mu} P_{\alpha}   =  &\nabla_{\mu} \left ( \xi^{\beta} \W_{\beta\alpha}
%\right ) 
%= \frac{1}{2} \left ( \F_{\mu}{}^{\beta} + \ov{\F}_{\mu}{}^{\beta}
%\right ) \W_{\beta\alpha} + \xi^{\beta} \nabla_{\mu} \W_{\beta\alpha} =
%\frac{R}{2} \W_{\mu}{}^{\beta} \W_{\beta\alpha} +
%\frac{1}{2R} \ov{F}_{\mu}{}^{\beta} \F_{\beta\alpha} + \\
%& + \J \xi^{\beta} \left ( P_{\mu} \W_{\beta\alpha}
%+ 4 \xi^{\nu} \I_{\nu\mu\beta\alpha} \right ).
%\end{align*}
%Expression (\ref{nablaP})  follows from the definitions of $t_{\alpha\beta}$ and
%$\I_{\nu\mu\beta\alpha}$. 
The second and third equations follow similarly from (\ref{derchi2}) and (\ref{dereta2}) after a straightforward calculation. 
The fourth is obvious from $\nabla_{\mu}\xi_{\alpha} = \frac{1}{2} \left (
\F_{\mu\alpha} + \ov{\F}_{\mu\alpha} \right )$.
$\hfill \qed$

\begin{lemma}
The principal null directions $k$ and $\ell$ satisfy the following relations
\begin{align}
2\nabla_{\mu}k_{\beta} & = k_{\beta}\left(JP_{\mu}+\overline J \overline P_{\mu}-\ell^{\alpha}\nabla_{\mu}k_{\alpha}\right)+(J+\overline J)\left[\xi_{\beta} k_{\mu}-g(\xi,k)g_{\beta\mu}\right]+i(J-\overline J) \xi^{\nu}k^{\alpha}\eta_{\nu\alpha\mu\beta},
\label{nablak}\\
2\nabla_{\mu}\ell_{\beta} & = \ell_{\beta}\left(JP_{\mu}+\overline J \overline P_{\mu}-k^{\alpha}\nabla_{\mu}\ell_{\alpha}\right)-(J+\overline J)\left[\xi_{\beta} \ell_{\mu}-g(\xi,\ell)g_{\beta\mu}\right]-i(J-\overline J) \xi^{\nu}\ell^{\alpha}\eta_{\nu\alpha\mu\beta}.
\label{nablaell}
\end{align}
\end{lemma}

\noindent {\it Proof:} Taking the real part of (\ref{nablaW_1}) one derives
$$
2\nabla_{\mu}W^{-}_{\alpha\beta}=(JP_{\mu}+\overline J \overline P_{\mu})W^{-}_{\alpha\beta}-i(JP_{\mu}-\overline J \overline P_{\mu})W^{+}_{\alpha\beta}+
(J+\overline J) (\xi_{\alpha}g_{\mu\beta}-\xi_{\beta}g_{\mu\alpha})+i(J-\overline J)\xi^{\nu}\eta_{\nu\mu\alpha\beta}
$$
whose left-hand side equals by the definition (\ref{RealImW})
$$
2\left(\ell_{\alpha}\nabla_{\mu}k_{\beta} +k_{\beta}\nabla_{\mu}\ell_{\alpha}-
k_{\alpha}\nabla_{\mu}\ell_{\beta} +\ell_{\beta}\nabla_{\mu}k_{\alpha}\right).
$$
Contracting then with $k^{\alpha}$, and with $\ell^{\alpha}$, gives the result directly. $\hfill \qed$

\vs

Equation (\ref{nablaP}) implies directly that
\begin{eqnarray*}
\nabla_{\alpha} P_{\beta} - \nabla_{\beta} P_{\alpha} = 0, \quad
\end{eqnarray*}
so that the one-form $P_{\alpha}$ is closed. This will be used in the proof of the next Lemma.

\begin{lemma}\label{lem:chinonzero}
The Ernst one-form $\chi_{\alpha}$ (or equivalently $P_{\alpha}$)
is non-zero almost everywhere and in 
fact vanishes only at points where $\xi$ vanishes. Moreover, 
$\xi$ is null at most on sets with empty interior and the function
$\J$ satisfies the equation
\begin{eqnarray}
\nabla_{\mu} \J = \J^2 P_{\mu}.
\label{nablanablaFsq_a}
\end{eqnarray}
\end{lemma}

\vs

\noindent {\it Proof:}  From the first identity in (\ref{propchi}) 
and $\F^2 \neq 0$ it follows that
$\chi_{\mu}$ vanishes at one point if and only if $\xi$ vanishes there.
In particular $\chi_{\alpha}$ (or $P_{\alpha}$)
cannot vanish on any non-empty open set.
For the second statement, assume that 
$\xi$ were null on a non-empty open set $V \subset \M$, i.e. $N=0$ on 
$V$. The last expression in (\ref{defchi}) would imply
$\chi^{\alpha}\chi_{\alpha} =0$ which combined with (\ref{DN+twist})
would lead to $\chi_{\alpha} = i \omega_{\alpha}$ being null %on $V$, a contradiction.
so that necessarily $\chi_{\mu}=2A\xi_{\mu}$ for some function $A$ such that $A=-\ov A$. But then the first in (\ref{propchi}) would provide
$$
2A\xi^{\rho}\F_{\rho\mu}=A\chi_{\mu}=2A^{2}\xi_{\mu}=2R^{2}\xi_{\mu}
$$
so that $A=\pm R$, that is to say $\chi_\mu =\pm 2R \xi_\mu$ with $R=-\overline R$. Thus $P_\mu =\pm \xi_\mu$ and the left-hand side of (\ref{divP}) is real while
its right-hand side is purely imaginary (since $N=0$). So $R=0$
and thus $\chi_\mu $ would vanish on $V$, a contradiction.

%Similarly, $\eta_\mu =-2R^2 \xi_\mu$. Equivalently, $q_\mu = \xi_\mu$ and $P_\mu =\pm \xi_\mu$, and thus from (\ref{xi+q},\ref{P+P}) all these vector fields would be parallel to either $k_\mu$ or $\ell_\mu$. But this would be incompatible with (\ref{divP}) plus $N=0$ as its lefthand side is real but the righthand side is purely imaginary, so that the only possibility would be to have $R=0$ and thus $\chi_\mu =0$ on $V$, a contradiction.

Using $\F^2 = -4 R^2$ and (\ref{nablaFsq_a}) equation (\ref{DeltaFsq_2})
transforms  into 
\begin{eqnarray}
P_{\mu} \nabla^{\mu} \J  = \J^2 N. \label{nablaJP}
\end{eqnarray}
Taking the exterior derivative in (\ref{nablaFsq_a}) 
and using the fact that $P_{\alpha}$ is closed
one derives the
existence of a smooth function 
$G : V_0 \longrightarrow \mathbb{C}$ on the
open dense set $V_0 := \{ p\in \M ;  \bm{\chi} |_p \neq 0 \}$
satisfying $\nabla_{\mu} \J = G P_{\mu}$. Inserting this into 
(\ref{nablaJP}) implies $N \left ( G - J^2 \right )=0$.
Since $N$ vanishes almost nowhere, we conclude $G= \J^2$ on $V_0$ and
consequently (\ref{nablanablaFsq_a}) holds on the open, dense set $V_0$. Both the right- and the left-hand sides in this equation are smooth one-forms on $\M$.
Since they agree on an open, dense set, they agree everywhere. $\hfill \qed$.

\vs 

The next lemma shows
that $P_\mu$ and $\chi_{\mu}$ are not only closed but in fact exact,
and provides explicit expressions for $J$, $R$ and the
Ernst potential $\chi$ satisfying $\chi_{\alpha} = \nabla_{\alpha} \chi$.
\begin{lemma}
\label{Functions_of_P}
Assume that $Q$ is not identically zero or $\Lambda \neq 0$. Then,
the one-form
$P_{\alpha}$ is exact on $\M$ and the complex function
$P : \M \longrightarrow \mathbb{C}$ satisfying $P_{\alpha} = \nabla_{\alpha} P$
can be chosen to be non-zero everywhere. Moreover, there exist smooth 
complex functions $\j,r,x$ defined on $\mathbb{C} \setminus \{0\}$ 
such that  $\J$ and $R$ satisfy
$\J = \j \circ P$, $ R = r \circ P$ and
$\chi \defi x \circ P$ satisfies (i) $\nabla_{\alpha} \chi = \chi_{\alpha}$ 
and (ii) $\chi + \ov{\chi} = 2N$ (in particular, the Ernst one-form is exact).
Moreover,
$\{\j,r,x\}$ belong to one of the two following classes,
\begin{itemize}
\item[(A)] $\displaystyle{\j (\zeta)=0, \quad r (\zeta) = - \Lambda \zeta  ,
\quad x(\zeta) = c - \Lambda \zeta^2, \quad \quad 
c \in \mathbb{R}}, \quad \quad \Lambda \neq 0.$
\item[(B)] $\displaystyle{\j(\zeta) = - \frac{1}{\zeta}, \quad
r(\zeta)
= \frac{b}{2 \zeta^2 } - \frac{\Lambda}{3} \zeta , \quad
x(\zeta) = c - \frac{b}{\zeta} - \frac{\Lambda}{3} \zeta^2, \quad \quad 
b = b_1 + i b_2, \quad b_1,b_2, c \in \mathbb{R}.}$
\end{itemize}
\end{lemma}

\vs

\noindent {\it Proof:} Assume first that $\J$ vanishes everywhere on $\M$. 
Then, $\Lambda \neq 0$ (otherwise
we would also have $Q=0$ against hypothesis).
Equation (\ref{nablaFsq_a})
implies $P_{\alpha} = -\nabla_{\alpha} \left ( \frac{R}{\Lambda} 
\right )$, and hence $P \defi -\frac{R}{\Lambda}$ satisfies
$\nabla_{\alpha} P = P_{\alpha}$ and vanishes
nowhere, as claimed. From $\chi_{\alpha} = 2R P_{\alpha}$ it follows
$\chi_{\alpha} = - \frac{1}{\Lambda} \nabla_{\alpha} R^2$ which implies 
$\chi_{\alpha} = \nabla_{\alpha} \chi$ for $\chi = - \frac{R^2}{\Lambda} + c$
for a constant $c$ which can be taken to be real without loss
of generality. Moreover, from (\ref{DN+twist})
%\begin{eqnarray*}
%\chi_{\alpha} + \chibar_{\alpha} = 2 \xi^{\beta} \left ( \F_{\beta\alpha}
%+ \ov{\F}_{\beta\alpha} \right ) = 4 \xi^{\beta} \nabla_{\beta } \xi_{\alpha} =
% 2 \nabla_{\alpha} N
%\end{eqnarray*}
it follows that $c$ can always be adjusted so that 
$ \chi + \chibar = 2 N$. This proves the Lemma for class (A).

Assume now that $\J$ is not identically
zero. We first show that $\J$ vanishes nowhere. 
Let  $V_1 \defi \{ p \in \M; \J (p) \neq 0 \}$ and assume that
$V_1 \not = \M$. Let $p$ be a point in the topological boundary
$\partial V_1$. Sufficiently near $p$, $P_{\alpha}$ is exact 
$P_{\alpha} = \nabla_{\alpha} P$ for some smooth function $P$.
Since
$P_{\mu}$ vanishes if and only if $\xi_\mu$ vanishes and the set
of fixed points of a Killing vector is at least of
co-dimension two, $p$ can be chosen so that $P_{\mu} |_p \neq 0$.
Restricting the neighbourhood of $p$ if necessary we can assume
$P_{\mu}$ nowhere zero there and hence there exists a
smooth complex function $\j$  defined on a neighbourhood
of $P_0\defi P(p) \in \mathbb{C}$ 
such that $\J = \j \circ P$ and
$\j' = \j^2$ (this follows from (\ref{nablanablaFsq_a})). 
Uniqueness  of solutions of ODE
implies $\j \equiv 0$
in a neighbourhood of $p$, against
hypothesis. Thus $V_1 = \M$ and $\J \neq 0$ everywhere. 
We can define $P = - \frac{1}{\J}$ which satisfies 
$\nabla_{\alpha} P = P_{\alpha}$ (hence $P_{\alpha}$ is exact) and vanishes nowhere.

Equation (\ref{nablaFsq_a}) becomes 
\begin{eqnarray*}
\nabla_{\alpha} R = - \left ( \frac{2R}{P} + \Lambda \right ) \nabla_{\alpha} P
\end{eqnarray*}
which integrates to $R = r \circ P$ with
$$r=\frac{b}{2\zeta^2} - \frac{\Lambda}{3} \zeta, \quad b \in \mathbb{C}.
$$
Finally, $\chi_{\alpha} = 2 R P_{\alpha}$ reads
\begin{eqnarray*}
\chi_{\alpha} = -\nabla_{\alpha} \left ( \frac{b}{P} + \frac{\Lambda}{3} P^2
\right ),
\end{eqnarray*}
which proves that $\chi_{\alpha}$ is exact $\chi_{\alpha} = \nabla_{\alpha} \chi$
with $\chi = x \circ P$ and $x: \mathbb{C} \setminus \{ 0 \} \rightarrow
\mathbb{C}$ given by 
$x(\zeta) = c - b/\zeta - (\Lambda/3) \zeta^2$, where $c$
is an arbitrary constant. Again without loss of generality
$c$ can be chosen  to be real and adjusted so that 
$\chi + \ov{\chi} = 2 N$.  $\hfill \qed$.

\vs 

{\bf Remark}: In the rest of the paper all results will split into the two cases (A) and (B) of this theorem. We will simply write (A) and (B) without further mention, e.g. in Proposition \ref{prop:2ndKilling} below.

\vs

The complex function $P$ will be decomposed in real and imaginary parts
as 
$$
P = y + i Z
$$
where $y,Z: \M \rightarrow \mathbb{R}$ are smooth.
The condition $P_{\alpha}P^{\alpha} = N$ (\ref{Psquare}) imposes
\begin{eqnarray}
\nabla_{\alpha} y\nabla^{\alpha} y - \nabla_{\alpha} Z \nabla^{\alpha} Z = N
\quad \quad \nabla_{\alpha} y \nabla^{\alpha} Z = 0. 
\label{nablaynablay}
\end{eqnarray}
From $g(\xi, P) = g(q,P)= 0$ and $g(k,P) = g (\xi,k)$ (\ref{gk})
and taking real and imaginary parts one has
\begin{eqnarray}
\xi (y) = 0, \quad  \xi (Z)=0, \quad q (y) = 0, \quad q(Z)=0,\quad k(y) = g(\xi,k),  \quad
k(Z) = 0. \label{deryZ}
\end{eqnarray}

From (\ref{nablaq}) and the formulas for case (A) in Lemma \ref{Functions_of_P} one sees that, in this case, $\nabla_{(\mu}q_{\nu)}=0$, so that $q$ is another Killing vector field on $\M$. The existence of a second Killing vector is proven in general in the next proposition
\begin{Proposition}
\label{prop:2ndKilling}
The following vector fields
\begin{itemize}
\item[(A)] $\varsigma = A\xi +C q$,
\item[(B)] $\varsigma = (A -CP^{2}-C \overline P^{2})\xi +2C P \overline P q$.
\end{itemize}
are Killing vectors on $(\M,g)$  for arbitrary real constants $A$ and $C$.
Moreover, any two vector fields in this class commute. 
\end{Proposition}

\vs

\noindent {\it Proof:} 
Let $\varsigma =F\xi + G q$ for some real functions $F$ and $G$ to be determined. A direct calculation using (\ref{nablaq}) and that $\xi$ is Killing provides
\bean
\nabla_{\mu}\varsigma_{\nu}+\nabla_{\nu}\varsigma_{\mu}=
\xi_{\mu}\left(\nabla_{\nu}F-G\overline J P_{\nu}-G J \overline P_{\nu} \right)+
\xi_{\nu}\left(\nabla_{\mu}F-G\overline J P_{\mu}-G J \overline P_{\mu} \right)\\
+q_{\mu}\left(\nabla_{\nu} G+GJ P_{\nu} +G\overline J \overline P_{\nu}\right)+
q_{\nu}\left(\nabla_{\mu} G+GJ P_{\mu} +G\overline J  \overline P_{\mu}\right)
\eean
so that $\varsigma$ is a Killing vector
provided the following two equations are satisfied
$$
\nabla_{\mu}F=G\overline J P_{\mu}+G J \overline P_{\mu},Ê\quad \quad
\nabla_{\mu} G=-GJ P_{\mu} -G\overline J \overline P_{\mu} .
$$
Using the formulas in Lemma \ref{Functions_of_P}
 we have that their general solution is $F=A$, $G=C$, where $A$ and $C$ are real constants in case (A). For case (B), the second equation integrates easily to
$$
G=2C P\overline P
$$
and then the first can be solved to give
$$
F=-C(P^{2}+\overline P^{2}) +A.
$$ 
The commutation follows directly from (\ref{Lie}) and $\xi (P)=\xi^{\mu}P_{\mu}=0$.
%For the commutation it suffices to show $[\xi,q]=0$ in case (A) 
%and $[\xi, 2 P \ov{P} q - (P^2 + \ov{P}^2) \xi]=0$ in case (B).
%Both follow as a straightforward consequence 
%of (\ref{nablaq})-(\ref{nablaxi}). 
\hfill $\qed$

\vs
{\bf Remark.}  In the case of vanishing cosmological constant and
assuming that the Killing vector $\xi$ is timelike so that it makes sense
to pass locally to the quotient manifold defined by the
orbits of $\xi$, a vector field equivalent to the one in case (B) of this
Proposition was introduced by Perj\'es \cite{Perjes} as
a useful tool to characterize locally the vacuum strictly stationary spacetimes
with vanishing Simon tensor. This vector field was shown to leave invariant the metric components \cite{Perjes}, so that it is 
a Killing vector field of the spacetime, see also \cite{Simon2}.

\vs 

This Proposition
provides two linearly independent Killing vectors {\em except} in some special situations where $q$ and $\xi$ happen to be co-linear. This leads to the analysis of the special cases arising from (\ref{xiwedgeq}).

\subsection{Special cases}
\label{sec:special}

\vs 
From Eq.(\ref{xiwedgeq}) we know that the set $\{\xi,q,P,\ov P\}$ constitutes a basis on the tangent spaces {\em unless} $P$ and $\ov P$ are collinear. 
These special cases are characterized by the condition $\bm{P}\wedge \ov{\bm P} =0$.
First note that at any $p \in \M$ and for
any $a \in \mathbb{C}$,
\begin{eqnarray}
\ov{\bm{P}} |_p =  a \bm{P}|_{p} \quad \Longleftrightarrow \quad q |_p = 
a \xi|_p, \label{equiv}
\end{eqnarray}
as a direct consequence of (\ref{PxiW}). Unless $\xi|_p=0$ (and then $q|_p = \bm{P}|_p =0$)
it follows that $a$ is real and actually $\bm{P}=a\ov{\bm{P}}=a^{2} \bm{P}$ at $p$ so that either $a=\pm 1$ or $\bm{P}|_{p}=0$.
%Assume $\bm{P} \wedge \ov{\bm P} |_p =0$
%at a point $p$. This is equivalent to 
%$dy \wedge d Z |_p =0$. If $dZ|_p \neq 0$ we necessarily have $dy |_p =0$.
%Indeed, there must be a constant $a_0$ such that $dy|_p = a_0 dZ |_p$. 
%The ortogonality $(dy,dZ)|_p =0$ forces $a_0 (dZ, dZ) |_p =0$
%but $(dZ,dZ)|_p$ can only vanish if $dZ|_p=0$, because $dZ|_p$ is ortogonal
%to $k|_p$ and $\ell|_p$. Hence $a_0=0$ as claimed. 
Thus, $\bm{P} \wedge
\ov{\bm{P}} |_p=0$ is equivalent to either $\ov{\bm{P}|}_p = \epsilon
\bm{P} |_p$ with $\epsilon = \pm 1$ or $\bm{P}|_{p} =0$. In each
case one has, respectively, $q |_p = \epsilon \xi|_p$ or $q |_p=0$ and $\xi|_{p}=0$, the last as a consequence of Lemma \ref{lem:chinonzero}.
Conversely, if $q|_p$ is proportional to $\xi |_p \neq 0$,
then necessarily $q |_p = \pm \xi|_p$ because of (\ref{equiv}).
Moreover, $\xi |_p =0$ implies $q |_p =0$ and $\bm P|_{p}=0$.

If $\bm{P} \wedge \ov{\bm{P}} =0$
on an open connected set, then there is $\epsilon \in \{ +1,-1 \}$
such that $\bm{P} = \epsilon \ov{\bm{P}}$ everywhere because of Lemma 
\ref{lem:chinonzero}  and
%the set of point where $\bm{P}$ vanishes cannot separate the open set.
we then have $q = \epsilon \xi$. Conversely, $\bm{\xi} \wedge \bm{q} =0$ 
on an open set implies $q = \epsilon \xi$ for $\epsilon \in \{-1,1\}$
constant and then $\bm{P} = \epsilon \ov{\bm{P}}$.
% or equivalently $\bm P= \epsilon \ov{\bm{P}}$ for some $\epsilon$.  From 
%$N=P_\mu P^\mu = \epsilon^2 \ov P_\mu \ov P^\mu = \epsilon^2 N$ one necessarily% gets
% $\epsilon =\pm 1$ (i.e. 
%$P_\mu$ is either real or purely imaginary).
%In both cases one has from (\ref{xiwedgeq}) that $\xi_\mu$ and $q_\mu$ (or $\et%a_\mu$) are collinear, and actually 
%$$
%\xi_\mu = \epsilon q_\mu \, .
%$$
%Similarly, if one starts with the assumption that $\bm{\xi}\wedge \bm{q} =0$ on%e arrives at exactly the same conclusions. All this holds at any point $p\in \M%$ where $\bm{P}\wedge \ov{\bm P}|_{p} =0$.
From (\ref{divP}) it also follows
\be
R - \epsilon \ov R = N(J - \epsilon \ov J) \label{RpmR}
\ee
which, upon using the expressions in Lemma \ref{Functions_of_P}, becomes
\begin{align}
& \mbox{(A):} \quad 
  y (\epsilon -1 ) - i Z (\epsilon + 1) =0  \label{specialA} \\
& \mbox{(B):} \quad 
(\epsilon - 1) \left (  b_1 - 2 c y + \frac{4 \Lambda}{3} y^3 \right )
+ i (\epsilon + 1) \left ( b_2 - 2 c Z - \frac{4 \Lambda}{3} Z^3 \right ) =0.
\label{specialB}
\end{align}
Further information can be obtained for each value of $\epsilon$ separately

\begin{enumerate}
\item $\bm{P} |_p \neq 0$ purely imaginary ($\epsilon =- 1$):
From  (\ref{xi+q},\ref{P+P}) and (\ref{gk},\ref{gell}) one gets at $p$
\be
g(k,P)=g(k,q)=g(k,\xi)=g(\ell,P)=g(\ell,q)=g(\ell,\xi)=0 \label{orthogonal}
\ee
so that $\left<\xi,P\right>$ is at $p$ a plane orthogonal to the plane generated by the principal null directions $\left<k,\ell\right>$. One also has
\be
2\, \bm\xi \wedge \bm P |_{p}= N(\ov{\bm \W}-\bm \W) |_{p}=+2N i \bm{W}^{+}|_{p} . 
\label{xiwedgeP}
\ee
%as well as
%\be
%R+\ov R = N(J + \ov J) \, .\label{R+R}
%\ee
%As $P_\mu =\nabla_\mu P$ is purely imaginary the function $P$ takes the form\mnotex{J: Some rewriting and additions in what follows}
%\be
%P=l  + i Z \label{ele}
%\ee
%where $l $ is a real constant and $Z$ is a smooth real function. Using the info%rmation (\ref{R+R}) in the formulas of Lemma \ref{Functions_of_P} 
%one gets the following relation between the constants for the case (B) 
%\be
%\mbox{case (B):} \hspace{1cm} b+\ov b = 4l  \left(c-2\frac{\Lambda}{3} l^2 \rig%ht) \label{constants}
%\ee
%while in case (A) one gets simply
%\be
%\mbox{case (A):} \hspace{1cm} l=0 \label{l=0} .
%\ee
\item $\bm{P} |_p \neq 0$ real ($\epsilon = 1$):
From (\ref{xi+q},\ref{P+P}) and (\ref{gk},\ref{gell}) one gets
\be
\bm P|_p=\overline{\bm P}|_p= (g(\ell,\xi)\bm k -g(k,\xi)\bm\ell)|_p , \quad \bm \xi|_p=\bm q |_p=(-g(\ell,\xi)\bm k -g(k,\xi)\bm\ell )|_p
\label{noortogonal}
\ee
so that $\left<\xi,P\right>$ is at $p$ the plane generated by the principal null directions $\left<k,\ell\right>$. One also has 
\be
2 \bm\xi \wedge \bm P |_p= 2 N\bm k \wedge \bm \ell |_p=
- N ( \bm{\W} + \ov{\bm \W} )|_p=  -2 N \bm{W}^- |_p
\label{xiwedgeP_bis}
\ee
%as well as
%\be
%R-\ov R = N(J - \ov J) \, .\label{R-R}
%\ee
%As $P_\mu =\nabla_\mu P$ is real the function $P=f+i\delta$ where $\delta$ is a% real constant and $f$ is a smooth real function. Using the information (\ref{R-R}) in the formulas of Lemma \ref{Functions_of_P} for case (B) one gets the following relation between the constants
%$$
%b-\ov b = 4i\delta \left(c+2\frac{\Lambda}{3} \delta^2 \right).
%$$
\item 
 If $\bm{P}|_p=0$, all equations (\ref{orthogonal})-(\ref{xiwedgeP_bis})
 hold trivially as $\xi |_p=0$ and $q|_p=0$ too.
\end{enumerate}

\section{Solving the field equations}
\label{sec:solving}

Equation (\ref{nablaP}) will be the key
for the integration of the field equations. This integration will proceed by 
identifying a natural Riemannian submersion and integrating the equations first
on the quotient space and then going up into the total space. 
In the generic case, namely when the Killing vector
$\xi$ is not simultaneously orthogonal to both null eigenvectors
$k$ and $\ell$ and when $dZ \neq 0$,
 we will be able to identify enough structure in the quotient
so that its geometry can be fully determined. In the special
cases when either $dZ=0$ or when
$\xi$ is orthogonal to both null eigenvectors, the problem will 
become harder and the fundamental equations for Riemannian submersions 
will have to be used. In fact, different submersions will have to be
introduced in order 
to deal, respectively, with the cases (i) $dZ=0$ and (ii) $\xi$ orthogonal
to both $k$ and $\ell$. In subsection \ref{submersion}
we identify the conformal rescaling 
in the spacetime that will allow is to define the Riemannian
submersions in the following subsections.
In subsection \ref{generic} we assume that $\xi$ and $k$
are not orthogonal and introduce a two-dimensional distribution
in the spacetime which is proven to be integrable and to define a Riemannian
submersion. The geometry of this Riemannian submersion 
is analyzed in detail with the additional assumption that
$dZ \neq 0$. In subsection \ref{special}   we study this
distribution in the case when $dZ=0$ and  we also introduce a
second distribution capable of dealing with the situation
when $\xi$ is simultaneously orthogonal to $k$ and $\ell$.
This second distribution is also shown to be two-dimensional
and integrable and to define a Riemannian submersion.  Despite the
very different nature of both distributions,
we introduce a notation that allows us to work with them simultaneously.
Once the geometry of the Riemannian submersions is understood,
we proceed by integrating up the field equations and 
determining the most general class of spacetime metrics satisfying our 
characterization  hypotheses. This is done in subsection \ref{integrationup}
where the main results of this section are stated and proved. Theorem
\ref{non-orthogonal} determines the  spacetime metric when the Killing vector
is not orthogonal to the eigenspace of the 
self-dual Killing form and Theorem \ref{th:orthogonal}
is the analogous result when 
$\xi$ is simultaneously orthogonal to $k$ and $\ell$.

\subsection{Conformal rescaling}
\label{submersion}

Let us first 
recall that a Riemannian submersion between two semi-Riemannian
manifolds (of arbitrary signature) 
$(\M,g)$ and $(S,h)$ is a smooth surjective map
$\pi : \M \rightarrow S$ of rank = $\mbox{dim} (S)$
such that, for all $p \in S$ and all $q \in V_p \defi \pi^{-1} (p)$
the tangent space $T_q \M$ admits a direct sum decomposition
$T_q \M = T_q V_p \oplus N_q V_p$, where $T_q V_p$ is the tangent
space of $V_p$ (this is a submanifold because $\pi$ is of maximum rank)
and $N_q V_p = ( T_q V_p )^{\bot}$ (orthogonal with respect to the metric $g$)
satisfying the property that $\pi_{\star} :N_q V_p \longrightarrow T_p S$ is
an isometry. For details on Riemannian submersions see \cite{ONeil}.

Consider now an integrable distribution $\D$
in $\M$ of dimension $m$ and assume that
at each point $p \in \M$ the induced metric at $\D_p \subset T_p \M$
is non-degenerate. Let $\{ \L_{\alpha} \}$ be the collection of maximal
integrable manifolds of the distributions (which exists by the
Frobenius theorem). This collection defines a foliation in $\M$ (as usual,
each integrable manifold is called a ``leaf'').
For any subset $A \subset \M$ we define an equivalence relation:
$p_1, p_2 \in A$ are related $p_1 \sim p_2$ if and only if the there exists a
smooth path $\gamma$ from $p_1$ to $p_2$ fully contained in $A$
with tangent vector $\gamma'$ everywhere tangent to $\D$. The quotient
space will be denoted by $A / \sim$.

We can construct the orthogonal projector to $\D_p$, i.e. the one-one tensor
$h^{\alpha}_{\,\,\,\beta}$
satisfying $h |_{\D_p} = 0$ and $h |_{\D_p^{\bot}} = \mbox{Id} |_{\D_p^{\bot}}$.
Let $h_{\alpha\beta} = g_{\alpha\mu} h^{\mu}_{\,\,\,\beta}$ (which is well-known to be symmetric) and 
let $X$ be an arbitrary  vector field on $\M$ tangent to $\D$ (i.e.
$X|_p \in \D_p, \forall p \in \M$). Then the following result is well-known
(cf. Theorem 1.2.1 in \cite{Gromoll})
\begin{lemma}
\label{existencesubmersion}
If $\pounds_{X} h_{\alpha\beta} = 0$ for all vector fields $X$ tangent to $\D$, then
for any $p \in \M$ there exists an open connected neighbourhood $U_p$
of $p$  such that $U_p / \sim$ is a smooth manifold and the projection
$\pi: U_p \rightarrow U_p / \sim$ is a Riemannian submersion.
\end{lemma}
The condition for $\pounds_X h_{\alpha\beta}$ can be rewritten in terms of
covariant derivatives of $X_{\alpha}$ as follows
\begin{lemma}
\label{Liecov}
For any vector field $X$ tangent to $\D$ the following expression holds
\begin{eqnarray*}
\pounds_X h_{\alpha\beta} = h^{\mu}_{\,\,\,\alpha} h^{\nu}_{\,\,\,\beta}
\left ( \nabla_{\mu} X_{\nu}
+ \nabla_{\nu} X_{\mu} \right ).
\end{eqnarray*}
\end{lemma}
\vs
\noindent {\it Proof:} Let $Y$ be any vector field tangent to the
distribution $\D$. It follows that $\pounds_{X} Y$ is also tangent to the
distribution. Let $V_{\alpha}$
be any one-form normal
to the distribution, then $\pounds_{X} V_{\alpha}$ is also normal
to the distribution because, for any tangent vector $Y$, we have
\begin{eqnarray*}
\left ( \pounds_X V_{\alpha} \right ) Y^{\alpha} = \pounds_{X}
\left ( V_{\alpha} Y^{\alpha} \right ) - V_{\alpha} [X,Y]^{\alpha} =0.
\end{eqnarray*}
Also $(\pounds_{X} h^{\alpha}_{\,\,\,\mu}) V_{\alpha} =0$, because
\begin{eqnarray*}
\left ( \pounds_{X} h^{\alpha}_{\,\,\,\mu} \right ) V_{\alpha}  = 
\pounds_{X} \left (  h^{\alpha}_{\,\,\,\mu} V_{\alpha} \right ) 
- h^{\alpha}_{\,\,\,\mu} \pounds_{X} V_{\alpha}  =
\left ( \delta^{\alpha}_{\,\,\,\mu} - h^{\alpha}_{\,\,\,\mu} \right )
\pounds_{X} V_{\alpha} = 0.
\end{eqnarray*}
It it clear then that $(\pounds_{X} h^{\mu}_{\,\,\,\alpha}) h_{\mu\nu} =0$
(because its contraction in the index $\nu$ with a tangent vector or with a
normal vector vanishes).  
Using $h_{\alpha\beta} = h^{\mu}_{\,\,\,\alpha} h^{\nu}_{\,\,\,\,\beta} g_{\mu\nu}$,
\begin{equation*}
\pushQED{\qed}
\pounds_{X} h_{\alpha\beta} =
\pounds_{X} \left ( 
h^{\mu}_{\,\,\,\alpha} h^{\nu}_{\,\,\,\,\beta} g_{\mu\nu} \right ) =
h^{\mu}_{\,\,\,\alpha} h^{\nu}_{\,\,\,\,\beta}  \pounds_{X} g_{\mu\nu}  =
h^{\mu}_{\,\,\,\alpha} h^{\nu}_{\,\,\,\,\beta} \left ( \nabla_{\mu} X_{\nu}
+ \nabla_{\nu} X_{\mu} \right ). \qedhere
\popQED
\end{equation*}

The Riemannian submersions we will define involve a 
suitable conformal rescaling of $g$. We start
with the following Lemma concerning the principal null
direction $k$. 
\begin{lemma}
\label{LieDerivative}
Let $\D$ be an integrable distribution in $\M$ such that 
$k |_p \in  \D_p$ $\forall p \in \M$ and let $h_{\alpha\beta}$ be the corresponding
projector. Then
\begin{align}
% h^{\alpha}_{\,\,\,\mu} h^{\beta}_{\,\,\,\nu} \left (
%\nabla_{\alpha} k_{\beta} - \nabla_{\beta} k_{\alpha} 
%\right ) & = - i g (\xi,k) \left ( \ov{\J} - \J 
%\right ) W^{+}_{\alpha\beta}
% h^{\alpha}_{\,\,\,\mu} h^{\beta}_{\,\,\,\nu}. 
h^{\alpha}_{\,\,\,\mu} h^{\beta}_{\,\,\,\nu} \left (
\nabla_{\alpha} k_{\beta} - \nabla_{\beta} k_{\alpha} 
\right ) & =  i \left (\J - \ov{\J}
\right ) \xi^{\sigma}k^{\rho}\eta_{\sigma\rho\alpha\beta}
 h^{\alpha}_{\,\,\,\mu} h^{\beta}_{\,\,\,\nu}, \label{exterior}\\
\pounds_{k} h_{\alpha\beta} & = 
- g (\xi,k) \left ( \J + \ov{\J} \right ) h_{\alpha\beta}.
\label{Lieprojector}
\end{align}
Moreover, if $g(\xi,k) \neq 0$ everywhere then
$ \displaystyle{  k^{\alpha} \nabla_{\alpha} k_{\beta} 
= \frac{k(g(\xi,k))}{g(\xi,k)} k_{\beta}}$.
\end{lemma}
%{\bf Remark.} Using the symmetry
%$k \longleftrightarrow l$, $R \longleftrightarrow -R$,  which induces the transformation
%$J \longleftrightarrow -J$, we conclude that for any integrable
%distribution in $\M$ with $l$ tangent to the leaves we have
%\begin{eqnarray*}
%\pounds_{k} h_{\alpha\beta} = 
% g (\xi,\ell) \left ( \J + \ov{\J} \right ) h_{\alpha\beta}
%\end{eqnarray*}
%where $h_{\alpha\beta}$ is the corresponding projector.

\noindent {\it Proof.}
Projecting (\ref{nablak}) and using the fact that 
$h^{\alpha}_{\,\,\,\beta} k_{\alpha} =0$ one gets
\begin{eqnarray*}
2h^{\alpha}_{\,\,\,\mu} h^{\beta}_{\,\,\,\nu} \nabla_{\alpha} k_{\beta} =
- g (\xi,k) \left ( \J  + \ov{\J}\right ) h_{\mu\nu}+i(J-\ov J) \xi^{\sigma}k^{\rho}\eta_{\sigma\rho\alpha\beta}
 h^{\alpha}_{\,\,\,\mu} h^{\beta}_{\,\,\,\nu}.
\end{eqnarray*}
The symmetric and anti-symmetric part of this relation give (\ref{Lieprojector}) and (\ref{exterior}), respectively. For the
last statement, (\ref{nablak}) gives $2k^{\alpha} \nabla_{\alpha} k_{\beta} =
-(\ell^{\rho} k^{\sigma}\nabla_{\sigma}k_{\rho}) k_{\beta}$. Contracting with $\xi^{\beta}$ fixes
\begin{equation*}
\pushQED{\qed}
-\ell^{\rho} k^{\sigma}\nabla_{\sigma}k_{\rho} =\frac{2\, k \left ( g (\xi,k \right ) )}{
g ( \xi,k)}. \qedhere
\popQED
\end{equation*}

\vs 
This Lemma  suggests the metric transformation required in order
to define a Riemannian submersion. Let $s : \mathbb{C} \setminus \{0\}  \rightarrow \mathbb{C} \setminus \{0\}$
be a solution
of the ODE $s' = -s \j$ and define $S \defi s \circ P$ and 
$\Omega \defi S \ov{S}$. The latter is  a positive real function satisfying
\begin{eqnarray}
\nabla_{\alpha} \Omega = - \Omega \left ( J P_{\alpha} + \ov{J}
\ov{P}_{\alpha} \right ). \label{nablaOmega}
\end{eqnarray}
The conformally related metric
$$
\hat{g}_{\alpha\beta} \defi \frac{1}{\Omega} g_{\alpha\beta}
$$  will play
a fundamental role in  the integration of the field equations.
Without loss of generality we choose $s$ (and hence $\Omega$)
as follows in each of the classes defined in Lemma \ref{Functions_of_P}:
\begin{itemize}
\item[(A)] $\displaystyle{ s(\zeta) = 1, \quad \quad \Omega =1}$.
\item[(B)] $\displaystyle{s(\zeta) = \zeta, \quad \quad \Omega = y^2 + Z^2.}$
\end{itemize}

\begin{lemma}
\label{invariance}
Let $\D$ be a distribution satisfying the same hypothesis as in Lemma
\ref{LieDerivative}. Let $\hat{h}_{\alpha\beta}$ be the corresponding projector
with the spacetime metric $\hat{g}$. Then
\begin{eqnarray*}
\pounds_{k} \hat{h}_{\alpha\beta} = 0. 
\end{eqnarray*}
\end{lemma}

\vs

\noindent {\it Proof:}
First note that $k
(\Omega) = - \Omega \left ( J + \ov{J} \right ) 
g (\xi,k)$. Since 
$\hat{h}_{\alpha\beta} = \frac{1}{\Omega} h_{\alpha\beta}$ , it follows
\begin{equation*}
\pushQED{\qed}
\pounds_{k} \hat{h}_{\alpha\beta} = \frac{1}{\Omega} 
\left ( \pounds_{k} h_{\alpha\beta} 
- \frac{k(\Omega) }{\Omega} 
h_{\alpha\beta} \right ) =
\frac{1}{\Omega} 
\left ( \pounds_{k} h_{\alpha\beta} + g (\xi,k) \left ( \J + \Jbar \right )
h_{\alpha\beta} \right ) = 0. \qedhere
\popQED
\end{equation*}

\vs 

We adopt the convention that all spacetime indices are raised and lowered with the
metric $g$, unless contrarily indicated.
The covariant derivative with respect to $\hat{g}$ will be denoted by $\nablahat$. 
The standard
transformation law for covariant derivatives under conformal rescaling allows us to obtain
$\nablahat_{\mu} P_{\alpha}$ from (\ref{nablaP}). The result is
\begin{eqnarray}
\nablahat_{\mu} P_{\alpha}  = \frac{R}{2} g_{\mu\alpha} - 
\frac{1}{R} t_{\mu\alpha} - \J \left ( \xi_{\mu}\xi_{\alpha} + \frac{N}{2} g_{\mu\alpha}
\right ) - \frac{\Jbar}{2} \left ( \Pbar_{\mu} P_{\alpha}
+ P_{\mu} \ov{P}_{\alpha} - g_{\mu\alpha} P_{\nu} \ov{P}^{\nu} \right )
\label{nablahatP}
\end{eqnarray}
whose imaginary part implies the following equation for $Z$: 
\begin{align}
2 i \nablahat_{\mu} \nablahat_{\alpha} Z = &
\left ( R - \Rbar \right ) 
\left ( \frac{1}{2} g_{\mu\alpha}+ \frac{1}{R \ov{R}}
t_{\mu\alpha} \right )+ \nonumber \\
& + \left ( \J - \Jbar \right )
\left ( - \xi_{\mu} \xi_{\alpha} + \nabla_{\mu} y \nabla_{\alpha} y  + 
\nabla_{\mu} Z \nabla_{\alpha} Z - g_{\mu\alpha} \left ( N + \nabla_{\nu} Z \nabla^{\nu} Z
\right ) \right ). \label{HessZ1}
\end{align}

\subsection{Riemannian submersion in the generic case}
\label{generic}

In this subsection we assume that $g(\xi,k) \neq 0$
and introduce a distribution shown to be a Riemannian submersion.
Several of its properties, required later, are stated and proven.
We start will the following Proposition.
\begin{Proposition}
\label{geometryquotient}
Assume that $\xi$ satisfies $g(\xi,k) \neq 0$ everywhere and define 
$\D_p = \mbox{span} \{ \xi, k \} |_p$ $\subset T_p \M$.
This collection of vector subspaces defines
a two-dimensional integrable distribution with timelike leaves. 
Let $h^{\mu}_{\,\,\,\alpha}$ be the corresponding orthogonal
projector and $\hat{h}_{\alpha\beta} = \hat{g}_{\mu\alpha} h^{\mu}_{\,\,\,\alpha}$. Then, for any vector
field $X$ tangent to the distribution we have $\pounds_{X} \hat{h}_{\alpha\beta} =0$ and the following
equation holds
\begin{eqnarray}
h^{\mu}_{\,\,\,\nu} h^{\alpha}_{\,\,\,\beta} \nablahat_{\mu} \nablahat_{\alpha} Z = U \hat{h}_{\nu\beta},
\quad \quad \mbox{with} \quad \quad 
U \defi \frac{\Omega}{2 i} \left ( R - \ov{R} - N \left ( \J - \Jbar \right ) \right ).
\label{HessianZ}
\end{eqnarray}
\end{Proposition}

{\bf Remark:} Substituting the explicit expressions in classes (A) and (B)
defined in Lemma \ref{Functions_of_P}, it follows that $U = u \circ Z$,
where  $u : \mathbb{R} \setminus \{ 0 \} \rightarrow \mathbb{R}$ is given by
\begin{itemize}
\item[(A)] $\displaystyle{ u = - \Lambda \zeta }$.
\item[(B)] $\displaystyle{ 
u =  \frac{b_2}{2} - c \, \zeta - 
\frac{2 \Lambda}{3} \zeta^3
}$.
\end{itemize}

\vs

{\it Proof:} The condition $g (\xi,k) \neq0$ implies that $\xi$ and $k$ are linearly independent everywhere, so that $\D_p$ is two-dimensional. To show that it is integrable it suffices to
show that $[\xi,k] |_p \in \D_p$, but this is immediate because $k$ is a simple eigenvector
of a tensor $\F_{\alpha\beta}$ which is invariant under $\xi$ and hence
$[\xi, k] \propto k$. The induced metric on $\D_p$ is non-degenerate because clearly no linear
combination of $\xi$ and $k$ can be orthogonal to $\xi$ and $k$ simultaneously. 
Being non-degenerate, it is obviously timelike because it contains a null direction. 
For the statement  $\pounds_{X} \hat{h}_{\alpha\beta} =0$, it suffices to check that
$\pounds_{\xi} \hat{h}_{\alpha\beta} =0$ and
$\pounds_{k} \hat{h}_{\alpha\beta} =0$. The first is immediate from the fact that 
$\xi$ is a Killing vector for $\hat{g}$, due to $\xi(\Omega)=0$, and the second
has been proven in Lemma \ref{invariance}. 

In order to establish (\ref{HessianZ}), we define $V_{\alpha} \defi h^{\mu}_{\,\,\,\alpha}
\nabla_{\mu} y$. 
Note that $h^{\mu}_{\,\,\,\alpha}  \nabla_{\mu} Z = 
\nabla_{\alpha} Z$ because of (\ref{deryZ}). 
From (\ref{EMTensor}) it follows $t_{\mu\alpha} h^{\nu}_{\,\,\,\nu} 
h^{\alpha}_{\,\,\,\beta} = 
\frac{R \ov{R}}{2} h_{\nu\alpha}$ and (\ref{HessZ1}) implies
\begin{eqnarray*}
2 i  h^{\mu}_{\,\,\,\nu} h^{\alpha}_{\,\,\,\beta} \nablahat_{\mu} \nablahat_{\alpha} Z =
\left ( R - \ov{R}  - N \left ( \J - \Jbar \right ) \right ) h_{\nu\beta} 
+ \left ( \J - \Jbar \right )
\left ( 
V_{\nu} V_{\beta} + \nabla_{\nu} Z \nabla_{\beta} Z - h_{\nu\beta} \nabla_{\rho} Z
\nabla^{\rho} Z 
\right ).
\end{eqnarray*}
Since $h_{\nu\beta} = \Omega \hat{h}_{\nu\beta}$ it only remains to show that
$V_{\nu} V_{\beta} + \nabla_{\nu} Z \nabla_{\beta} Z - h_{\nu\beta} \nabla_{\rho} Z
\nabla^{\rho} Z$ vanishes. From (\ref{nablaynablay}) one has $V_{\alpha} \nabla^{\alpha} Z =0$. Moreover, it is immediate to check
that the projector $h_{\alpha\beta}$ can be written as
\begin{eqnarray}
h_{\alpha\beta} = g_{\alpha\beta} - \frac{N}{g(\xi,k)^2} k_{\alpha} k_{\beta}
- \frac{1}{g(\xi,k)} \left ( k_{\alpha} \xi_{\beta} + k_{\beta} \xi_{\alpha}
\right )
\label{DecomProjector}
\end{eqnarray}
and hence, using (\ref{deryZ}),
\begin{align*}
V_{\alpha} V^{\alpha} & =  h^{\alpha\beta} \nabla_{\alpha} y \nabla_{\beta} y =
\left ( g^{\alpha\beta} - \frac{N}{g(\xi,k)^2} k^{\alpha} k^{\beta}
- \frac{1}{g(\xi,k)} \left ( k^{\alpha} \xi^{\beta} + k^{\beta} \xi^{\alpha}
\right ) \right) \nabla_{\alpha} y \nabla_{\beta} y = \\
& = \nabla_{\alpha} y \nabla^{\alpha} y - N = 
\nabla_{\alpha} Z \nabla^{\alpha} Z
\end{align*}
The claim now follows from the fact that  $\D_p^{\bot}$ is two-dimensional.
$\hfill \qed$

%The following Proposition will be used later for the integration of the
%spacetime metric from the geometry of the quotient.
%\begin{Proposition}
%Assume that $g(\xi,k)\neq0$ and define $D_p$ as in the previous Proposition.
%Assume that $\nabla_{\alpha}Z \neq0 $ everywhere. Fix the normalization
%of $k$ so that $g(\xi,k)$ is constant. Then the following equations
%hold
%\begin{eqnarray*}
%k^{\mu} \nabla_{\mu} k_{\alpha} =0 \quad \quad \Longleftrightarrow
%\quad \quad \pounds_{k} k_{\alpha} =0, \\
%h^{\alpha}_{\beta} h^{\mu}_{\nu}
%\left ( \nabla_{\mu} k_{\alpha} - \nabla_{\alpha} k_{\mu} \right ) 
%=  \left ( \Jbar - J \right ) g (\xi,k) \eta_{\alpha\beta}
%\end{eqnarray*}
%\end{Proposition}

The direct sum decomposition $T_p \M = \D_p \oplus \D_p^{\bot}$
allows to decompose any vector $X$ in $T_p \M$ as $X = X^{\VV} + X^{\HH}$, where
${\cal V}$ stands for ``vertical'' (i.e. along $\D_p$) and ${\cal H}$ for
``horizontal'' (i.e. along $\D_p^{\bot}$). The following lemma relates
the orientation of the quotient space to the orientation of the ambient
spacetime.

\begin{lemma}
\label{volume}
Assume that $g(\xi,k)\neq 0$ everywhere,
define $\D_p = \mbox{span} \{ \xi, k \} |_p$ $\subset T_p \M$
and assume that $\M/\sim$ is a manifold.
Then there exists a volume form $\hat{\bm \eta}$ on $\M / \sim$ such that
$\hat{\eta}_{\mu\nu} \defi \pi^{\star} (\hat{\bm \eta})_{\mu\nu}$ satisfies
\begin{eqnarray*}
\Omega^{-1} \eta_{\alpha\beta\rho\sigma} h^{\alpha}_{\,\,\,\mu} h^{\beta}_{\,\,\,\nu} k^{\rho}
\ell^{\sigma} = \Omega^{-1} W^{+}_{\mu\nu} = \hat{\eta}_{\mu\nu}.
\end{eqnarray*}
In particular $\displaystyle{
 h^{\alpha}_{\,\,\,\mu} h^{\beta}_{\,\,\,\nu} \left (
\nabla_{\alpha} k_{\beta} - \nabla_{\beta} k_{\alpha} 
\right )  = i g (\xi,k) \left (J - \ov{J}\right )
 \Omega \hat{\eta}_{\mu\nu}}$.
\end{lemma}

\vspace{5mm}

{\it Proof:} The two-form $W^{+}_{\alpha\beta} = \eta_{\alpha\beta\rho\sigma}
k^{\rho} \ell^{\sigma}$ is orthogonal to $k$ so that, from
(\ref{DecomProjector}), 
\begin{eqnarray*}
W^{+}_{\alpha\beta} W^{+}_{\mu\nu} h^{\alpha\mu} h^{\beta\nu} = 
W^{+}_{\alpha\beta} W^{+}{}^{\alpha\beta} = 2,
\end{eqnarray*}
where in the last equality we used (\ref{squareW}).
Consequently the two-form 
$W^{+}_{\alpha\beta} h^{\alpha}_{\,\,\,\mu} h^{\beta}_{\,\,\,\nu}$ 
vanishes nowhere. Let $\hat{\eta}$ be
the volume-form on $\M/\sim$ such that
$W^{+}_{\alpha\beta} h^{\alpha}_{\,\,\,\mu} h^{\beta}_{\,\,\,\nu}$ is proportional
to $\hat{\eta}_{\mu\nu} = \pi^{\star} (\hat{\bm \eta})_{\mu\nu}$ with a positive
proportionality factor (existence follows because both are two-forms on
the two-dimensional vector space $\D_p^{\bot}$ and both are
everywhere non-zero). Squaring
\begin{eqnarray*}
W^{+}_{\alpha\beta} h^{\alpha}_{\,\,\,\mu} h^{\beta}_{\,\,\,\nu} = f \hat{\eta}_{\mu\nu},
\quad \quad f>0
\end{eqnarray*}
with the metric $\hat{g}$, and using $\hat{\eta}_{\mu\nu}
\hat{\eta}_{\rho\sigma} \hat{g}^{\mu\rho} \hat{g}^{\nu\sigma} =
2$ ($\hat{g}^{\alpha\beta}$ is the inverse of 
$\hat{g}_{\alpha\beta}$ and {\it not} the tensor obtained by raising
indices with $g^{\alpha\beta}$)
we conclude $f^2 = \Omega^2$ and hence $f= \Omega$, as claimed.

Expanding the vertical part of $\ell$  in the basis $\{\xi,k\}$ we find
$\ell^{\VV} = -\frac{1}{g (\xi,k)} \xi + h_0 k$ where $h_0$ is a function
whose explicit from does not concern us. It follows
that 
\begin{eqnarray}
\eta_{\alpha\beta\rho\sigma}  h^{\alpha}_{\,\,\,\mu} h^{\beta}_{\,\,\,\nu}
k^{\rho} \ell^{\sigma} = \frac{1}{g (\xi,k)}
\eta_{\alpha\beta\rho\sigma}  h^{\alpha}_{\,\,\,\mu} h^{\beta}_{\,\,\,\nu}
\xi^{\rho} k^{\sigma}
\label{relationorientations}
\end{eqnarray}
so that the last statement follows directly from (\ref{exterior}) in 
Lemma \ref{LieDerivative}.$\hfill \qed$
 
\vs 

\noindent {\bf Remark.}
This Lemma implies that, for any pair of
horizontal vectors $X_1 =
X_1^{\HH}$ and $X_2 = X_2^{\HH}$ in $T_p\M$, the pair
$\{ \pi_{\star} (X_1), \pi_{\star} (X_2)\}$ is positively oriented
if and only if $\{ k,\ell,X_1, X_2 \}$ is positively oriented in $T_p \M$, and also if and only if $\frac{1}{g (\xi,k)} \eta_{\alpha\beta\rho\sigma} X_1^{\alpha}
X_2^{\beta} \xi^{\rho} k^{\sigma} > 0$. This will be used below.

The function $Z$ is constant along the leaves of $\{ {\cal D} \}$. 
Hence there exists a function $\hat{Z}$ such that
$Z = \pi^{\star} (\hat{Z})$. Let $\star_{\hat{h}} d \hat{Z}$ be the Hodge dual of $d\hat Z$, in components
\begin{eqnarray*}
(\star_{\hat{h}} d \hat{Z})_A = \hat{\eta}_{AB} D^B \hat{Z},
\end{eqnarray*}
where $D$ is the covariant derivative of $\hat{h}$ and all indices ($A,B,\dots $)
on $\M/\sim$ are raised and lowered with $\hat{h}_{AB}$. 
It is immediate that $\{ \star_{\hat{h}} d\hat{Z},
d \hat{Z} \}$
is either identically zero or defines a positively oriented basis.

\begin{lemma}
\label{VstardZ}
Under the same conditions as in Lemma
\ref{volume}, assume further that $dZ \neq 0$ everywhere. Then,
the one-form $V_{\alpha} \defi h^{\beta}_{\,\,\,\alpha} \nabla_{\beta} y$ is given by
$\displaystyle{V_{\alpha} = \pi_{\star} ( \star_{\hat{h}} d \hat{Z} )_{\alpha}}$.
\end{lemma}

{\it Proof:} We already know the properties 
$V_{\alpha} V^{\alpha} = 
\nabla_{\alpha} Z \nabla^{\alpha}Z$ and
$V_{\alpha} \nabla^{\alpha} Z= 0$. Since both $V_{\alpha}$
and $\pi_{\star} ( \star_{\hat{h}} d \hat{Z} )_{\alpha}$
are horizontal, it must happen that $V_{\alpha} =
\epsilon_1 \pi_{\star} ( \star_{\hat{h}} d \hat{Z} )_{\alpha}$, 
with $\epsilon_1 = \pm 1$. To decide
the sign, it is necessary to analyze
the orientation of $\{ \xi^{\alpha}, k^{\beta}, V^{\alpha}, 
\nabla^{\alpha} Z\}$. We compute
\begin{eqnarray*}
\eta_{\alpha\beta\mu\nu} V^{\alpha} \nabla^{\beta} Z \xi^{\mu} k^{\nu} =
\eta_{\alpha\beta\mu\nu} \nabla^{\alpha} y \nabla^{\beta} Z \xi^{\mu} k^{\nu} =
- \frac{i}{2} 
\eta_{\alpha\beta\mu\nu} k^{\alpha} \ov{P}^{\beta} \xi^{\mu} P^{\nu} =
- \frac{1}{2} \left ( N q_{\nu} - P_{\alpha} \ov{P}^{\alpha} \xi_{\nu} 
\right ) k^{\nu}
\end{eqnarray*}
where in the last equality we have used (\ref{xiwedgeq}). 
Since $q_{\nu} k^{\nu} = g (\xi,k)$ (cf. (\ref{xi+q})) and $N - 
P_{\alpha} \ov{P}^{\alpha} = P_{\alpha} (P^{\alpha} - \ov{P}^{\alpha}
) = - 2 \nabla_{\alpha} Z \nabla^{\alpha} Z$ we conclude that
\begin{eqnarray*}
\frac{1}{g (\xi,k)} \eta_{\alpha\beta\mu\nu} V^{\alpha} 
\nabla^{\beta} Z \xi^{\mu} k^{\nu} = \nabla_{\alpha} Z \nabla^{\alpha} Z \geq 0.
\end{eqnarray*}
Hence, in view of (\ref{relationorientations}),
 $\{ V^{\alpha}, \nabla^{\alpha} Z, k^{\alpha}, \ell^{\alpha} \}$ is positively
oriented  provided $\nabla^{\alpha} Z \neq 0$.
Since $\{ \star_{\hat{h}} d \hat{Z}, d \hat{Z}\}$ is also positively
oriented in the quotient, it follows that $\epsilon_1 =1$.
$\hfill \qed$

\subsection{Riemannian submersions in the special cases}
\label{special}

%
%An important ingredient in the integration of the field equations is the
%determination of the geometry of 

%. In the case
%when $g(\xi,k)\neq 0$ and $\nabla_{\alpha} Z \neq 0$, we proceed
%by identifying a positively oriented orthogonal basis of vectors
%in the quotient constructed out from the scalar functions $Z$ and $y$.

The results  in the previous subsection will be 
sufficient to determine fully the geometry of the quotient
space $(\M /\sim, \hat{h})$ of the Riemannian submersion when $dZ \neq 0$.
The key for this will be equation (\ref{HessianZ}) in Proposition
\ref{geometryquotient}.
 However, when $Z$ is a constant this equation gives no information
at all, so an alternative method must be used.  In addition, 
when $\xi$ is orthogonal to both $k$ and $\ell$ the 
distribution $\mbox{span} \{ \xi,\ell\}$ in the previous subsection
is no longer non-degenerate, and hence does not define 
a Riemannian submersion. In order to deal with this situation we 
need to introduce an alternative distribution and show that 
it defines a Riemannian submersion. 
Note that  the situation $Z = \mbox{const.}$
corresponds precisely to the special case 
$P_{\alpha} = \ov{P}_{\alpha}$ in Subsection \ref{sec:special}
while the case $g(\xi,k) = g(\xi,\ell)=0$ corresponds to the other
special case $P_{\alpha} = - \ov{P}_{\alpha}$, see (\ref{P+P}).

The next lemma introduces the second distribution and proves 
it to define a Riemannian submersion.
\begin{lemma}
Assume $g(\xi,k)=g(\xi,\ell)=0$ with $\xi \neq 0$ everywhere. Then
the distribution $\D^{-}_p  \defi  \mbox{span} \{ \xi |_p, \gradZ |_p\}$
with  $(\gradZ)^{\alpha} = \nabla^{\alpha} Z$
is two-dimensional, spacelike and integrable.
Let $\hminus{}^{\mu}_{\,\,\,\alpha}$ be its
orthogonal projector 
and $\hminusdown_{\alpha\beta} \defi \hat{g}_{\alpha\mu} \hminus{}^{\mu}_{\,\,\,\alpha}$.
Then,
\begin{eqnarray*}
\pounds_{\xi} \hminusdown_{\alpha\beta} = \pounds_{\gradZ} \hminusdown_{\alpha\beta} =0.
\end{eqnarray*}
\end{lemma}

{\it Proof:} From $P_{\alpha} = -\ov{P}_{\alpha}$ it follows $y=\mbox{const}$
and hence, using (\ref{nablaynablay}),  $\nabla_{\alpha} Z \nabla^{\alpha} Z =
\xi_{\alpha} \xi^{\alpha} = - N > 0$ (recall that $P_{\alpha} = i \nabla_{\alpha} Z$
can vanish  only if $\xi$ vanishes and
$\xi^{\alpha} \nabla_{\alpha} Z =0$). This proves that the distribution
$\{\D^{-}\}$ is two-dimensional and spacelike and,  moreover, that the metric 
$g_{\mu\alpha}$ can be decomposed as
\begin{eqnarray}
g_{\mu\alpha} 
= - k_{\mu} \ell_{\alpha}
- k_{\alpha} \ell_{\mu} 
- \frac{1}{N} \left ( \nabla_{\mu} Z \, \nabla_{\alpha} Z + \xi_{\mu} \xi_{\alpha} \right )
= H_{\mu\alpha} 
- \frac{1}{N} \left ( \nabla_{\mu} Z \, \nabla_{\alpha} Z +
\xi_{\mu} \xi_{\alpha} \right ).
\label{decommetric}
\end{eqnarray}
$\{ \D^{-} \}$ is also integrable because
$[\xi,\gradZ]=0$ as a consequence of 
$\xi$ being  a Killing vector. $\pounds_{\xi} \hminus_{\alpha\beta}=0$ for the same
reason. To show $\pounds_{\gradZ} \hminus_{\alpha\beta} =0$ we use Lemma 
\ref{Liecov}. First note, from (\ref{HessZ1}) and 
$\hminus{}^{\mu}_{\,\,\,\alpha} \nabla_{\mu} Z = 
\hminus{}^{\mu}_{\,\,\,\alpha} \xi_{\nu} = 0$,
\begin{eqnarray*}
2 i 
\hminus{}^{\mu}_{\,\,\,\nu}  \hminus{}^{\alpha}_{\,\,\,\beta} \hat{\nabla}_{\mu} \hat{\nabla}_{\alpha} Z
= \left ( R - \ov{R} \right ) 
\hminus{}^{\mu}_{\,\,\,\nu}  \hminus{}^{\alpha}_{\,\,\,\beta} 
\left ( \frac{1}{2} g_{\mu\alpha}
+ \frac{1}{R \ov{R}} t_{\mu\alpha} \right ) = 0
\end{eqnarray*}
where in the last equality we used (\ref{EMTensor}) so that
\begin{eqnarray*}
\label{g+EM}
\frac{1}{2}g_{\mu\alpha} + \frac{1}{R \ov{R}} t_{\mu\alpha} =
g_{\mu\alpha} + k_{\mu}\ell_{\alpha} + 
k_{\alpha} \ell_{\mu} = - N^{-1} \left ( \nabla_{\mu} Z \nabla_{\alpha} Z
+ \xi_{\mu} \xi_{\alpha} \right ).
\end{eqnarray*}
Applying now Lemma \ref{Liecov} (observe that
$(\gradZ)^{\alpha} \hat{g}_{\alpha\beta} 
 = \Omega^{-1} \hat{\nabla}_{\beta} Z$)
\begin{equation*}
\pushQED{\qed}
\pounds_{\gradZ} \hminusdown_{\alpha\beta} = 
\hminus{}^{\mu}_{\,\,\,\alpha} \hminus{}^{\nu}_{\,\,\,\beta} \left ( 
\hat{\nabla}_{\mu} \left ( \Omega^{-1} \hat{\nabla}_{\nu} Z 
\right )+ 
\hat{\nabla}_{\nu} \left ( \Omega^{-1} \hat{\nabla}_{\mu} Z 
\right ) \right ) = 
\frac{2}{\Omega} 
\hminus{}^{\mu}_{\,\,\,\alpha} \hminus{}^{\nu}_{\,\,\,\beta} \hat{\nabla}_{\mu}
\hat{\nabla}_{\nu} Z = 0. \qedhere
\popQED
\end{equation*}

In order to deal with the special cases we 
exploit the relationship between the Riemann tensor on the quotient and
suitable components of the Riemann tensor on the ambient manifold
valid for any Riemannian submersion
\begin{Proposition}[\cite{ONeil}]
\label{IdentitySubmersion}
Let $\pi : (\M, \hat{g}) \rightarrow (S,\hat{h})$ be a 
Riemannian submersion and denote by 
${}^{\hat{g}} R$, ${}^{\hat{h}} R$ the respective Riemann tensors. 
Let $\{X,Y,W,Z\}$ be horizontal vector fields.  Then
\begin{align*}
{}^{\hat{h}}R( \pi_{\star}(X),  \pi_{\star}(Y), \pi_{\star}(W), \pi_{\star}(Z))= &
{}^{\hat{g}} R (X,Y,W,Z) 
+\frac{1}{2} \hat{g} \left ( [X,Y]^{\VV}, [W,Z]^{\VV} \right ) +
\nonumber \\
& +\frac{1}{4} \hat{g} \left ( [X,W]^{\VV}, [Y,Z]^{\VV} \right )
-\frac{1}{4} \hat{g} \left ( [X,Z]^{\VV}, [Y,W]^{\VV} \right ).
\end{align*}
\end{Proposition}
In our situation, we know the Riemann tensor of 
$(\M,g)$ and the conformal transformation between $g$ and $\hat{g}$, so 
we will be able to compute the first term in the right-hand side.
For the remaining terms, we need to compute the vertical part of the 
commutator of two arbitrary horizontal vectors. We have two distributions
to consider.
In order to deal with both of them in parallel and given their
characterization as $\ov{P}_{\alpha} = \epsilon P_{\alpha}$, $\epsilon = \pm 1$, 
we introduce the following notation (we work always away from fixed points of
$\xi$). The distribution $\{ \D^{\epsilon} \}$ is defined
as $\mbox{span}  \{ k, \ell \}$ if $\epsilon =1$ and
$\mbox{span}  \{ \xi, \gradZ \}$ if $\epsilon = - 1$. 
The corresponding
orthogonal projectors are denoted by $\hepsilon^{\mu}_{\,\,\,\alpha}$ 
(so that, in the notation above
$\hepsilon^{\mu}_{\,\,\,\alpha} =  h^{\mu}_{\,\,\,\alpha}$ if 
$\epsilon = 1$ 
and   $\hepsilon^{\mu}_{\,\,\,\alpha} = H^{\mu}_{\,\,\,\alpha}$ if 
$\epsilon = -1$) and we also write
$\hepsilondown_{\alpha\beta} = \hepsilon^{\mu}_{\,\,\,\alpha} \hat{g}_{\mu \beta}$.
Note that
\begin{eqnarray}
\hepsilon^{\mu}_{\,\,\,\alpha} \hepsilon^{\nu}_{\,\,\,\beta}
\W_{\mu\nu} = - \epsilon \sqrt{-\epsilon} W^{\epsilon}_{\alpha\beta} 
\label{ProyWW}
\end{eqnarray}
where, naturally, $W^{\epsilon}_{\alpha\beta}$ is $W^{+}_{\alpha\beta}$ when $\epsilon = +1$
and $W^{-}_{\alpha\beta}$ when $\epsilon = -1$
and the square root is defined by
$\sqrt{-\epsilon} = i$ when $\epsilon =1$
and $\sqrt{-\epsilon} = 1$ when $\epsilon = - 1$.

We have the following result for the vertical parts of the commutator
of horizontal fields.
\begin{lemma}
\label{vertical}
Assume both $\ov{P}_{\alpha} = \epsilon P_{\alpha}$ and $\xi \neq 0$ everywhere.
With the notation above, let $X,Y$ be any
pair of horizontal vector fields of the distribution $\{ \D^{\epsilon} \}$. Then, 
the vertical part of their commutator is 
\begin{eqnarray*}
[X,Y]^{\VV}
 = \sqrt{-\epsilon} \left ( \ov{J} - \epsilon J \right ) 
X^{\mu} Y^{\nu} W^{\epsilon}_{\mu\nu} \xi.
\end{eqnarray*}
\end{lemma}
{\it Proof.} 
We first note that for any pair of horizontal vectors
$X,Y$ and a vertical vector $t$ we have, using the orthogonality
of $t$ and $X,Y$,
\begin{eqnarray*}
g ( [ X,Y], t) = t_{\nu} \left ( X^{\mu} \nabla_{\mu} Y^{\nu}
- X^{\mu} \nabla_{\mu} Y^{\nu} \right ) = 
- X^{\mu} Y^{\nu} h^{\alpha}_{\,\,\,\mu} h^{\beta}_{\,\,\,\nu}
\left (  \nabla_{\alpha} t_{\beta}  -  \nabla_{\beta} t_{\alpha}  \right ). 
\end{eqnarray*}
We first prove that $[X,Y]^{\VV}$ is proportional to $\xi$. It suffices to
show that its scalar product with a vertical, nowhere zero vector orthogonal
to $\xi$ vanishes. 

For the case $\epsilon = +1$, 
 the condition $\nabla_{\alpha} Z =0$ implies $V_{\alpha} =0$ so that, using
the projector (\ref{DecomProjector})
\begin{eqnarray*}
0 = V_{\alpha} = h^{\alpha}_{\,\,\,\beta} \nabla_{\alpha} y =
\nabla_{\alpha} y - \xi_{\alpha} - \frac{N}{g(\xi,k)} k_{\alpha}.
\end{eqnarray*}
It follows that $t \defi \xi + \frac{N}{g(\xi,k)} k$ is vertical, nowhere
zero, orthogonal to $\xi$ and
satisfies $\nabla_{\alpha} t_{\beta}  -\nabla_{\beta} t_{\alpha} =0$.
Consequently
$[X,Y]^{\VV}$ is orthogonal to $t$ as required. For the case
$\epsilon = -1$ the claim follows immediately with the choice $t
= \gradZ$ which is again vertical, nowhere zero, orthogonal to $\xi$ and
satisfies $\nabla_{\alpha} t_{\beta}  -\nabla_{\beta} t_{\alpha} =0$.

In order to determine the proportionality factor between
$[X,Y]^{\VV}$ and $\xi$ we compute
\begin{align*}
g ( [ X,Y]^{\VV} , \xi) = & 
- X^{\mu} Y^{\nu} \, \hepsilon^{\alpha}_{\,\,\,\mu} \hepsilon^{\beta}_{\,\,\,\nu}
\left (  \nabla_{\alpha} \xi_{\beta}  -  \nabla_{\beta} \xi_{\alpha}  \right ) = \\
= & - X^{\mu} Y^{\nu} \, \hepsilon^{\alpha}_{\,\,\,\mu} \hepsilon^{\beta}_{\,\,\,\nu}
\left ( R \W_{\alpha\beta} + \ov{R} \ov{\W}_{\alpha\beta} \right ) = 
- \sqrt{-\epsilon} \left ( \ov{R} - \epsilon R \right ) 
X^{\mu} Y^{\nu} W^{\epsilon}_{\mu\nu} 
\end{align*}
where in the third equality we used (\ref{ProyWW}). The lemma
follows from (\ref{RpmR}) and $g(\xi,\xi) = -N$. $\hfill \qed$

We can now determine the geometry of the quotient 
in the special cases. 
\begin{Proposition}
\label{CurvaturePreal}
Assume both $\ov{P}_{\alpha} = \epsilon P_{\alpha}$ and $\xi \neq 0$
everywhere. 
Choose any point $p \in \M$ and a sufficiently small neighbourhood
$U_p$ of $p$ where Lemma \ref{existencesubmersion} applies. Then, the curvature
scalar of $(U_p /\sim, \hepsilondown)$ reads
\begin{itemize}
\item[(A)] $\displaystyle{R(\hepsilondown) = 2 \Lambda}$.
\item[(B)] $\displaystyle{R(\hepsilondown) = 
\left \{ \begin{array}{ll}
2 ( c + 2 \Lambda Z^2 ) & \mbox{ if  } \quad \epsilon = 1 \\
2 ( - c + 2 \Lambda y^2 ) & \mbox{ if } \quad \epsilon = - 1 
\end{array} \right . }$
\end{itemize}
\end{Proposition}

{\bf Remark.} In the case $\epsilon = 1$ we can equivalently write
$R(\hat{h}) = - 2 u' (Z)$
where $u$ is the real function defined in the Remark after 
Proposition \ref{geometryquotient}.

\vs 
{\it Proof:} We have already indicated the steps to prove this result. 
We start with the determination of
$R_{\alpha\beta\mu\nu} \hepsilon^{\alpha}_{\,\,\,\rho}
\hepsilon^{\beta}_{\,\,\,\sigma} \hepsilon^{\mu}_{\,\,\,\kappa} \hepsilon^{\nu}_{\,\,\,\delta}$. From (\ref{C=FF}) and (\ref{DefJ}) and recalling
(\ref{ProyWW}), 
\begin{eqnarray}
\C_{\alpha\beta\mu\nu} \hepsilon^{\alpha}_{\,\,\,\rho}
\hepsilon^{\beta}_{\,\,\,\sigma} \hepsilon^{\mu}_{\,\,\,\kappa} \hepsilon^{\nu}_{\,\,\,\delta} =
 \left ( 3 J R - \Lambda \right ) \left ( 
- \epsilon W^{\epsilon}_{\rho\sigma} W^{\epsilon}_{\kappa\delta}
+ \frac{1}{3} \left ( 
\hepsilon_{\rho\kappa} \hepsilon_{\sigma\delta} - \hepsilon_{\rho\delta} 
\hepsilon_{\sigma\kappa} \right ) \right ).
\label{WeylZconst}
\end{eqnarray}
Since 
$W^{\epsilon}_{\rho\sigma} W^{\epsilon}_{\kappa\delta}$ has the same
symmetries as a Riemann tensor in two-dimensions and
$W^{\epsilon}_{\alpha\beta} W^{\epsilon}{}^{\alpha\beta} = - \epsilon$ it follows
\begin{eqnarray*}
\epsilon W^{\epsilon}_{\rho\sigma} W^{\epsilon}_{\kappa\delta} = 
\hepsilon_{\rho\kappa} \hepsilon_{\sigma\delta} - \hepsilon_{\rho\delta} 
\hepsilon_{\sigma\kappa}. 
\end{eqnarray*}
Taking the real part in (\ref{WeylZconst}) and using (\ref{RiemLam}) it follows
\begin{eqnarray*}
R_{\alpha\beta\mu\nu} \hepsilon^{\alpha}_{\,\,\,\rho}
\hepsilon^{\beta}_{\,\,\,\sigma} \hepsilon^{\mu}_{\,\,\,\kappa} 
\hepsilon^{\nu}_{\,\,\,\delta} =
\left ( \Lambda - JR - \ov{J} \, \ov{R} \right )
\left ( \hepsilon_{\rho\kappa} \hepsilon_{\sigma\delta} - 
\hepsilon_{\rho\delta} \hepsilon_{\sigma\kappa} \right ).
\end{eqnarray*}
The next step is to perform the conformal rescaling $\hat{g} = \Omega^{-1}
g$ and obtain the corresponding Riemann tensor.  
The distribution $\{ \D^{\epsilon} \}$ is such that (recall
expression (\ref{EMTensor})  for $t_{\mu\alpha}$)
\begin{eqnarray*}
\hepsilon^{\mu}_{\,\,\,\nu} P_{\mu} =0, \quad \quad
\hepsilon^{\mu}_{\,\,\,\nu} \hepsilon^{\alpha}_{\,\,\,\beta} t_{\mu\alpha} =
\frac{\epsilon R \ov{R}}{2} \hepsilon_{\nu\beta}.
\end{eqnarray*}
Thus, equation (\ref{nablaP}) implies, using (\ref{RpmR}), 
\begin{eqnarray}
\hepsilon^{\mu}_{\,\,\,\nu} \hepsilon^{\alpha}_{\,\,\,\beta} 
\nabla_{\mu} P_{\alpha} 
= - \frac{1}{2} N \left ( J + \epsilon \ov{J} \right ) \hepsilon_{\nu\beta}.
\label{nablaepsilon}
\end{eqnarray}
Given that $\ov{P}_{\alpha}
 = \epsilon P_{\alpha}$, equations (\ref{nablaOmega}) and (\ref{nablaepsilon})
yield
\begin{align*}
& \nabla_{\alpha} \Omega  = - \Omega \left ( J + \epsilon
\ov{J} \right ) P_{\alpha}, \\
& \hepsilon^{\mu}_{\,\,\,\nu} \hepsilon^{\alpha}_{\,\,\,\beta} 
\nabla_{\mu} \nabla_{\alpha} \Omega  = \frac{1}{2} N \Omega \left ( J + 
\epsilon \ov{J} \right )^2 \hepsilon_{\nu\beta}.
\end{align*}
The standard transformation law for the Riemann
tensor under conformal
rescaling gives, after a straightforward calculation,
\begin{eqnarray*}
{}^{\hat{g}} R_{\alpha\beta\mu\nu} \hepsilon^{\alpha}_{\,\,\,\rho}
\hepsilon^{\beta}_{\,\,\,\sigma} \hepsilon^{\mu}_{\,\,\,\kappa} 
\hepsilon^{\nu}_{\,\,\,\delta} =
\frac{1}{\Omega} \left ( \Lambda + \frac{1}{4} N \left ( J + \epsilon 
\ov{J}
\right )^2 - J R - \ov{J} \ov{R} \right )
\left ( \hepsilon_{\rho\kappa} \hepsilon_{\sigma\delta} - 
\hepsilon_{\rho\delta} \hepsilon_{\sigma\kappa} \right ).
\end{eqnarray*}
Inserting this into Proposition \ref{IdentitySubmersion}
and using Lemma \ref{vertical} gives  
%($A,B,...$ are tensor indices in $U_p / \sim$)
\begin{align*}
{}^{\hepsilondown} R_{ABCD} = &  
\Omega \left ( \Lambda + \frac{N}{4}  \left ( J + \epsilon
\ov{J} \right )^2 - 
J R - \ov{J} \, \ov{R} + \frac{3N }{4} \left ( \ov{J} - \epsilon J \right )^2
\right ) 
\left ( \hepsilondown_{AC} \hepsilondown_{BD} - \hepsilondown_{AD} 
\hepsilondown_{BC} \right ).
\end{align*}
The Proposition now follows by explicit substitution of the formulae in 
Lemma \ref{Functions_of_P}
in this expression and  using (\ref{specialA},\ref{specialB}).
$\hfill
\qed$

\vspace{5mm}

\subsection{Spacetime geometry}
\label{integrationup}

We are ready to prove our main results of this section, 
namely
to determine
the spacetime metric under our characterization hypotheses. 
As already discussed, the strategy is to use the information
in the previous two subsections concerning the Riemannian 
submersions and integrate the field equations from the quotient
up to the spacetime. We need 
to distinguish two cases depending on whether or not 
$\xi$ is orthogonal to $k$ and $\ell$.

\begin{theorem}
\label{non-orthogonal}
Let $(\M,g)$ be  a $\Lambda$-vacuum, non-locally flat, spacetime admitting a Killing vector
$\xi$. Let $\F_{\mu\nu}$ be the self-dual Killing form of $\xi$ and assume
that $\F^2 \neq 0$ everywhere on $\M$ 
and that the self-dual Weyl tensor $\C_{\alpha\beta\mu\nu}$ satisfies
(\ref{C=FF}). Let $k$ and $\ell$ be the two eigenvectors of $\F_{\alpha\beta}$
and let $p \in  \M$ be such that $\xi |_p \not \in \mbox{span} \{ k|_p,
\ell|_p \}^{\bot} \subset T_p \M$.  Then, there exists an open, connected
neighbourhood
$U_p$ of $p$ with $\xi = \partial_v$ and
%and coordinates $\{v,y,x^A\}$ on $U_p$ 
such that $g$ takes the form
\begin{eqnarray*}
g = - N \left ( dv - \hat{\bm{w}}  \right )^2
+ 2 \left (dy - \star_{\hat{h}} dZ \right )\left ( dv - \hat{\bm{w}} \right )
+ \Omega \hat{h}.
\end{eqnarray*}
where $\hat h$ is a
Riemannian metric on a two-dimensional
manifold $S_p$ with  Gaussian curvature 
\be
K(\hat{h})  = - u' \circ Z, \label{scalarcurvature}
\ee
$\hat{\bm{w}}$ is a one-form on the metric $\hat h$ satisfying the equation
\be
\hat d \hat{\bm w} = (f \circ Z) \hat{\eta},  \label{Eqs}
\ee
and $Z$ is a function on the metric $\hat h$ whose Hessian satisfies
\be
\mbox{\rm Hess}\,  Z = (u \circ Z)  \hat{h}. \label{Hessian} 
\ee
Here, $u'$ is the derivative of $u$, 
$\hat{\eta}$ and $\star_{\hat{h}}$ denote the 
volume 2-form
and its corresponding Hodge dual of the metric $\hat{h}$
and $\hat d$ is the exterior differential on $S_p$. 
The functions $N,\Omega: \M \rightarrow \mathbb{R}$ and
$u,f :\mathbb{R} \rightarrow \mathbb{R}$ belong to
one of the following two classes 
\begin{itemize}
\item[(A)] $\displaystyle{ N =  c - \Lambda \left ( y^2 - Z^2 \right ) 
, \quad \Omega = 1, \quad u(\zeta) =  - \Lambda \zeta,
\quad f=0, \quad \Lambda \neq 0}$.
\item[(B)] $\displaystyle{ N = 
c - \frac{\Lambda}{3} \left ( y^2 - Z^2 \right )
- \frac{b_1 y +b_2 Z}{y^2 + Z^2}
, \quad \Omega = y^2 + Z^2
, \quad u(\zeta) = \frac{b_2}{2} - c \, \zeta - 
\frac{2 \Lambda}{3} \zeta^3, \quad f = 2 \zeta
}$.
\end{itemize}
where $b_1, b_2, c$ are arbitrary real constants. 
\end{theorem}

{\it Proof:}
Since $\xi|_p $ is not simultaneously orthogonal 
to $k|_p$ and $\ell|_p$ we can assume, after interchanging $\ell$ and $k$
if necessary, that $g(\xi,k) \neq 0$ at $p$.
Choose $U_p$ an open, connected neighbourhood
of $p$ where $g(\xi,k) \neq 0$ everywhere. Without loss
of generality we can assume $g(\xi,k)=1$ on $U_p$.
From 
\begin{eqnarray*}
\xi_{\alpha} [\xi,k]^{\alpha} = 
\xi_{\alpha} \left ( \xi^{\mu} \nabla_{\mu} k^{\alpha} - 
k^{\mu} \nabla_{\mu} \xi^{\alpha} \right ) = 
\xi^{\mu} \nabla_{\mu} g (\xi,k) = 0,
\end{eqnarray*}
together with $[\xi, k] \propto k$ we conclude 
$[\xi,k]=0$. Let $\{ \D \}$ be the distribution
defined in Proposition \ref{geometryquotient} and restrict
$U_p$ further, if necessary, so that Lemma \ref{existencesubmersion}
can be applied to the spacetime $(\M,\hat{g})$. Denote by
$\pi$ the corresponding submersion and $\hat{h}$
the metric of $S_p \defi U_p /\sim$. 
We can assume without loss of generality
that $S_p$ is connected
and $U_p = I_1 \times I_2 \times S_p$, where $I_1$ and $I_2$ are open
intervals of the real line. 
After restricting $S_p$ if necessary, let
us introduce coordinates $\{\hat{x}^A \}$ in $S_p$.
As in \cite{Mars2} we introduce
a coordinate system in $U_p$ as follows.
Since the bundle $(U_p,S_p,\pi)$ is trivial, select
a global section $\hat{\sigma}: S_p \longrightarrow U_p$. 
The sets $\Sigma_{y_0} \defi \{ y = y_0 \}$ are smooth hypersurfaces
(because $\nabla_{\alpha} y \neq 0$ everywhere) and transversal to 
the fibers $\pi^{-1} (\hat{x}), \hat{x} \in S_p$ (because
$k(y) \neq 0$ everywhere). Consider an
arbitrary point $s \in U_p$ and let $y_{s} = y(s)$. The point $s' 
= \hat{\sigma} (\pi(s))$ belongs to the same fiber as $s$. Consider the integral
line of $k$ passing through $s'$ and let $r$ be the intersection
of this line with $\Sigma_{y_{s}}$ (this intersection exists and it is unique,
possibly after restricting $U_p$ further, because $k$ is everywhere
transverse to $\{ \Sigma_{y} \}$). The points $s$ and $r$ belong to the same
fibre and to the same hypersurface $\Sigma_{y_{s}}$. Since $\xi$
is tangent to the fibers and tangent to $\Sigma_{y_{s}}$, there exists
an integral line of $\xi$ connecting $r$ and $s$. Let $v$ be
the natural parameter of this curve starting at $r$ (i.e.
$\xi(v)=1$ with $v(r)=0$) and define $v_{s} = v(s)$. By construction, it follows
that the set of values $\{ v(s), y(s), x^A \defi \hat{x}^A (\pi(s))\}$
define a coordinate system on $U_p$. Moreover, $\xi = \partial_v$
and $k = \partial_y$ in these coordinates (recall that we
have chosen $g(\xi,k)=1$ and hence $k(y)=1$).
Given that $k$ is null, it follows that the one-form $\bm{k}$ reads
\begin{eqnarray}
\bm{k} =  dv - \bm{w}
\label{Exprek}
\end{eqnarray}
for some one-form $\bm w$ that satisfies $\pounds_{\xi}\bm w=0$ due to $\pounds_{\xi} \bm{k} =0$.
The last statement in Lemma \ref{LieDerivative} implies also that 
$\pounds_{k} \bm{k} = 0$ so that $\pounds_{k}\bm w =0$ too. Thus, there is a 
one-form $\hat{\bm{w}}$ on $(S_{p},\hat h)$ satisfying
$\bm w =\pi^{\star}\hat{\bm w}$.
The explicit form of the projector  (\ref{DecomProjector}) and the definition
$h^{\alpha}_{\,\,\beta} \nabla_{\alpha} y  = V_{\beta}$ implies
\begin{eqnarray*}
\xi_{\beta} = \nabla_{\beta} y - N k_{\beta} - V_{\beta}
\end{eqnarray*}
which, inserted in (\ref{DecomProjector}), provides
\begin{eqnarray}
g_{\alpha\beta} = \Omega \hat{h}_{\alpha\beta} 
- N k_{\alpha} k_{\beta} + k_{\alpha} \left ( \nabla_{\beta} y - V_{\beta}
\right ) + k_{\beta} \left ( \nabla_{\alpha} y - V_{\alpha} \right ).
\label{metric1}
\end{eqnarray}
From Lemma \ref{VstardZ}
we have  $\bm{V} = \pi_{\star} (\star_{\hat{h}} d \hat{Z})$.
Since in the coordinates $\{v,y,x^A\}$ the submersion $\pi$
takes a trivial form,
it is safe to use the same notation for objects on the quotient and
for corresponding objects on the spacetime (this applies in particular
to $\hat{h}$, $\hat{\bm{w}}$ and  to $\hat{Z}$). 
Thus the metric  $g$ in (\ref{metric1}) takes the form claimed in the Theorem.
It only remains to show that the field equations
(\ref{scalarcurvature}), (\ref{Eqs}) and (\ref{Hessian}) hold. 

From  Proposition \ref{geometryquotient} we have
\begin{eqnarray*}
D_A D_B Z = (u \circ Z) \hat{h}_{AB}
\end{eqnarray*}
where as before $D$ denotes the covariant derivative in $(S_p,\hat{h})$.
This proves (\ref{Hessian}). The Ricci identity applied to this
equations implies
\begin{eqnarray*}
R (h) D_A \hat{Z}  = -2 u'(\hat{Z}) D_A \hat{Z}.
\end{eqnarray*}
Since $S_p$ is connected, $d \hat{Z}$ is either non-zero on an open dense set
or identically zero
In the former case we can drop $D_A \hat{Z}$ and
we conclude (\ref{scalarcurvature}). The case $dZ=0$ has been dealt with in 
Proposition \ref{CurvaturePreal} (see in particular the Remark after this
Proposition) 
and also leads to (\ref{scalarcurvature}).  Finally,
(\ref{Exprek})  and Lemma \ref{volume}
imply 
$$
\hat d \hat{\bm{w}} = i \left ( \ov{J} - J \right ) \Omega
\bm{\hat{\eta}}.
$$
In case (A) we have $J=0$ so (\ref{Eqs}) holds with $f=0$
and in case (B) we have 
$J = - 1 /P$ and $\Omega = P \ov{P}$.
Hence,  $i (\ov{J} - J ) \Omega = 2 Z$, as claimed. $\hfill \qed$

Having dealt with the case when 
$\xi$ is not orthogonal to the eigenspace of $\F_{\alpha\beta}$ we 
study next the remaining case when 
$\xi$ is orthogonal to both $k$ and $\ell$.

\begin{theorem}
\label{th:orthogonal}
Let $(\M,g)$ be a $\Lambda$-vacuum, non-locally
flat, spacetime admitting a Killing vector
$\xi$ with no fixed points. Let $\F_{\mu\nu}$ be the self-dual Killing form of $\xi$ and assume that $\F^2 \neq 0$ everywhere on $\M$ 
and that the self-dual Weyl tensor $\C_{\alpha\beta\mu\nu}$ satisfies
(\ref{C=FF}). Let $k$ and $\ell$ be the two real null eigenvectors of $\F_{\alpha\beta}$ and assume that $g(\xi,k)=g(\xi,\ell)=0$ on $\M$.  Then, locally the metric $g$ adopts the form
$$
g = \frac{1}{-N} dZ^2-N \left(dv - \hat{\bm{w}}\right)^{2}
+ \Omega \hat{H}.
$$
where $\xi =\partial_v$, $\hat H$ is a 2-dimensional Lorentzian metric of constant curvature $\kappa$,
$\hat{\bm{w}}$ is a one-form on the metric $\hat H$ satisfying the equation
$$
\hat d \hat{\bm w} = 2y  \hat{\bm{\eta}}
$$
where $y $ is an arbitrary real constant and $\hat d$ and $\hat{\bm\eta}$ are the exterior differential and volume form of  the metric $\hat{H}$.
The functions $N$ and $\Omega$ belong to one of the following two classes:
\begin{align}
  & (A)\quad y=0, \quad N = c+\Lambda Z^2 , \quad \Omega  =1, \quad
\kappa = \Lambda, \quad \Lambda \neq 0.
\nonumber  \\
&  (B) \quad N = \frac{1}{y^2 + Z^2} \left ( c (Z^2 -y^2) - b_2 Z 
+ \frac{\Lambda}{3} \left(Z^4 +3 y^4 \right) \right ),
\quad  \Omega =y^2 +Z^2, \quad
\kappa = 2\Lambda y^2 - c.\label{caseB}
\end{align}
where $b_2, c$ are arbitrary (real) constants.
\end{theorem}

{\it Proof:} Let $p \in \M$ be any point
and choose $U_p$ an open, connected neighbourhood so that
Lemma \ref{existencesubmersion}
can be applied to the distribution $\{ \D^{-}\}$ and with respect to 
the metric $(\M,\hat{g})$. Denote by
$\pi$ the corresponding submersion and $\hat{H}$ the (Lorentzian) metric of
$S_p \defi U_p /\sim$. Proposition \ref{CurvaturePreal} shows that
$\hat{H}$ is a two-dimensional metric of constant
curvature $\kappa$ as given in cases (A) and (B) of the
Theorem. Let $\bm{\hat{\eta}}$ be a volume form 
of $(S_p,\hat{H})$ (the orientation will be chosen later)
and define $\hat{\eta}_{\alpha\beta} = \pi^{\star} (\hat{\bm{\eta}} )_{\alpha\beta} $.
Since $W^{-}_{\alpha\beta}$ is a horizontal two-form, it must be proportional
to $\hat{\eta}_{\alpha\beta}$. Its square norm in the metric $\hat{g}$ is
$- 2 \Omega^2$ (cf. (\ref{squareW})) which forces
$W^{-}_{\alpha\beta} = \pm \Omega \hat{\eta}_{\alpha\beta}$. Choose the 
orientation in $(S_p,\hat{H})$ so that the plus sign holds.

In order to construct the metric, the  starting point is 
(\ref{decommetric}). $\nabla_{\alpha} Z$ is nowhere zero (because $\xi$ has no fixed point
in $\M$), so $Z$ can be used as a coordinate on $U_p$ and we can write
\begin{eqnarray*}
g = - \frac{1}{N} \left ( \bm{\xi} \otimes \bm{\xi} + dZ^2 
\right ) + \Omega \hat{H}. 
\end{eqnarray*}
It only remains to find $\bm{\xi}$. From $d \bm{\xi} = \bm{\F} +
\ov{\bm{\F}} = R \bm{\W} + \ov{R} \ov{\bm \W}$
and $\nabla_{\alpha} N = R P_{\alpha} + \ov{R} \ov{P}_{\alpha}$
(recall that $2 N= (\chi + \ov{\chi})$ and 
$\nabla_{\alpha} \chi = 2 R P_{\alpha}$) the following general identity
follows
\begin{eqnarray*}
d \left ( N^{-1} \bm{\xi}  \right ) = 
N^{-1} \left ( R \bm{\W} + \ov{R} \ov{\bm \W}
+ N^{-1} \bm{\xi} \wedge \left ( R \bm{P} + \ov{R} \ov{\bm{P}}
\right )\right ).
\end{eqnarray*}
In the present case $\ov{\bm{P}} = - \bm{P}$ and (\ref{xiwedgeP}) holds, so that
\begin{eqnarray}
d \left ( N^{-1} \bm{\xi} \right ) = 
N^{-1} \left ( R + \ov{R} \right ) \bm{W^{-}} = 
\left ( J + \ov{J} \right ) \Omega \bm{\hat{\eta}},
\label{dxiN}
\end{eqnarray}
where in the second equality we used (\ref{RpmR}) 
and the relationship between $\bm{W^{-}}$ and $\bm{\hat{\eta}}$
just discussed. In either case $(A)$ or $(B)$ we have
$( J + \ov{J} ) \Omega = -2 y$. Let $\hat{\bm{w}}$ be a one-form on $(S_p,\hat{H})$
satisfying
\begin{eqnarray*}
\hat{d} \hat{\bm{w}} = 2 y \bm{\hat{\eta}}.
\end{eqnarray*}
It follows from (\ref{dxiN}) that 
$d \left ( N^{-1} \bm \xi + \pi^{\star} (\hat{\bm{w}}) \right ) =0$, 
and consequently, the existence (restricting $U_p$ further if necessary)
of a function $v: U_p \rightarrow \mathbb{R}$ such that
\begin{eqnarray*}
\bm{\xi} = N \left ( d v - \pi^{\star} ( \hat{\bm{w}} ) \right ).
\end{eqnarray*}
With the slight abuse of notation of naming $\pi^{\star} (\hat{\bm{w}})$ 
still as $\hat{\bm{w}}$, the metric given in the Theorem follows.
Finally, expressions (\ref{caseB}) are a consequence of the general
formulae in Lemma \ref{Functions_of_P}
after using (\ref{specialA})-(\ref{specialB}) with
$\epsilon =-1$. $\hfill \qed$

\section{Semiglobal considerations}
\label{sec:semiglobal}

So far we have found the local form of the metric under the assumption  $\F^2 \neq 0$ everywhere and we have 
split the analysis depending on whether $\xi \not \in \mbox{span} \{ k, \ell \}^{\perp}$
 everywhere or $\xi \in \mbox{span} \{ k, \ell \}^{\perp}$ everywhere.
In a given
spacetime  it may happen a priori 
that $\F^2$ becomes zero and/or that those different situations 
can occur in disjoint open sets. We analyze in this section whether this is
possible or not. 

Throughout this section we assume that $(\M,g)$ is a smooth $\Lambda$-vacuum spacetime admitting
a Killing vector $\xi$ such that (\ref{C=FF}) holds with a smooth
proportionality function $Q$. Note that smoothness of $Q$ is an assumption,
because in principle it may be the case that $Q$ diverges somewhere,
while
$\F_{\alpha\beta} \rightarrow 0$ at the same place so that
the right-hand side in (\ref{C=FF}) stays smooth.
However, this is not what is meant when saying that 
$\C_{\alpha\beta\mu\nu}$ and $\F_{\alpha\beta} \F_{\mu\nu}
- \frac{1}{3} \F^2 \I_{\alpha\beta\mu\nu}$ are proportional everywhere.
We have the following result concerning $\F^2$.
\begin{Proposition}
Define $\M_{\F^2} \defi \{ p \in \M; \F^2|_p \neq 0\}$ and 
assume $\M_{\F^2} \neq \emptyset$ and that there is $p \in \M_{\F^2}$ with 
$Q(p) \neq 0$. Then $\M_{\F^2} = \M$.
\end{Proposition}

\vs 

\noindent {\bf Remark:} If $Q=0$ everywhere on $\M_{\F^2}$, then
the spacetime on this open set
is locally isometric to the Minkowski ($\Lambda =0$), de Sitter  ($\Lambda >0$)
or Anti-de Sitter ($\Lambda <0$) spacetimes. In the case
$\Lambda \neq 0$, these spacetimes admit Killing vectors for which 
$\F^2$ is not identically zero and vanishes somewhere (an example is the
Killing vector $\xi = y (\partial_t + \partial_x) + (t-x) \partial_y$
in the de Sitter space in conformally flat coordinates $\{t,x,y,z\}$, for which
$\F^2 \propto (t-x)^2$ with a nowhere zero proportionality factor). Thus, the 
condition on the existence of $p$ is necessary when $\Lambda \neq 0$. 
On the other hand, if $\Lambda =0$, then the spacetime on $\M_{\F^2}$
would be locally Minkowski for which $\F^2$ is constant for
any Killing vector. So automatically we would have $\M_{\F^2}
= \M$ in this case. This shows that the condition on the existence of 
$p$ can be dropped in the $\Lambda =0$ case.

\vs

{\it Proof:} Consider the connected component $\M^0_{\F^2}$
of $\M_{\F^2}$ containing $p$ and assume $\M^0_{\F^2} \neq \M_{\F^2}$.
Let $q \in \partial \M^0_{\F^2}$ and consider the smooth function $Q \F^2$ on $\M$,
which  on $\M^0_{\F^2}$ takes the form
\begin{eqnarray}
- \frac{1}{4} Q \F^2 = R^2 Q = 3 J R - \Lambda.
\label{limit_q}
\end{eqnarray}
The left-hand side approaches zero when we approach $q$.
We know that $J$ is either identically zero or
nowhere zero on $\M^0_{\F^2}$. In the former case, the right hand-side
of (\ref{limit_q}) takes the constant value $-\Lambda$ which 
cannot approach $0$ at $q$ unless $\Lambda=0$, but then
$Q$ vanishes identically on $\M^0_{\F^2}$ against hypothesis.

For the remaining case $J\neq 0$ on $\M^0_{\F^2}$, we use the expressions
in Lemma \ref{Functions_of_P} to compute the right-hand side of
(\ref{limit_q}) as
\begin{eqnarray*}
R^2 Q = \left .  \left (  3 J R - \Lambda \right ) \right |_{\M^0_{\F^2}} =-
\frac{3 b}{2 P^3}
\end{eqnarray*}
Since $Q(p) \neq 0$, the constant $b$ cannot be zero. Hence, it must be
$P \rightarrow \infty$ when we approach $q$. Since $R = b/(2P^2) -
\Lambda P/3$ and $R \rightarrow 0$ at $q$, we necessarily have $\Lambda =0$.
But then  $Q = \frac{3J}{R} = \frac{- 6 P}{b}$ which diverges at $q$
against hypothesis. $\hfill  \qed$

\vs 
Concerning the possibility that the functions $g(\xi,k)$ and $g(\xi,\ell)$ vanish at a point $p\in \M$, we can prove the following general result concerning the special cases of subsection \ref{sec:special}.
\begin{Proposition}
\label{prop:bifurcation}
Assume that $\bm{\xi} \wedge {\bm q} |_{p}=0$ at $p\in \M$.  Then, 
there exists $\epsilon \in \{-1,1\}$ such that either
\begin{enumerate}
\item $q =\epsilon \xi$  everywhere, or
\item $p$ belongs to a 2-dimensional connected and totally geodesic surface ${\cal B}$,  
such that $q|_{\cal B}=\epsilon \xi|_{\cal B}$ on all ${\cal B}$,
\end{enumerate}
depending on whether $R-\epsilon \ov R -N (J-\epsilon \ov J)$
vanishes (case 1) or not (case 2) at $p$. In the latter case, $\cal B$ is spacelike or timelike if $\epsilon$ is $-1$ or $1$, respectively.
\end{Proposition}

\noindent {\bf Remark:} At $p$ one has, from the results of
subsection \ref{sec:special}, that $\bm{P} |_p = \epsilon \ov{\bm{P}}|_p$
(allowing also the case when they vanish identically), ergo
$dy|_{p}=0$ (if $\epsilon =-1$)
or $dZ|_{p}=0$ (if $\epsilon =1$)
From (\ref{divP}), we have that 
$$
(R-\epsilon \ov R -N (J-\epsilon \ov J))|_{p}=\left\{
\begin{array}{ccl} 
\nabla_{\mu}\nabla^{\mu}y |_{p} & \mbox{if}& \epsilon =-1 \\
\\
i \nabla_{\mu}\nabla^{\mu}Z |_{p} & \mbox{if} & \epsilon =1
\end{array}
\right.
$$
so one can reformulate the proposition in terms of the vanishing or not of the differential and the D'Alembertian of $y$ ($\epsilon =-1$) or $Z$ ($\epsilon =1$) at $p$.

\vs

\noindent {\it Proof:}
From the results in Subsection \ref{sec:special} we know that $\bm\xi \wedge \bm q |_p =0$ implies that there is $\epsilon \in \{ -1,1\}$ such that
$q |_p = \epsilon \xi|_p$ (if $\xi|_p=0$ then $q|_p=0$ and the two choices
of $\epsilon$ are allowed). 
From Proposition \ref{prop:2ndKilling} 
there is a Killing vector $\varsigma$
that vanishes at $p$, given by the constants $C \neq 0$ and $A=-\epsilon C$ in case (A), and $A=C(P|_{p}-\epsilon \ov P|_{p})^{2}$ in case (B). 
Thus $p$ is a fixed point of this Killing $\varsigma$. The
connected set of fixed points of $\varsigma$ containing $p$
is a totally geodesic, smooth surface with the same
dimension and causal character as the vector subspace 
${\cal V}_p (\varsigma)= \{ u \in T_p \M; d \varsigma|_p (u, \cdot)=0 \}$ (see
e.g. Lemma 5 in \cite{MarsReiris}).
So, we must compute the  Killing 2-form $\nabla_{\mu}\varsigma_{\nu}$ at $p$. 
By using formulas (\ref{nablaq}) and $\nabla_{\mu}P=P_{\mu}=\epsilon \ov P_{\mu}$
and  recalling (\ref{xiwedgeP}) and (\ref{xiwedgeP_bis}), a somewhat
long  calculation leads to
$$
\nabla_{\mu}\varsigma_{\nu}|_{p}= \sqrt{-\epsilon} G|_p 
\left(R-\epsilon \ov R-N (J-\epsilon \ov J) \right)W^{\epsilon}_{\mu\nu} |_{p}
$$
where $G$ is as in the proof of Proposition \ref{prop:2ndKilling}.
Thus, if $R-\epsilon \ov R-N (J-\epsilon\ov J)$ vanishes at $p$ the Killing vector $\varsigma$ vanishes everywhere on $\M$ and we have (using again the notation
in the proof of Prop. \ref{prop:2ndKilling})
$q = -G^{-1} F  \xi$ everywhere with $G^{-1} F |_{p} = -\epsilon$. The function 
$-G^{-1} F$ is smooth in $\M$ and according to Subsect. \ref{sec:special} it must be constant 
and equal to $\epsilon$ and we are in case 1 of the Proposition.

If on the other hand
\be
 \left(R-\epsilon \ov R-N (J-\epsilon\ov J)\right)|_p \neq 0 \label{condi}
\ee 
then the vector subspace ${\cal V}_p(\varsigma)$ is two-dimensional and spacelike if $\epsilon =-1$ or timelike if 
$\epsilon =1$. 
To finish the proof, observe that the Killing vector $\varsigma$ vanishes on all of ${\cal B}$, and therefore $\bm\xi \wedge \bm q |_{{\cal B}}=0$. Invoking again Subsection \ref{sec:special} we know that this can only happen if,
at each point of $\cal B$, either $q=\pm \xi$ or $q=0=\xi$. If $\xi|_{\cal B} \neq 0$ everywhere
then $q_{{\cal B}} =\epsilon \xi|_{{\cal B}}$ on all ${\cal B}$. 
If on the other hand there is a $p' \in {\cal B}$ such that $\xi |_{p'}=0$ (and thus $q |_{p'}=0$ too), the condition (\ref{condi}) may hold or not at $p'$. If it does not for either value of $\epsilon$ then the same reasoning as above, applied now to the point $p'$, proves that $q =\epsilon \xi$ everywhere
for that choice of $\epsilon$. Otherwise, (\ref{condi}) at $p'$ ---where $N=0$--- provides 
$R|_{p'}-\epsilon \ov R |_{p'} \neq 0$ for both choices of $\epsilon$, so that $R$ can be neither real nor purely imaginary at $p'$. But then $p'$ is a fixed point of $\xi$ whose Killing 2-form is given by (\ref{nablaxi}), and thus dim${\cal V}_p' (\xi) =0$, so that $p'\in {\cal B}$ is isolated and $\xi$ does not vanish around $p'$ on ${\cal B}$. Thus, again there is an $\epsilon\in \{-1,1\}$ such that $q|_{\cal B}=\epsilon \xi|_{\cal B}$.

%If on the other hand there is a $p\in \cal B$ such that $\xi|_p =0$ (and thus $q|_p=0$ too), condition (\ref{condi}) becomes simply $R|_p-\epsilon \ov R |_p \neq 0$ for both choices of $\epsilon$, so that $R$ can be neither real nor purely imaginary at $p$. But then $p$ is a fixed point of $\xi$ whose Killing 2-form is given by (\ref{nablaxi}), and thus dim${\cal V}_p(\xi) =0$, so that $p$ is isolated and $\xi$ does not vanish around $p$ on ${\cal B}$. Thus, again there is an $\epsilon\in \{-1,1\}$ such that $q|_{\cal B}=\epsilon \xi|_{\cal B}$.
$\hfill \qed$

\vs

From item 2 in the Proposition and (\ref{equiv})
we have, on ${\cal B}$,
$\ov{\bm{P}} = \epsilon \bm{P}$. For $\epsilon =1$ this means
$dZ |_{\cal B}=0$ and for $\epsilon = -1$, $dy |_{\cal B} =0$. Given that
${\cal B}$ is connected it follows
$$
y|_{\cal B}=y_{p} \,\,\, \mbox{if} \,\,\, \epsilon =-1; \quad \quad 
Z|_{\cal B}= Z_{p}\,\,\,  \mbox{if} \,\,\, \epsilon =1 .
$$

As an immediate consequence of the previous Proposition we have
\begin{corollary} 
\label{coro:bifurcation}
Assume that $g(\xi,k)|_{p}=g(\xi,\ell)|_{p}=0$ at $p\in \M$. Then, either 
\begin{enumerate}
\item $g(\xi,k)=g(\xi,\ell)=0$ everywhere on $\M$, or
\item $p$ belongs to a 2-dimensional connected and totally geodesic spacelike surface ${\cal B}$ such that $g(\xi,k)$ and $g(\xi,\ell)$ vanish on all ${\cal B}$,
\end{enumerate}
depending on whether $R+\ov R -N (J+\ov J)$
vanishes (case 1) or not (case 2) at $p$.
\end{corollary}

\section{Alternative forms of the metrics and characterization results}
\label{sec:characterization}

The main theorems in the previous sections involve field equations 
on two dimensional Riemannian manifolds. The following Lemma solves
them.

\begin{lemma}
\label{Solving2sim}
Let $(S,\hat{h})$  be a two-dimensional oriented
Riemannian manifold with smooth metric
$\hat{h}$ and volume form $\hat{\eta}$.
Let  $Z$ be a scalar and $\hat{\bm{w}}$ a one-form on $S$ satisfying
\begin{eqnarray}
\mbox{\rm Hess}\, Z = (u \circ Z)  \hat{h}, \hspace{2cm} 
\hat{d} \hat{\bm w} = (f \circ Z) \hat{\eta}. \hspace{2cm}
\label{Eqs2} 
\end{eqnarray}
where $u,f : I \subset \mathbb{R} \longrightarrow \mathbb{R}$ are smooth real
functions. Then, on the open set $\{ p \in S; dZ |_p \neq 0 \}$ 
(if non-empty) there
exist local coordinates $\{Z,\x\}$ such that 
\begin{eqnarray}
\hat{h} = \frac{dZ^2}{V(Z)} + V(Z) d\x^2,
\hspace{2cm} \hat{\bm w} = F(Z) d \x + \hat{\bm w}_0, \label{sol}
\end{eqnarray}
where $\displaystyle{\frac{dV}{dZ} = 2 u(Z)}$, $\displaystyle{\frac{dF}{dZ} = f(Z)}$
and $\hat{\bm w}_0$ is a closed one-form. Moreover
$\star_{\hat{h}} dZ = - V d \x$.
\end{lemma}

\vs

{\it Proof: } From $D_A D_B Z = u(Z) \hat{h}_{AB}$ it follows
$D_{A} \left ( (dZ,dZ)_{\hat{h}} \right ) = 2 u(Z) D_A Z$. On any connected
component $U^0$ of $\{ p \in S; dZ |_p \neq 0 \}$ this implies the existence
of a function $V : I \longrightarrow \mathbb{R}$ satisfying $V' =2u$
and $(dZ,dZ) = V \circ Z$ (observe that this is positive).
The Hodge dual $\star_{\hat{h}} dZ$ is nowhere
zero on $U^0$ and hence defines a distribution, which is always integrable in
two dimensions. Locally, there exists a function
$\x : U^0 \longrightarrow 
\mathbb{R}$ such that
$\star_{\hat{h}} dZ = - V(Z) H d\x$, where $H(Z,\x)$ is
a positive function to be determined (the sign is chosen so that
$\{dZ,d\x\}$ is positively oriented).
Given that $(\star_{\hat{h}} dZ, \star_{\hat{h}} dZ )_{\hat{h}} = (dZ, dZ)_{\hat{h}} =
V(Z)$, and $\star_{\hat{h}} dZ$ is orthogonal to $dZ$, the metric on
$U^0$ takes
the local form
\begin{eqnarray*}
\hat{h} = \frac{dZ^2}{V(Z)} + V(Z) H^2 d\x^2.
\end{eqnarray*}
The `$\x\x$' component of $(\mbox{\rm Hess } Z)_{AB} =
u(Z) \hat{h}_{AB}$ reads $\partial_{Z} H = 0$. Redefining $\x$
as a function of itself, we can set $H=1$ without loss of generality. 
This gives the metric in (\ref{sol}). The volume form is $\hat{\bm \eta} = dZ \wedge
d \x$, so equation (\ref{Eqs2}) is simply $\hat{d} \hat{\bm w} = f(Z) dZ \wedge
d \x$. With $F(\zeta)$ being any integral of $f(\zeta)$,
the general solution of this equation is as in  (\ref{sol}). The last
statement is obvious from the previous considerations. 
$\hfill \qed$.

\vs 

The following theorem summarizes the main results so far in a
self-contained form.
\begin{theorem}
\label{full}
Let $(\M,g)$ be a $\Lambda$-vacuum spacetime admitting a Killing vector
$\xi$ and corresponding self-dual two-form  $\F_{\alpha\beta}$.
Assume 
\begin{eqnarray}
\C_{\alpha\beta\mu\nu} = Q \left (\F_{\alpha\beta}
\F_{\mu\nu} - \frac{1}{3} \F^2 \I_{\alpha\beta\mu\nu} \right )
\label{WeylRepetido}
\end{eqnarray}
for a smooth function
 $Q : \M \rightarrow \mathbb{C}$ and that $\exists \, p \in \M$ such that
$Q |_p \neq 0$ and $\F^2|_p \neq 0$. Then
$\F^2 \neq 0$ everywhere and 
$$
q_{\mu} \defi 4 (\F^2 \ov{\F}^2)^{-1/2} \xi^{\alpha} \ov{\F}{}_{\alpha}^{\,\,\,\,\beta} \F_{\beta\mu}
$$ 
exists globally.

\begin{enumerate}
\item[(A)]
If $Q\F^2  - 4 \Lambda =0$ at one point, then it vanishes
everywhere and $q$ is a Killing vector field 
satisfying $[\xi,q]=0$.  The metric is locally reducible
$g = h_{-} + h_{+}$
where $h_{-}$ (resp. $h_{+}$) is a two-dimensional
Lorentzian (resp. Riemannian) metric
of constant curvature $\Lambda$ and $\xi$
is any of the Killing  
vectors of $h_{-}$ or $h_{+}$ as long as it satisfies $d{\bm \xi} \neq 0$
everywhere.

%\begin{itemize}
%\item[(A.i)] If $\mbox{\rm{dim}} ({\cal L}) = 2$, $\exists \, c,k \in \mathbb{R}$ 
%such that, away from 
%a collection of totally geodesic, nowhere null,
%codimension-two surfaces where a Killing vector in
%${\cal L}$ vanishes, the metric is, locally,
%\begin{eqnarray*}
%ds^2 = - N dv^2 + 2 \left ( dy + V d\x \right ) dv  + \frac{dZ^2}{V} 
%+ V d\x^2,
%\end{eqnarray*}
%where $\xi =\partial_v$, 
%$N = c - \Lambda (y^2 - Z^2)$ and $V= k - \Lambda Z^2$. 
%\item[(A.ii)] If $\mbox{\rm{dim}} ({\cal L}) = 1$, then, locally, the metric is reducible
%$g = h_{-} + h_{+}$
%where $h_{-}$ (resp. $h_{+}$) is a two-dimensional
%Lorentzian (resp. Riemannian) metric
%of constant curvature $\Lambda$ and $\xi$
%is any of the Killing  
%vectors of $h_{-}$ or $h_{+}$ as long as it satisfies $d{\bm \xi} \neq 0$
%everywhere.
%\end{itemize}

\item[(B)]
If $Q \F^2 -4 \Lambda \neq 0$ at one point then it does not vanish anywhere on $\M$ and
%the Ernst one-form 
$\chi_{\alpha} \defi 2 \xi^{\beta} \F_{\beta\alpha}$ is exact
$\chi_{\alpha} = \nabla_{\alpha} \tilde\chi$ with
$$
\tilde \chi = 6 \F^2 \frac{ Q \F^2 + 2 \Lambda}{ (Q \F^2 - 4 
\Lambda)^{2}}.
$$
Moreover $\exists \, b_1,b_2,c \in \mathbb{R}$
such  that
\begin{equation}
36 Q ( \F^2)^{\frac{5}{2}} = \left ( -b_2 + i b_1  \right )
( Q \F^2 - 4 \Lambda )^3 \label{idenb1b2}
\end{equation}
and $N - \mbox{Re}(\tilde \chi) =  c$, where
$N \defi - g (\xi,\xi)$ and the vector field
\begin{eqnarray*}
\varsigma^{\mu} = \frac{4}{|Q\F^2-4\Lambda|^2}\xi^\sigma\overline\F_\sigma{}^\rho\F_{\rho}{}^\mu+\mbox{{\rm Re}}\left(\frac{\F^2}{(Q\F^2-4\Lambda)^2}Ê\right)\xi^\mu
%8 | Q \F^2 - 4 \Lambda |^{-2} \xi^{\alpha} \ov{\F}{}_{\alpha}^{\,\,\,\beta} \F_{\beta\mu} 
%- \left ( \F^2 (Q \F^2 - 4 \Lambda)^{-2}
%+  \ov{\F^2}(\ov{Q}  \ov{\F^2} - 4 \Lambda)^{-2}  \right ) \xi_{\mu}
\end{eqnarray*}
is Killing and commutes with $\xi$. Let ${\cal L} = 
%\mbox{\rm{span}} \{ \xi,q \}= 
\mbox{\rm{span}} \{\xi,\varsigma\}$.
\begin{itemize}
\item[(B.i)] If $\mbox{dim}({\cal L}) =2$, then $\exists k \in \mathbb{R}$
such that, away from totally geodesic, 
codimension-two, non-degenerate surfaces where a Killing vector in ${\cal L}$
vanishes, the metric is, locally
\begin{align*}
ds^2 & = - N \left ( dv - Z^2 d\x \right )^2
+ 2 \left ( dy + V d\x \right ) \left ( dv - Z^2 d\x \right ) 
+ (y^2 + Z^2) \left ( \frac{dZ^2}{V} + V d\x^2 \right ) \\
  \xi & = \partial_v, \quad N = c - \frac{\Lambda}{3} \left (y^2 - Z^2\right )
- \frac{b_1 y + b_2 Z}{y^2 + Z^2}, 
\quad V =k +b_2 Z - c Z^2 - \frac{\Lambda}{3} Z^4.
\end{align*}
\item[(B.ii)] If $\mbox{dim}({\cal L}) =1$, then $\exists \epsilon
\in \{ -1,1\}$ and $\kappa, \v,n \in \mathbb{R}$ 
such that, away from fixed points of $\xi$,
the metric is, locally, 
\begin{align}
ds^2 & = \left \{ \begin{array}{lr}
       - V \left ( dv - \hat{\bm{w}} \right )^2 + 2 d\y \left ( dv - 
\bm{\hat{w}} \right ) + (\v^2+ \y^2) h_{+} & \mbox{if } \hspace{2mm} 
\epsilon =1 \nonumber \\
     V^{-1}  d\y^2+ V \left ( dv - \hat{\bm{w}} \right )^2 
+ (\v^2+ \y^2) h_{-} & \mbox{if } \hspace{2mm}\epsilon = -1
\end{array} \right . \nonumber \\
V & = (\v^2 + \y^2)^{-1}   \left ( 
- \frac{\Lambda}{3} \left ( \y^4  + 6 \v^2 \y^2 - 3 \v^4 \right ) - \kappa
\left ( \v^2 - \y^2 \right ) + n \y  \right ), \nonumber \\
\xi & = \partial_v, \quad \quad
\hat{d} \hat{\bm{w}} = 2 \v \bm{\eta_{\epsilon}} \label{dd}
\end{align}
where $h_{\epsilon}$ is a metric of constant curvature
$\kappa$, signature $\{\epsilon,1\}$ and volume form
$\bm{\eta_{\epsilon}}$.
\end{itemize}
\end{enumerate}
\end{theorem}

\vs

Concerning the result in case (A), note that any Killing vector on the sphere satisfies $d{\bm \xi}  \neq 0$
everywhere, but this is not the case for the rest of two-dimensional
spaces of constant curvature (Riemannian or Lorentzian),
so the condition $d{\bm \xi} \neq 0$ is a restriction for the
choice of Killing in those cases. Nevertheless, in any 2-dimensional
space of constant curvature
there is an open subset of the Killing algebra satisfying
$d{\bm \xi} \neq 0$ everywhere.

\vs

\noindent {\it Proof:}  
From (\ref{DefJ}) we note that $Q \F^2 -4 \Lambda =0$
is equivalent to $J=0$. Since $J=0$
at one point if and only if $J=0$ everywhere, the first claim follows.
So we are in case (A) (in the notation above). From Proposition
\ref{prop:2ndKilling}
we have that $q$ is a Killing vector which commutes with $\xi$, hence
the Lie algebra they generate has dimension one or
two. The former case corresponds, by definition, to the so-called special cases
with $J=0$. So either
we are in case (A) of Theorem \ref{non-orthogonal} with
$Z=0$ (because of (\ref{specialA}) with $\epsilon =1$) or in case
 (A) of Theorem \ref{th:orthogonal}. The one-form
$\hat{\bm{w}}$ is closed, hence locally exact
and a redefinition of $v$ and $\x$, respectively, makes it identically
zero. Define $h_{-} = - N(y) dv^2 + 2 dy dv$, $h_+ = \hat{h}$ in 
the case $\xi + q =0$ and $h_{-} = \hat{H}$, $h_{+} = N^{-1}(Z)dZ^2 + N(Z) d
\x^2$ in the case $\xi - q=0$, where $N(\zeta)  \defi c - \Lambda \zeta^2$.
It is immediate to check that, in all cases, $h_{-}$ 
and $h_{+}$ are of constant curvature $\Lambda$.

In the other case ---when the Lie algebra is two-dimensional---, from 
Theorem \ref{non-orthogonal}, Proposition \ref{prop:bifurcation} and
Lemma \ref{Solving2sim} $\exists \, c,k \in \mathbb{R}$ 
such that, away from 
a collection of totally geodesic, nowhere null,
codimension-two surfaces where a Killing vector in
$\mbox{\rm{span}} \{ \xi,q \}$ vanishes, the metric is, locally,
\begin{eqnarray*}
ds^2 = - N dv^2 + 2 \left ( dy + V d\x \right ) dv  + \frac{dZ^2}{V} 
+ V d\x^2,
\end{eqnarray*}
where $\xi =\partial_v$, 
$N = c - \Lambda (y^2 - Z^2)$ and $V= k - \Lambda Z^2$. 
Performing the trivial change $d\tilde x =d(x+v)$ this can be rewritten as
$$
ds^2 = - \tilde N(y) dv^2 + 2 dy dv  + \frac{dZ^2}{V} + V d\tilde x^2,
$$
where now $\tilde N(y) =N +V = c+k-\Lambda y^{2}$ and the claim follows again.

To conclude the first part of the theorem, 
it remains to show that (\ref{WeylRepetido})
holds for any Killing vector $\xi$ of $h_{\pm}$
for which $d \bm{\xi}$ is nowhere zero. Let $\bm{\eta}_{\pm}$
be the volume form of $h_{\pm}$, so we have $d \bm{\xi} = 
2 S \bm{\eta}_{\pm}$ with $S$ nowhere zero. 
The curvature tensor of a product manifold
inherits a product structure. As a consequence, it turns out that 
the tensor relation (\ref{WeylRepetido})
contains just one independent equation, which can
be computed to be $Q S^2 =  \pm \Lambda$ showing 
the proportionality claimed  for such $\xi$.

Concerning the case $Q \F^2 - 4 \Lambda \neq 0$, by
computing $\tilde\chi$ as defined in the theorem and using Lemma
\ref{Functions_of_P}
it follows $\tilde\chi = \chi -c$, from which the statement
$N - \mbox{Re} (\tilde\chi) = c$ follows immediately. The expression
involving $-b_2 + i b_1 = i b$ also follows by direct computation from
Lemma \ref{Functions_of_P}. The Killing vector is simply
$\varsigma $ in case (B) of Proposition \ref{prop:2ndKilling} 
with $A=0$ and $C= 1/36$, after rewriting $P$ in terms of $\F^2$
as
\begin{eqnarray*}
P = \frac{6 i \sqrt{\F^2}}{Q \F^2 - 4\Lambda}.
\end{eqnarray*}
The form of the metric  in case (B.i) 
is a direct consequence of Theorem \ref{non-orthogonal}, 
Proposition \ref{prop:bifurcation} and
Lemma \ref{Solving2sim}. Regarding case (B.ii), this is achieved from case (B) in Theorem \ref{th:orthogonal} (for $\epsilon =-1$) and from case (B) with $Z=$const.\  in Theorem \ref{non-orthogonal} (for $\epsilon =1$; recall that then (\ref{specialB}) implies $b_2=2cZ+4\Lambda Z^3/3$) together with the
name substitutions 
$\{ y \rightarrow \v, Z \rightarrow \y, 
c \rightarrow 2 \Lambda \v^2 - \kappa, b_2 \rightarrow n, N \rightarrow -V\}$
in the case $\epsilon =-1$ and 
$\{ Z \rightarrow \v, y \rightarrow \y, 
c \rightarrow - 2 \Lambda \v^2 + \kappa,
b_1 \rightarrow -n, N \rightarrow V\}$ in the case $\epsilon = 1$.
$\hfill \qed$.

\vs

We can proceed to the identification of the metrics and consequently obtain
characterizations thereof.
\begin{theorem}
\label{Plebanski}
\begin{enumerate}
\item The metric in case (A) corresponds to the uncharged Bertotti-Robinson metric, and in particular it is the Nariai metric for $\Lambda >0$.
\item At points where  $\nabla_{\alpha} y \nabla^{\alpha} y \neq 0$,
the metric in case (B.i) in Theorem \ref{full} is
locally isometric to the uncharged Plebanski metric.
\item At points where the Killing vector $\xi$ is not null the metric in case (B.ii) with $\epsilon =1$ is locally isometric to a spacetime determined by Cahen and Defrise \cite{CD} having a 4-dimensional group of isometries acting on timelike (respectively spacelike) hypersurfaces on the regions where $\xi$ is spacelike (resp.\ timelike). In particular, for $\kappa >0$ they are locally isometric to the Taub-NUT-(A)de Sitter spacetime.
\item The metric in case (B.ii) with $\epsilon =-1$ are the $\Lambda$-vacuum type-D solutions of Kundt's class. They happen to have both principal null directions expansion- and twist-free, and a 4-dimensional group of isometries acting on spacelike hypersurfaces.
\end{enumerate}
\end{theorem}

{\it Proof:} 
Point 1 follows directly because the generalized Bertotti-Robinson is the general metric product of two 2-dimensional metric of constant curvature, and due to the fact that Nariai's solution corresponds to the case with a positive curvature Riemannian part.

To prove point 2, in the (B.i) metric we easily find
$\nabla_{\alpha} y \nabla^{\alpha} y = (y^2 + Z^2 ) W(y)$
with
\begin{eqnarray*}
W(y)= k - b_1 y + c y^2 - \frac{\Lambda}{3} y^4.
\end{eqnarray*}
On any domain  where $\nabla_{\alpha} y \nabla^{\alpha} y \neq 0$
we can define new coordinates $\tau$ and $\sigma$
by the coordinate change
\begin{eqnarray*}
d\tau = dv - \frac{y^2}{W(y)} dy, \quad
d \sigma = d \x + \frac{1}{W(y)} dy.
\end{eqnarray*}
After elementary manipulations the line-element in (B.i) becomes
\begin{align*}
ds^2 = \frac{1}{y^2+Z^2} \left ( V(Z) \left ( d \tau + y^2 
d \sigma \right )^2 - W(y) \left ( d \tau - Z^2 d \sigma \right )^2 \right )
+ (y^2 + Z^2) \left ( \frac{dZ^2}{V(Z)} + \frac{dy^2}{W(y)}\right ).
\end{align*}
which is exactly the form of the Plebanski metric as
given in formula (21.16) in \cite{Exact} with the electric
and magnetic charges set to zero.

Similarly, to prove point 3, on any domain where $g(\xi,\xi)=-V(z)\neq 0$, one can define a new coordinate $t$ by means of $dt = dv-dz/V(z)$ so that the line-element in case (B.ii) with $\epsilon =1$ becomes
$$
ds^{2}=-V(dt-\bm{\hat{w}})^{2}+\frac{1}{V} dz^{2} +(k^{2}+z^{2})h_{+} \, .
$$
This form corresponds to the $\Lambda$-vacuum solutions presented in formulae [(2.6) plus (4.33)] or [(2.8) plus (4.43)] of \cite{CD}, the former for the case with $V<0$ having a 4-dimensional group of motions acting on timelike hypersurfaces, the latter for $V>0$ and has a 4-dimensional group of motions acting on spacelike hypersurfaces (see alternatively [(13.9) plus (13.48)] without charge in \cite{Exact}). The Taub-NUT case with a cosmological constant corresponds to the case where $h_{+}$ has positive constant curvature.

Finally, to prove point 4, and as the case (B.ii) with $\epsilon =-1$ corresponds to the special case with $g(\xi,k)=g(\xi,\ell)=0$, it is easily checked from (\ref{nablak}) and (\ref{nablaell}) that both principal null directions have zero expansion and twist. Hence, they belong to the $\Lambda$-vacuum Petrov type-D Kundt class. They actually exhaust this class, as can be seen by simply comparing the metric with that in the discussion in section 7.2.1 of \cite{GP2} ---alternatively, with expression (31.61) in \cite{Exact} without the electromagnetic charges plus the $\Lambda$ term mentioned in p.484 of that reference.
These solutions were presented also in \cite{CD} as [(2.11) plus (4.61)] and thus they possess a 4-dimensional group of motions acting on spacelike hypersurfaces everywhere.
$\hfill \qed$.

We can finally obtain the characterization of the Kerr-NUT-(A)dS metric
as defined in the Introduction. 
\begin{theorem}
\label{Kerr-NUT-AdS}
With the same hypothesis as in Theorem  \ref{full}, assume that there is
one point $p \in \M$ where $Q\F^2 - 4 \Lambda \neq 0$ (and hence everywhere), \textcolor{blue}{and that at least at one point $\xi$ is not orthogonal to the plane spanned by the two real null eigenvectors of $\F_{\alpha\beta}$}.
Let $b_1,b_2,c$ be as in Theorem \ref{full} and
define $P := \frac{6 i \sqrt{\F^2}}{Q \F^2 - 4 \Lambda}$ and
$y,Z : \M \rightarrow \mathbb{R}$ by $P := y + i Z$. Then the function
\begin{eqnarray}
k = (y^2 + Z^2) \nabla_{\alpha}Z \nabla^{\alpha} Z - b_2 
\textcolor{blue}{Z}+ c Z^2 +
\frac{\Lambda}{3} Z^{\textcolor{blue}{4}} \label{defk}
\end{eqnarray}
is constant on $\M$. If the polynomial 
\begin{eqnarray}
V(\zeta) := k + b_2 \, \zeta - c \, \zeta^2 - \frac{\Lambda}{3} \zeta^4
\label{defV}
\end{eqnarray}
admits two zeros $\zeta_0 \leq \zeta_1$ such that the factor
polynomial $\hat{V} \equiv V (\zeta - \zeta_0)^{-1} (\zeta_1 - \zeta)^{-1}$
is strictly positive on $[\zeta_0, \zeta_1]$
and $Z$ takes values in $[\zeta_0,\zeta_1]$
then the spacetime $(\M,g)$
is locally isometric to the
Kerr-NUT-(A)dS with parameters $\{\Lambda, m, a, l \}$ where
\begin{eqnarray*}
m = \frac{b_1}{2 v_0 \sqrt{v_0}}, \quad
\quad a = \frac{\zeta_1 - \zeta_0}{2 \sqrt{v_0}},
\quad  \quad l = 
\frac{\zeta_1 + \zeta_0}{2 \sqrt{v_0}}
\end{eqnarray*}
and $v_0 := \hat{V}(\frac{\zeta_0+ \zeta_1}{2})$.
\end{theorem}

{\bf Remark.} It is immediate to check that
the characterization  Theorem \ref{intro} stated
in the introduction is a combination
of Theorems \ref{full} and \ref{Kerr-NUT-AdS}.

\vs

{\it Proof:} If $Z$ is constant on $\M$, the constancy
of $k$ is trivial. Otherwise, its
constancy has been shown in the proof of Lemma \ref{Solving2sim} 
with $u$ as given in Theorem \ref{non-orthogonal}.
The  conditions $V(\zeta_0)= V(\zeta_1)=0$ and 
the definitions of $l,a, v_0$ imply
\begin{eqnarray}
b_2= 2 v_0 \sqrt{v_0} l  ( 1 + \frac{\Lambda}{3} ( a^2 - 4 l^2 )  ), \quad
k= v_0^2 (a^2 -l^2 )(1 - l^2 \Lambda  ), \quad
c = v_0  ( 1 - \frac{\Lambda}{3}( a^2 + 6 l^2 ) ),
\label{rename}
\end{eqnarray}
and the polynomial $\hat{V}$ reads $\hat{V} = v_0 
( 1 + \frac{\Lambda}{3} (\frac{\zeta}{\sqrt{v_0}} - 
l) (\frac{\zeta}{\sqrt{v_0}}  + 3 l) ).$
Assume  first that $\zeta_1 - \zeta_0 >0$ (i.e. $a \neq 0$).
We first show that there is no point where $dZ$ and $\nabla_{\mu}
\nabla^{\mu} Z$ vanish simultaneously. Indeed, if this were
the case, then $Z$ would be constant on $\M$ (see
the Remark after Proposition \ref{prop:bifurcation}),
which, given the definitions
(\ref{defk}), (\ref{defV}) and the hypothesis $\mbox{Image} (Z) \subset
[\zeta_0,\zeta_1]$, can only happen if $Z = \zeta_i$ ($i=0,1$) 
everywhere. In either case, expression (\ref{specialB}) with
$\epsilon =1$ becomes
$2 a  v_0 \sqrt{v_0}  \hat{V}(\zeta_i) =0$, against hypotheses. Thus, 
we are in case (B.i) of Theorem \ref{full}. Moreover,
if $Z$ takes somewhere the values $\zeta_0$ or $\zeta_1$, it does so 
(by Proposition \ref{prop:bifurcation})
on a totally geodesic 2-dimensional timelike surface where a Killing vector
vanishes. 
Away from these points we can perform the coordinate
changes 
\begin{eqnarray*}
Z = \sqrt{v_0} \left ( l + a \cos \theta \right ) , \quad \quad 
y = \sqrt{v_0} r, \quad \quad
v =  \frac{u}{\sqrt{v_0}} + v_0 (a + l)^2 x, \quad \quad 
x = - \frac{\phi}{v_0 \sqrt{v_0} a},
\end{eqnarray*}
which brings the metric of case (B.i) in 
Theorem \ref{full} into (\ref{KNUTaDS}).

If on the other hand $\zeta_0 = \zeta_1$ (i.e. $a=0$),
then $Z = \zeta_0 = \sqrt{v_0} \, l$ 
is constant and we are in case (B.ii)
of Theorem \ref{full} with $\epsilon = 1$. Given 
the relation $c =  - 2 \Lambda v_0 l^2 + \kappa$
(see the proof of Theorem \ref{full}) 
it follows $\kappa = v_0 > 0 $ and hence $h_+ = (1/v_0) \gamma$
where $\gamma$ is the standard metric of the sphere. 
Since $\beta = Z = \sqrt{v_0} l$, equation 
(\ref{dd}) becomes $\hat{d} \hat{\bm{w}} = (2 l /\sqrt{v_0}) \bm{\eta_{\gamma}}$, 
where $\bm{\eta_{\gamma}}$ the volume form of $\gamma$. The general solution 
to this equations is, in standard spherical coordinates $\{ \theta,\phi\}$,
$\hat{\bm{w}} = (4 l/\sqrt{v_0}) \sin^2 (\theta/2)
d\phi + df_0$, where $f_0$ is any smooth function
on the sphere. With the change of variables 
$y = \sqrt{v_0} r$ and $v  = f_0 + u/\sqrt{v_0}$ and the redefinition
$n = - 2 v_0 \sqrt{v_0} m$, the metric in case (B.ii) ($\epsilon =1$)
of Theorem \ref{full} becomes
(\ref{KNUTaDS}) with $a=0$. $\hfill \qed$

\subsection{Comments on Theorem \ref{Kerr-NUT-AdS} and remarks concerning other related metrics}

As we have seen, the spacetime characterization of the Kerr-NUT-(A)dS metric given in 
Theorem \ref{Kerr-NUT-AdS} requires some extra conditions, apart from our main assumption (\ref{WeylRepetido}). Even though those conditions may look somehow artificial, they are actually required and generalize similar previous conditions used to characterize other simpler metrics. For instance, Theorem \ref{Kerr-NUT-AdS} includes and extends the characterization
of the Kerr metric \textcolor{blue}{---and of the Kerr-NUT metric \cite{Mars3}--- as originally given in \cite{Mars1,Mars2} and recently complemented and corrected in \cite{MSnull} with the necessary, but only implicitly assumed in \cite{Mars1,Mars2}, condition that $\xi$ must be non-orthogonal to the 2-plane spanned by the real null eigenvectors of $\F_{\alpha\beta}$ somewhere. Observe that this omitted assumption is actually redundant whenever $\xi$ is timelike somewhere. Taking this assumption into account we note that}, in the $\Lambda =0$ case, the condition that the polynomial
(\ref{defV}) has two zeros $\zeta_0$, $\zeta_1$ and the factor polynomial
$\hat{V}$ is positive between them is simply equivalent to $c >0$, which
is the characterization of the Kerr-NUT metric as given
in Theorem 3 in \cite{Mars3}. The Kerr subcase corresponds
to the NUT parameter $l$ being zero and $m \neq 0$,
or equivalently $b_1 \neq 0$ and $b_2=0$
(see (\ref{rename})). Since in the case $\Lambda =0$
we have from (\ref{idenb1b2}):
$$\Lambda =0 \, \, \, \Longrightarrow \, \, \, \F^2 Q^4 (-b_2 + i b_1 )^2= 36^2
$$
the conditions $b_1 \neq 0=b_2$ can be equivalently written by demanding that
the constant $\F^2 Q^4$ is real and negative, which agrees with the statement
of Theorem 1 in \cite{Mars2}.

\vs

The question arises about metrics for which the polynomial (\ref{defV}) does not satisfy the conditions of Theorem \ref{Kerr-NUT-AdS} ---while still satisfying the main assumption (\ref{WeylRepetido}). We do not intend to give an exhaustive
discussion here. 
We just mention two known classes solutions which do
satisfy our main characterization hypothesis (\ref{WeylRepetido}) but not
the conditions on the zeroes of the polynomial $V(\zeta)$ in
Theorem \ref{Kerr-NUT-AdS}. More precisely, in 
\cite{KMV} two families of metrics  with negative cosmological constant 
are constructed. First, the family of metrics (10) in \cite{KMV}
is obtained by analytic continuation of the
parameters of the Kerr-de Sitter metric with vanishing NUT parameter $l=0$. Second, the family (43) in \cite{KMV} is built 
by direct choice of parameters in the Plebanski solution. Thereby, these
two classes of $\Lambda$-vacuum solutions also satisfy our main algebraic
constraint (\ref{WeylRepetido}), thus they must belong to the general family 
in Theorem \ref{full}. It turns out that the two classes have $b_2=0$, and are then distinguished as follows (keeping $\Lambda <0$)
\begin{itemize}
\item
The first class, (10) in \cite{KMV}, corresponds to the situation when furthermore 
the constant $k$ defined as in Theorem \ref{Kerr-NUT-AdS} is negative.
%(in order to set the NUT parameter equal to zero) 
In this case, 
the polynomial $V(\zeta)$ has precisely
two real zeros $- \zeta_1 = \zeta_2 > 0$ and $V(\zeta) \geq 0$ on the set ${\cal D}_0
:=(-\infty,
- \zeta_2] \cup [\zeta_2, \infty)$. The solution is recovered with $Z : {\cal M} \rightarrow \mathbb{R}$ taking values in 
either of the two connected components of ${\cal D}_0$.
%, the class ofmetrics (10) in \cite{KMV} is recovered.
Not surprisingly, the analytic continuation of the parameters performed in \cite{KMV}
changes drastically the structure of the domains where the polynomial
$V(\zeta)$ is positive (in this case, it
transforms a compact interval into
a non-compact closed interval). It is precisely
this domain structure that plays a crucial role in  determining
the local form of the metric. This is why the statement
of Theorem \ref{Kerr-NUT-AdS} characterizing the Kerr-NUT-(A) de Sitter
metric requires fixing one such domain structure.

\item The second class, (43) in \cite{KMV}, has $V(\zeta)>0$ everywhere, so that $V(\zeta)$ has no zeros, and moreover $c=0$ which implies also that $k>0$. Thus, $V(\zeta)$ has a unique minimum at the origin. This class of metrics, have been considered in the context of the AdS/CFT correspondence as viable models of an holographic description for the Quark-Gluon plasma \cite{Mc1,Mc2}.
\end{itemize}

Actually, a generalization of the last case has also been recently analyzed in \cite{McT}. The metric (21) in \cite{McT} is simply the case (keeping $\Lambda <0$, $b_2=0$) with no zeros for $V(\zeta)$, which has $k>0$ necessarily and is defined by $c^2 +4k\Lambda/3 <0$, plus the condition $c>0$, ergo $V(\zeta)$ has a local maximum at the origin.

It seems therefore advisable to perform a complete analysis of the different qualitative possibilities for the function (\ref{defV}) according to its roots and local extrema, see \cite{GP} pp. 309 and following,
where several results along these lines can be found.

\vs

\section*{Acknowledgements}
The authors wish to thank Walter Simon for comments on the manuscript
and for providing useful references.
MM acknowledges financial support under the project FIS2012-30926 (MICINN).
JMMS  is supported by grants
FIS2010-15492 (MICINN), GIU12/15 (Gobierno Vasco) and UFI 11/55 (UPV/EHU). The authors acknowledge support from project P09-FQM-4496 (J. Andaluc\'{\i}a---FEDER).

\vs

\section*{Appendix}
\label{Appen}
In this Appendix we collect some of the formulas that are used in the main text.
\begin{align}
& \F_{\mu}{}^{\rho}t_{\nu\rho}=\frac{1}{8}\F^{2}\overline\F_{\mu\nu} ,\\
&\xi^\mu t_{\mu\nu}=-\frac{1}{4} \eta_{\nu}, \hspace{1cm}
\chi^\mu t_{\mu\nu}=\frac{1}{8}\F^2 \overline\chi_\nu , \hspace{1cm}  \eta^\mu t_{\mu\nu}=-\frac{1}{16}\F^2 \overline\F^{2}\xi_\nu ,\\
&\xi^\mu \xi^\nu t_{\mu\nu}=\frac{1}{8}\chi_\rho \overline\chi^\rho , \hspace{1cm} \xi^\mu \chi^\nu t_{\mu\nu}=0, \hspace{1cm} \xi^{\mu}\eta^\nu t_{\mu\nu}=\frac{N}{16}\F^2 \overline\F^{2}, \hspace{1cm} \chi^\mu \eta^\nu t_{\mu\nu}=0, \\
& \chi^\mu \chi^\nu t_{\mu\nu}=\frac{1}{8}\F^{2}\chi_\rho \overline\chi^\rho , \hspace{1cm} \chi^{\mu}\overline\chi^{\nu}t_{\mu\nu}= -\frac{N}{8} \F^{2}\overline\F^{2}, \hspace{1cm} \eta^{\mu}\eta^\nu t_{\mu\nu}=\frac{1}{32}\F^2 \overline\F^{2}\chi_\rho \overline\chi^\rho \, .
\end{align}

For arbitrary (complex or real) 2-forms $A_{\mu\nu}$ and $B_{\mu\nu}$, the Lanczos identity for double 2-forms ---expression (A15) in p.2839 of \cite{S}--- applied to $A_{\alpha\beta}B_{\mu\nu}$ implies the 4-dimensional identity
$$
A_{\alpha\beta}B_{\mu\nu}+B^{\star}_{\alpha\beta}A^{\star}_{\mu\nu}=L_{\mu\alpha}g_{\beta\nu}-L_{\nu\alpha}g_{\beta\mu}-L_{\mu\beta}g_{\alpha\nu}+L_{\nu\beta}g_{\alpha\mu}
$$
where $L_{\mu\nu}\defi B_{\mu}{}^{\rho}A_{\nu\rho}-(1/4) g_{\mu\nu}\,  B^{\rho\sigma}A_{\rho\sigma}$. 
In particular, one has the identity
\be
\F_{\alpha\beta}\overline\F_{\mu\nu}+\overline\F_{\alpha\beta}\F_{\mu\nu}=2\left(t_{\alpha\mu}g_{\beta\nu}- t_{\alpha\nu}g_{\beta\mu}-t_{\beta\mu}g_{\alpha\nu}+t_{\beta\nu}g_{\alpha\mu}\right) . \label{identity}
\ee
Different contractions here lead to
\begin{align}
& 4t_{\beta[\mu}\xi_{\nu]}=g_{\beta[\mu}\eta_{\nu]}-\frac{1}{2}\chi_{\beta}\overline\F_{\mu\nu}-\frac{1}{2}\overline\chi_{\beta}\F_{\mu\nu}, \label{twedgexi}\\
& 2Nt_{\beta\mu}=\frac{1}{4} \chi_\rho \overline\chi^\rho g_{\beta\mu}+\xi_{(\beta}\eta_{\mu)}+\frac{1}{8}\chi_{(\beta}\overline\chi_{\mu)}, \label{Nt}\\
& 4t_{\beta[\mu}\chi_{\nu]}=\frac{1}{2}\F^{2}\left(\xi_{\beta}\overline\F_{\mu\nu}-g_{\beta[\mu}\overline\chi_{\nu]}\right)-\eta_{\beta}\F_{\mu\nu}, \label{twedgechi}\\
& 2 (\chi_\rho \overline\chi^\rho) t_{\beta\mu}= \eta_{\beta}\eta_{\mu}+\frac{1}{4}\left(\overline\F^{2} \chi_{\beta}\chi_{\mu}+\F^{2} \overline\chi_{\beta}\overline\chi_{\mu}\right)+\frac{1}{4}\F^{2}\overline\F^{2}\left(\xi_{\beta}\xi_{\mu} +Ng_{\beta\mu}\right) \label{chichit}\\
& 4t_{\beta[\mu}\eta_{\nu]}=\frac{1}{4}\F^{2}\overline\F^{2}g_{\beta[\mu}\xi_{\nu]}+\frac{1}{4}\overline\F^{2}\chi_{\beta}\F_{\mu\nu}+\frac{1}{4}\F^{2}\overline\chi_{\beta}\overline\F_{\mu\nu} \label{twedgeeta}
\end{align}

Contracting (\ref{twedgexi},\ref{twedgechi},\ref{twedgeeta}) with $k_{\pm}$ one gets
\begin{align}
& -2R\overline R\,  \bm{k}_{\pm}\wedge \bm{\xi}=\bm{k}_{\pm}\wedge\bm{\eta} \mp 2g(\xi,k_{\pm})\left(R\overline{\bm{\F}} +\overline R \bm{\F}\right), \label{kwedgexi}\\
& -R\overline R\, \bm{k}_{\pm}\wedge \bm{\chi}= R^{2}\, \bm{k}_{\pm}\wedge\overline{\bm{\chi}}-2R\, g(\xi,k_{\pm})\left(R\overline{\bm{\F}} +\overline R \bm{\F}\right), \label{kwedgechi}\\
& -R\overline R\, \bm{k}_{\pm}\wedge \bm{\eta} =2(R\overline R)^{2}  \bm{k}_{\pm}\wedge \bm{\xi}\mp 2R\overline R\, g(\xi,k_{\pm})\left(R\overline{\bm{\F}} +\overline R \bm{\F}\right).
\label{kwedgeeta}
\end{align}
Observe that (\ref{kwedgeeta}) is simply (\ref{kwedgexi}) multiplied by $R\overline R$.

Using again the identity (\ref{identity}) we derive
\be
4t_{\beta[\mu}k^{\pm}_{\nu]}=2R\overline R g_{\beta[\mu}k^{\pm}_{\nu]}\pm k^{\pm}_{\beta}\left(R\overline\F_{\mu\nu}+\overline R \F_{\mu\nu}\right) \label{twedgek}
\ee
and contracting here the $+$-equation with $k_-^\beta$
\be
R\overline\F_{\mu\nu}+\overline R \F_{\mu\nu}=-2R\overline R (k^+_{\mu}k^-_{\nu}-k^-_{\mu}k^+_{\nu})
\ee
or with $k^\mu_{-}$
\be
-2g(k_{+},k_{-})t_{\mu\nu}=R\overline R \left[2k^{+}_{\mu}k^{-}_{\nu}+2k^{-}_{\mu}k^{+}_{\nu}-g(k_{+},k_{-})g_{\mu\nu} \right] \label{tk+k-}
\ee
and both equations with $\xi$
\be
2g(k_{\pm},\xi)t_{\mu\nu}=R\overline R\,  g(k_{\pm},\xi)g_{\mu\nu}-\frac{1}{2}k^{\pm}_{\nu}\left(\eta_{\mu}+2R\overline R \xi_{\mu}\right)\mp \frac{1}{2}k^{\pm}_{\mu}\left(R\overline\chi_{\nu} +\overline R \chi_{\nu}\right) .\label{kxit}
\ee

Equations (\ref{kwedgexi}) and (\ref{kwedgechi}) ---or directly (\ref{kxit})--- imply necessarily
$$
\bm k_{\pm} \wedge \left(\bm\eta +2R\overline R \bm{\xi} \mp R\overline{\bm{\chi}}\mp \overline R \bm{\chi} \right)=0
$$
whose general solution reads
\be
\bm\eta +2R\overline R \bm{\xi} \mp R\overline{\bm{\chi}}\mp \overline R \bm{\chi} =A_{\pm}\bm{k}_{\pm}
\label{ks}
\ee
for real $A_{\pm}$ with
$$
g(k_+,k_-)A_{\pm}=8R\overline R g(k_\mp ,\xi), \hspace{1cm} A_{\pm}g(k_{\pm},\xi) =-2NR\overline R-\frac{1}{2} g(\chi ,\overline\chi).
$$
The above is valid for both the regular and singular cases.  
In this paper we are only concerned with the regular case, and thus $\F^{2}=-4R^{2 }\neq 0$. We can always normalize the null eigenvectors such that $g(k_{+},k_{-})=-1$ so that (\ref{tk+k-}) becomes
\be
2t_{\mu\nu}=R\overline R \left[g_{\mu\nu} +2k^{+}_{\mu}k^{-}_{\nu}+2k^{-}_{\mu}k^{+}_{\nu}\right] .\label{tk+k-2}
\ee


\begin{thebibliography}{9999}
\bibitem{Ionescu2}
S. Alexakis, A.D. Ionescu, S. Klainerman, ``Uniqueness of smooth
stationary black holes in vacuum: small perturbations of the Kerr spaces''
(2010) Commun. Math. Phys. {\bf 299} 89-127.
\bibitem{Ionescu3}
S. Alexakis, A.D. Ionescu, S. Klainerman, ``Hawking's
local rigidity theorem without analyticity'' (2010)
Geom. Funct. Anal. {\bf 20} 845-869.
\bibitem{Ionescu5}
S. Alexakis, A. D. Ionescu, S. Klainerman, ``Rigidity of stationary black
holes with small angular momentum on the horizon'' (2013)
arXiv:1304.0487 [gr-qc]
\bibitem{JuanThomas_a}
T. B\"ackdahl, J.A. Valiente Kroon, 
``On the construction of a
geometric invariant measuring the deviation from Kerr data'' (2010)
Ann. Henri Poincar\'e {\bf 11} 1225-1271.
\bibitem{CD} M. Cahen, L. Defrise, ``Lorentzian 4 dimensional manifolds with `local isotropy«'' , (1968)  Commun. Math. Phys. {\bf 11}  56-76.
\bibitem{Chen}
W. Chen, H. L\"u, C.N. Pope, ``General
Kerr-NUT-AdS metrics in all dimensions`` (2006) 
Class. Quant. Grav. {\bf 23} 5323-5340.
\bibitem{Chong}
Z-W. Chong, G.W. Gibbons, H. L\"u. C.N. Pope,
``Separability and Killing Tensors in Kerr-Taub-Nut-De Sitter
Metrics in Higher Dimensions'' (2005) Phys. Lett. B {\bf 609} 124-132.
\bibitem{LecturesDafermos}
M. Dafermos, I. Rodnianski, ``Lectures on black holes and linear waves'',
in Evolution equations, Clay Mathematics Proceedings, Vol. 17.
Amer. Math. Soc., Providence, RI, (2013) 97-205 (online
at http://www.arxiv.org/abs/0811.0354).
\bibitem{Debever}
R. Debever, N. Kamran, R.G. McLenaghan, ``Exhaustive
integration and a single expression for the general solution of the type D
vacuum and electrovac field equations with cosmological constant for a
nonsingular aligned Maxwell field'' (1984) J. Math. Phys. {\bf 25} 1955-1972.
\bibitem{Dias}
O.J.C. Dias, G.T. Horowitz, D. Marolf, J.E. Santos, ``On the
nonlinear stability of asymptotically anti-de Sitter solutions'', (2012)
Class. Quant. Grav. {\bf 29} 230519.
\bibitem{Dyatlov}
S. Dyatlov, ``Exponential energy decay for Kerr-de Sitter
black holes beyond event horizons'' (2011) Math. Res. Lett. {\bf 18} 1023-1035.
\bibitem{FerrandoSaez}
J. J. Ferrando,  J. A. S\'aez, 
``On the invariant symmetries of the D-metrics'',  (2007)
J. Math. Phys. {\bf 48} 102504. 
\bibitem{Garcia}
A. Garc\'{\i}a-D\'{\i}az, ``Electrovac type D solutions with cosmological
constant'' (1984) J. Math. Phys. {\bf 25} 1951-1954.
\bibitem{ParradoJose}
A. Garc\'{\i}a-Parrado G\'omez-Lobo, J.M.M. Senovilla, ``A set of invariant
quality factors measuring the deviation from the Kerr metric'' (2013)
Gen. Rel. Grav. {\bf 45}  1095-1127
\bibitem{ParradoJuan}
A. Garc\'{\i}a-Parrado G\'omez-Lobo, J.A. Valiente Kroon
``Kerr Initial data'' (2008)
Class. Quant. Grav. {\bf 25} 205018. 
\bibitem{GP2} J.B. Griffiths, J. Podolsk\'y, ``A new look at the Pelba\'nski-Demia\'nski family of solutions'', (2006)
Int. J. Mod. Phys. D {\bf 15} 335-370. 
\bibitem{GP3} 
J.B. Griffiths, J. Podolsk\'y, ``A
note on the parameters of the Kerr-NUT-(anti-)de Sitter spacetime'',
(2007) Class. Quant. Grav. {\bf 24} 1687-1689.
\bibitem{GP} J.B. Griffiths, J. Podolsk\'{y}, {\it Exact Space-Times in Eintein's General Relativity}, Cambridge Monographs on Mathematical Physics (Cambridge Univ. Press, Cambridge, 2009)
\bibitem{Gromoll} D. Gromoll, G. Walschap, ``Metric foliations and
curvature'', Progress in Mathematics {\bf 268} (Birkh\"auser Verlag AG, Basel,
2009).
\bibitem{Holzegel2010}
G. Holzegel,  ``On the massive wave equation
on slowly rotating 
Kerr-AdS spacetimes'', (2010) 
Commun. Math. Phys. {\bf 294} 169-197.
\bibitem{HolzegelSmulevici2011}
G. Holzegel, J. Smulevici, ``Decay Properties of Klein-Gordon Fields on
Kerr-AdS spacetimes'', (2011) arXiv:1110.6794 [gr-qc].
\bibitem{Houri_a}
T. Houri, T. Oota, Y. Yasui, 
``Closed conformal Killing-Yano tensor and Kerr-NUT-de Sitter
space-time uniqueness''
(2007) Phys. Lett. B {\bf 656} 214-216.
\bibitem{Houri_b}
T. Houri, T. Oota, Y. Yasui, ``Closed conformal Killing-Yano
tensor and the uniqueness of generalized Kerr-NUT-de Sitter spacetime'',
(2009) Clas. Quant. Grav. {\bf 26}  045015.
\bibitem{Ionescu1}
A.D. Ionescu, S. Klainerman, ``On the uniqueness of smooth,
stationary black holes in vacuum'', (2009) Invent. Mathem. {\bf 175} 
35-102.
\bibitem{Ionescu4}
A.D. Ionescu, S. Klainerman, ``On the local extension of Killing
vector-fields in Ricci flat manifolds'', (2013)
J. Amer. Math. Soc. {\bf 26} 563-593. 
\bibitem{Israel1970} W. Israel, ``Differential forms in general relativity'',
(1970) Commun. of the Dublin Institute for Advanced Studies, Series A, {\bf 19}
1-100.
\bibitem{KMV}
D. Klemm, V. Moretti, L. Vanzo,
``Rotating topological black holes'', (1998)  Phys. Rev. D {\bf 57}  6127-6137. Erratum, (1999) {\it ibid.} {\bf 60} 109902. 
\bibitem{Krtous}
P. Krtous, V.P. Frolov, D. Kubiznak,
``Hidden symmetries of higher-dimensional black holes and uniqueness
of the Kerr-NUT-(A)dS spacetime'', (2008) Phys. Rev. D {\bf 78}
064022.
\bibitem{Frolov}
D. Kubiznak, V.P. Frolov, ``The hidden
symmetry of higher dimensional Kerr-NUT-AdS spacetimes''
(2007) Class. Quant. Grav. {\bf 24} F1-F6.
\bibitem{Mars1} M. Mars, ``A spacetime characterization of the
Kerr metric'', (1999)  Class. Quantum. Grav. {\bf 16}  2507-2523.
\bibitem{Mars2}  M. Mars, ``Uniqueness properties of the Kerr metric'',
(2000) Class. Quantum Grav. {\bf 17} 3353-3373.
\bibitem{Mars3}  M. Mars, ``Spacetime Ehlers group:
transformation law for the Weyl tensor'', (2001)
Clas. Quant. Grav. {\bf 18} 719-738.
\bibitem{Mars4}  M. Mars, ``Wahlquist-Newman solution'', (2001)
Phys. Rev. D  {\bf 63} 064022.
\bibitem{MarsReiris} M. Mars, M. Reiris, ``Global and uniqueness properties of stationary and static spacetimes with outer trapped surfaces'', 
Commun. Math. Phys. (2013), (DOI) 10.1007/s00220-013-1739-5.
\bibitem{MSnull} \textcolor{blue}{M. Mars and J.M.M. Senovilla, ``Spacetime characterizations of $\Lambda$-vacuum metrics with a null Killing 2-form'' (2016) Class. Quantum Grav. {\bf 33} 195004}
\bibitem{Mc1} B. McInnes, ``Fragile black holes and an angular momentum cutoff in peripheral heavy ion collision'' (2012) Nucl. Phys. B {\bf 861} 236.
\bibitem{Mc2} B. McInnes, ``Universality of the holographic angular momentum cutoff'' (2012) Nucl. Phys. B {\bf 864} 722.
\bibitem{McT} B. McInnes and E. Teo, ``Generalised planar black holes and the holography of hydrodynamic shear'' (2013) arXiv:1309.2054.
\bibitem{ONeil} B. O'Neill, ``The fundamental equations
of a submersion'', (1966) Michigan Math. J. {\bf 13} 459-469. 
\bibitem{Papapetrou1966} A. Papapetrou, ``Champs gravitationnels 
stationnaires \`a sym\'etrie axiale'', (1966)
Annales de l'institut Henri Poincar\'e (A) Physique th\'eorique
{\bf 4}, 83-105.
\bibitem{Perjes}
Z. Perj\'es, ``An improved characterization of the Kerr metric'', 
(1985) KFKI-1984-115 preprint, in {\it Quantum Gravity 3},
Ed. M.A. Markov, World Scientific Publishing Co., Singapore.
\bibitem{S} J.M.M. Senovilla, ``Super-energy tensors'', (2000) 
Class. Quantum Grav.  {\bf 17}, 2799-2841.
\bibitem{Simon} W. Simon,  ``Characterizations of the
Kerr metric'', (1984) Gen. Rel.Grav. {\bf 16 } 465-476.
\bibitem{Simon2}
W. Simon, ``Nuts have no hair'', (1995) Class. Quantum Grav. {\bf 12}
L125-L130.
\bibitem{Exact}
H. Stephani, D. Kramer, M. MacCallum, C. Hoenselaers, E. Herlt, \emph{Exact
  solutions of {E}instein's field equations}, 2nd ed., Cambridge Monographs
  on Mathematical Physics (Cambridge University Press, Cambridge, 2003).
\bibitem{Wong} W.W-Y. Wong, ``A
space-time characterization of the Kerr-Newman metric'',
(2009) Ann. Henri Poincar\'e {\bf 10} 453-484.
\end{thebibliography}
\end{document}